\documentclass[aps, prd, groupedaddress, preprint, tightenlines,  eqsecnum, nofootinbib, showpacs]{revtex4}
\def\ma{a}   
\def\mb{b}   
\def\mc{c}   
\def\pd{d}
\def\pe{e}
\def\pf{f}

\def\K{K}   

\def\fDelta{f}  

\def\mtria{M_1^2}  
\def\mtrib{M_2^2}  
\def\mtric{M_3^2}  
\def\calR{R}   
\def\calM{M}   

\usepackage{epsfig}
\usepackage{setspace}
\usepackage{graphicx}
\usepackage{leqno}
\usepackage{cancel}
\usepackage{amsmath}
\usepackage{amsfonts}
\usepackage{amssymb}
\usepackage{enumerate}

\usepackage{listings}
\usepackage{mathtools}

\usepackage{axodraw, float, slashed, graphicx, amssymb, amsmath}
\usepackage[usenames, dvipsnames]{xcolor}
\usepackage{booktabs}

\usepackage[margin=2.2cm, a4paper, includefoot]{geometry}

\def\spa#1.#2{\left\langle#1\,#2\right\rangle}
\def\spb#1.#2{\left[#1\,#2\right]}
\def\eps{\epsilon}
\def\la{\langle}
\def\ra{\rangle}
\def\Mloop{M^{\oneloop}}

\def\NeqEight{\Neq8}
\def\NeqSix{\Neq6}
\def\NeqFour{{\Neq4}}

\newcommand{\oneloop}{\text{1-loop}}

\newcommand{\tree}{\text{tree}}
\newcommand{\Neq}[1]{\mathcal{N} = #1}

\newcommand\trunc{\text{trunc}}

\DeclareMathOperator{\Ord}{\mathcal{O}}

\newcommand\figref[1]{fig.~\ref{#1}}
\def\e{\epsilon}

\def\be{\begin{equation}}
\def\ee{\end{equation}}
 
\everydisplay{\medmuskip=0mu\relax}
\begin{document}

\hfill\today 

\title{The ${\cal N}$=4 Supergravity NMHV six-point one-loop amplitude}

\author{David~C.~Dunbar and Warren~B.~Perkins}

\affiliation{
College of Science, \\
Swansea University, \\
Swansea, SA2 8PP, UK\\
\today
}

\begin{abstract}
We construct the
six-point NMHV one-loop amplitude in $\NeqFour$ supergravity using unitarity and 
recursion.   The use of recursion requires the introduction of rational descendants of the cut-constructible pieces of the amplitude and the computation of
the non-standard factorisation terms arising from the loop integrals.

\end{abstract}

\maketitle

\section{Introduction}

Despite perturbative Quantum Gravity being a mature subject~\cite{Qgrav}, it is a very challenging area computationally.  Although great strides have been made in 
computing tree amplitudes~\cite{GravTrees}, there remain a very limited number of  loop calculations available to study.  
For  the four and five point amplitudes all 
one-loop graviton scattering amplitudes have now been calculated~\cite{GravOneLoop,Dunbar:1994bn,Bern:1993wt,Bern:1998sv,Dunbar:2010fy,Dunbar:2011xw,Alston:2012xd}.  For Maximal supergravity great progress has also be made at multi-loop level for the four point 
amplitude~\cite{Bern:1998ug,Bern:2007hh,Bern:2009kd}.  
These computations are by necessity amplitudes which are ``Maximally-Helicity-Violating'' (MHV).  MHV amplitudes are very special and have
many features not shared by non-MHV amplitudes. 
In this article we compute the six graviton  ``Next-to-MHV'' (NMHV)
scattering amplitude for $\NeqFour$ supergravity. (The first NMHV amplitude appears at six-points.) 
The six-graviton scattering amplitude has been computed for
$\NeqEight$ and $\NeqSix$ supergravity. 

This amplitude has considerable algebraic complexity
relative to the more supersymmetric cases including the appearance of rational terms.  
We construct the $M^\NeqFour_6(\ma^-,\mb^-,\mc^-,\pd^+,\pe^+,\pf^+)$\footnote{We
use the normalisation for the full physical amplitudes
${\cal  M}^\tree=i(\kappa/2)^{n-2} M^\tree$,
\break
${\cal  M}^\oneloop=i(2\pi )^{-2} (\kappa/2)^{n} M^\oneloop$.}  amplitude using unitarity and recursion augmented by  limited off-shell behaviour. 

One-loop amplitudes in a massless theory can be expressed as~\cite{Bern:1994cg}
\begin{equation}
\Mloop_n=\sum_{i\in \cal C}\, a_i\, I_4^{i} +\sum_{j\in \cal D}\,
b_{j}\, I_3^{j} +\sum_{k\in \cal E}\, c_{k} \, I_2^{k} +R_n +O(\eps),
\label{DecompBasis}
\end{equation}
where the $I_r^i$ are $r$-point scalar integral functions and the $a_i$
etc. are rational coefficients. $R_n$ is a purely rational term.
The box, triangle and bubble coefficients can be 
determined via unitarity methods~~\cite{Cutkosky:1960sp,
Bern:1994zx,Bern:1994cg,%
Bern:1997sc, BrittoUnitarity}
using four dimensional on-shell tree amplitudes. These contributions are termed cut-constructible. 
Progress has been made both via the two-particle cuts~\cite{Bern:1994zx,Bern:1994cg,Dunbar:2009ax}
and using 
generalisations of unitarity~\cite{Bern:1997sc} where, for example, 
triple~\cite{Bidder:2005ri,Darren,BjerrumBohr:2007vu,Mastrolia:2006ki} and quadruple cuts~\cite{BrittoUnitarity} 
are utilised to identify the triangle and box coefficients respectively.
The remaining purely rational term, $R_n$, may in principle
be obtained using unitarity but this requires a knowledge of $4-2\epsilon$ dimensional tree amplitudes~\cite{DUnitarity}. 
In this paper a recursive approach is adopted that generates the rational term from  four dimensional
amplitudes.

An important technique for computing tree amplitudes is Britto-Cachazo-Feng-Witten (BCFW)~\cite{Britto:2005fq} recursion which applies
complex analysis to amplitudes. By shifting the momenta\footnote{As usual,  a null momentum is represented as a
pair of two component spinors $p^\mu =\sigma^\mu_{\alpha\dot\alpha}
\lambda^{\alpha}\bar\lambda^{\dot\alpha}$. For real momenta
$\lambda=\pm\bar\lambda^*$ but for complex momenta $\lambda$ and
$\bar\lambda$ are independent~\cite{Witten:2003nn}.  
We are using a spinor helicity formalism with the usual
spinor products  $\spa{a}.{b}=\epsilon_{\alpha\beta}
\lambda_a^\alpha \lambda_b^{\beta}$  and 
 $\spb{a}.{b}=-\epsilon_{\dot\alpha\dot\beta} \bar\lambda_a^{\dot\alpha} \bar\lambda_b^{\dot\beta}$.} 
\begin{equation}
\bar\lambda_a\to \bar\lambda_a -z \bar\lambda_d \quad ,\qquad \lambda_d \to \lambda_d + z\lambda_a
\; , 
\end{equation}
the resultant amplitude $M_n(z)$ may be computed via Cauchy's theorem.
Loop amplitudes, as functions of complexified momentum, contain both poles and cuts so BCFW does not immediately apply to these. However by defining
\begin{align}
  {R_n}={M}_n-{M}_n^{cc}  \; , 
\end{align}
where ${M}_n^{cc}$ is the cut-constructible part of the amplitude, we can compute the purely rational $R_n$ 
from a knowledge of its singularities. 
$R_n$ has  singularities corresponding to the poles of the amplitude  but also induced singularities arising because $M_n^{cc}$ has singularities which are
not present in the amplitude and which must be cancelled by equal and opposite singularities in $R_n$. We refer to these contributions as the  
{\it rational descendants} of ${M}^{cc}_n$.

The particle content  of the $\Neq4$ graviton and matter
multiplets are shown in table~\ref{multiple:table}. 
\begin{table}
  \begin{tabular}{c|ccccccccc}
    \toprule
    Helicity  & \;\;$2$\;        &  \;\;$3/2$ \;   & 
    \;\;$1$ \; &   \;\;$1/2$ \;  &  \;\;$0$ \;  &  \;$-1/2$\;  & \;$-1$\;  & \;$-3/2$\;  & \;$-2$\; \cr \hline
    graviton   & $1$  & $4$ & $6$ & $4$ & $2$   & $4$ & $6$ & $4$ & $1$    \\ 
    matter  & $0$ & $0$ & $1$ & $4$ & $6$ & $4$ & $1$ & $0$ & $0$  \cr\bottomrule
  \end{tabular}
  \caption{Particle content of the multiplets}
\label{multiple:table}\end{table}
For convenience, we will calculate the one-loop amplitude using the
$\Neq4$ matter multiplet, which is related to the amplitude with  the graviton multiplet circulating in the loop by
\begin{equation}
M_n^{\NeqFour,\text{graviton}} = M_n^{\NeqEight}
-4M_n^{\NeqSix,\text{matter}}+2M_n^{\NeqFour,\text{matter}}.
\end{equation} 
The $\NeqEight$ and $\NeqSix$ components are considerably simpler and given in~\cite{BjerrumBohr:2005xx,Dunbar:2010fy}.
In particular $R_n=0$ for these two components.

\section{Cut-Constructible Pieces}
\def\rg{r_\gamma}
\def\e{\epsilon}
\def\Li{{\rm Li}}

The cut-constructable pieces consist of box-functions,  triangle functions and bubble integral functions.  The analytic form of these depends upon 
how many of the external legs have non-null momentum: these are referred to as massive legs although non-null is more correct. For the one-mass box 
the fourth leg is conventionally the massive leg and the integral function depends upon 
$S\equiv (k_1+k_2)^2$, $T\equiv (k_2+k_3)^2$ and $M_4^2\equiv K_4^2$. 
For the two-mass boxes there are two types: ``two-mass-easy'' ($2me$) where legs $2$ and $4$ are massive 
and ``two-mass-hard'' ($2mh$) where legs $3$ and $4$ are massive.   For six-point amplitudes the three and four mass boxes do not appear and for the
NMHV $M_6^{\NeqFour:{\rm matter}}$ amplitude there are no  two-mass-easy boxes. 

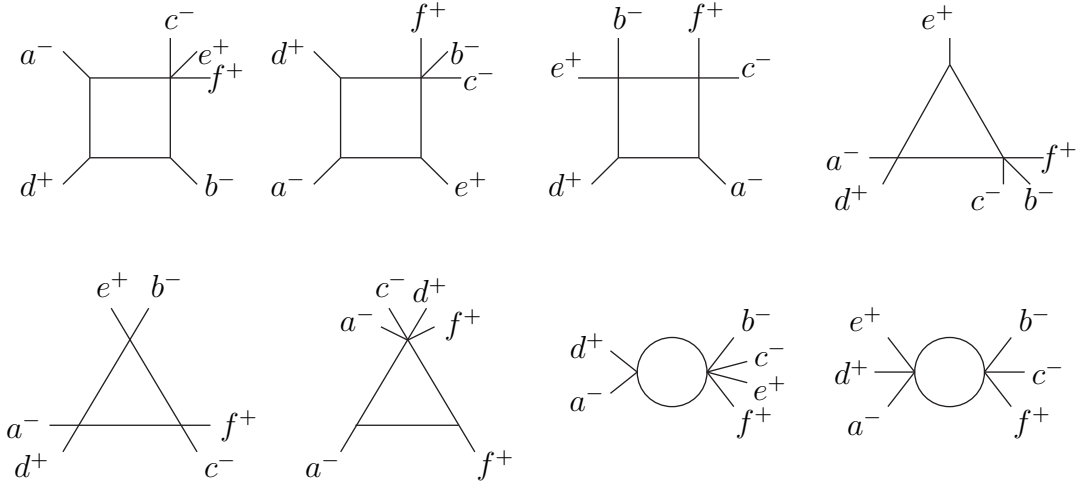
\begin{figure}[h]
    \begin{picture}(150,100)(-60,-30)  
      \Line(0,0)(0,30)
      \Line(0,30)(30,30) 
      \Line(30,30)(30,0)
      \Line(30,0)(0,0) 
      \Line(0, 30)(-10, 40)
      \Line(0,0)(-10, -10) 
      \Line(30,0)(40,-10) 
      \Line(30,30)(45,30)
      \Line(30,30)(40,40)
      \Line(30,30)(30,45)
      \Text(-13, 40)[r]{${a}^-$}
      \Text(-13,-10)[r]{$d^+$}
      \Text(55,-10)[r]{$b^-$}
      \Text(40, 53)[r]{$c^-$}
      \Text(53,40)[r]{$e^+$}
      \Text(58,30)[r]{$f^+$}
    \end{picture}
    \begin{picture}(100,100)(0,-30)  
      \Line(0,0)(0,30)
      \Line(0,30)(30,30) 
      \Line(30,30)(30,0)
      \Line(30,0)(0,0) 
      \Line(0, 30)(-10, 40)
      \Line(0,0)(-10, -10) 
      \Line(30,0)(40,-10) 
      \Line(30,30)(45,30)
      \Line(30,30)(40,40)
      \Line(30,30)(30,45)
      \Text(-13, 40)[r]{${d}^+$}
      \Text(-13,-10)[r]{$a^-$}
      \Text(55,-10)[r]{$e^+$}
      \Text(40, 53)[r]{$f^+$}
      \Text(53,40)[r]{$b^-$}
      \Text(58,30)[r]{$c^-$}
    \end{picture}
   \begin{picture}(100,100)(0,-30)  
      \Line(0,0)(0,30)
      \Line(0,30)(30,30) 
      \Line(30,30)(30,0)
      \Line(30,0)(0,0) 
      \Line(0,30)(-15,30)
      \Line(0,30)(0,45)
      \Line(0,0)(-10, -10) 
      \Line(30,0)(40,-10) 
      \Line(30,30)(45,30)
      \Line(30,30)(30,45)
      \Text(-13, 35)[r]{$e^+$}
      \Text(-13,-10)[r]{$d^+$}
      \Text(55,-10)[r]{$a^-$}
      \Text(40, 53)[r]{$f^+$}
      \Text(10,53)[r]{$b^-$}
      \Text(58,35)[r]{$c^-$}
    \end{picture}
    \begin{picture}(100,100)(0,-30)  
      \Line(-5,-10)(20,35)
      \Line(-10,0)(55,0) 
      \Line(40,0)(20,35)
      \Line(20,35)(20,45)  
      \Line(40,0)(40,-10)  
      \Line(40,0)(50,-10)
      \Text(-13, 0)[r]{${a}^-$}
      \Text(-10,-15)[r]{$d^+$}
      \Text(40,-15)[r]{$c^-$}
      \Text(60,-15)[r]{$b^-$}     
      \Text(22, 53)[r]{$e^+$} 
      \Text(68,0)[r]{$f^+$}
    \end{picture}
      \begin{picture}(160,100)(-60,-30)  
      \Line(-5,-10)(27,44)
      \Line(-10,0)(50,0) 
      \Line(45,-10)(13,44)    
      \Text(-13, 0)[r]{${a}^-$}
      \Text(-10,-15)[r]{$d^+$}
      \Text(60,-15)[r]{$c^-$}
      \Text(20, 53)[r]{$e^+$}
      \Text(40,53)[r]{$b^-$}
      \Text(68,0)[r]{$f^+$}
    \end{picture}
      \begin{picture}(100,100)(0,-30)  
      \Line(-5,-10)(27,44)
      \Line(0.5,0)(39,0) 
      \Line(45,-10)(13,44)    
      \Line(20,32)(10,37)
      \Line(20,32)(30,37)
      \Text(-5,-15)[r]{$a^-$}
      \Text(60,-15)[r]{$f^+$}      
      \Text(8, 40)[r]{${a}^-$}
      \Text(20, 53)[r]{$c^-$}
      \Text(35, 51)[r]{$d^+$}
      \Text(48,38)[r]{$f^+$}
    \end{picture}
   \begin{picture}(100,100)(0,-30)  
      \BCirc(15,20){13}
      \Line(2,20)(-8,28)
      \Line(2,20)(-8,12)
      \Line(28,20)(38,33)
      \Line(28,20)(43,24)
      \Line(28,20)(43,16)
      \Line(28,20)(38,7)
      \Text(-10, 30)[r]{${d}^+$}
      \Text(-10,10)[r]{$a^-$}
      \Text(53,40)[r]{$b^-$}
      \Text(58,27)[r]{$c^-$}
      \Text(58,14)[r]{$e^+$}
      \Text(53,1)[r]{$f^+$}      
    \end{picture}
    \begin{picture}(100,100)(0,-30)  
      \BCirc(15,20){13}
      \Line(2,20)(-8,33)
      \Line(2,20)(-13,20)
      \Line(2,20)(-8,7)
      \Line(28,20)(38,33)
      \Line(28,20)(43,20)
      \Line(28,20)(38,7)
      \Text(-10,40)[r]{$e^+$}
      \Text(-15, 20)[r]{${d}^+$}
      \Text(-10,1)[r]{$a^-$}
      \Text(53,40)[r]{$b^-$}
      \Text(58,20)[r]{$c^-$}
      \Text(53,1)[r]{$f^+$}      
    \end{picture}
    \caption{The  Integral functions appearing in the six-point NMHV amplitude}
    \label{thezoo}
\end{figure}

The various box, triangle and bubble contributions present in the six-point NMHV are shown in \figref{thezoo} together with the labelling of helicities which yield a non-zero coefficient. Permuting  
the positive and negative helicity legs separately gives eighteen   
one-mass boxes, thirty-six two-mass hard boxes,  nine  2:4 bubbles, nine 3:3 bubbles and six three-mass triangles.

\vskip 2.0 truecm 

\subsection{IR consistency and Choice of Integral Function Basis}
For one-loop amplitudes Infra-Red (IR) consistency imposes a system of constraints
on the rational coefficients of the integral functions.  For the
matter multiplets~\cite{Dunbar:1995ed} there are in fact no IR singular
terms in the amplitude, so the singular terms in the individual
integral functions cancel.  This gives enough information to fix the
coefficients of the one- and two-mass triangles in terms of the box
coefficients. The three-mass triangle is IR finite, so its coefficient
is not determined by these constraints.  It is convenient to combine
the boxes and triangles in such a way that these infinities are
manifestly absent.  There are several ways to do
this~\cite{Bern:1994zx,BBDP,Bidder:2005ri,Britto:2005ha,Dunbar:2008zz}, here we choose to
work with truncated box functions,
\begin{equation}
I_4^{\trunc}= I_4 - \sum_i \alpha_i { \frac{(-K_i^2)^{-\eps}}{\eps^2} }\,,
\end{equation}
where the $\alpha_i$ and $K^2_i$ are chosen to make $I_4^{\trunc}$ IR
finite.  This effectively incorporates the one- and two-mass triangle contributions
into the box contributions.  Using these truncated
boxes the amplitudes can be written as
\begin{equation}
\Mloop_n=\sum_{i\in \cal C}\, a_i\, I_4^{i,\trunc} +\sum_{j\in\cal D'}
b_{j}\, I_3^{j,\text{3-mass}} +\sum_{k\in \cal E}\, c_{k} \, I_2^{k}
+R_n\, .
\label{DecompBasisTruncated}
\end{equation} 
There is one remaining IR consistency constraint: $\sum c_k=0$.

The one-mass and two-mass-hard truncated integral functions are 
\begin{align}
I_{4}^{1{\rm m,trunc}}
& = -{2\rg\over ST}\biggl(
  \Li_2\left(1-{ K_4^2 \over S }\right)
 \ + \ \Li_2\left(1-{K_4^2 \over T}\right)
 \ +{1\over 2} \ln^2\left({ -S  \over -T}\right)\
+\ {\pi^2\over6}\ \biggr)\,,
\notag \\
 I_{4}^{\rm 2{\rm m}h,trunc}
&= -{2\rg\over ST}\biggl(
  \ -\ {1\over 2} \ln\left({ K_3^2/S }\right)\ln\left({ K_4^2/S }\right)
  \ +\ {1\over 2} \ln^2\left({ S/T }\right)
\notag \\ & \hskip 60 pt
  \ +\ \Li_2\left(1-{ K_3^2\over T}\right)
  \ +\ \Li_2\left(1-{ K_4^2\over T }\right) \biggr)\,,
\end{align}
where $S=(k_1+k_2)^2$ and $T=(k_1+K_4)^2$.

\subsection{Boxes}

The six-pt NMHV amplitude contains both one-mass and two-mass hard boxes. The  coefficients of these boxes are readily obtained 
using quadruple cuts~\cite{BrittoUnitarity}.

The coefficient of the first one-mass box in \figref{thezoo}  is
\begin{align}
a^{\rm 1m[\NeqFour]}_{\ma,\pd,\mb,\{\mc,\pe,\pf\}} &=  
{1\over 2}
{ 
s_{\ma\pd}^2s_{\mb\pd}^2   [\pd|\K|\mc\ra^4  [\mb|\K|\mc\ra [\ma|\K|\mc \ra
\over 
\spb{\ma}.{\mb}^4 \spa{\mc}.{\pe} \K^2 [\mb|\K|\pf\ra 
[\ma|\K|\pf\ra[\mb|\K|\pe\ra}
\times
\biggl({[\mc\mb][\pe\pf]   \over \la \pe\pf \ra }   
-     {[\pe\mb][\mc\pf] [\ma|\K|\mc\ra \over [\ma|\K|\pe\rangle
\la \mc\pf\ra
}
\biggr)\,,
\label{onemassbox:eq}
\end{align}
where $\K=\K_{\mc\pe\pf}$. 
The other one-mass box coefficients can be obtained from this by conjugation and relabelling. 
The coefficient of the two-mass-hard box is 
\begin{align}
a^{\rm 2mh[\NeqFour]}_{\ma,\pd,\{\mb,\pe\},\{\mc,\pf\}} &= - 
{1\over 2}
{ s_{\ma\pd}^2  \K_{\mc\pf\ma}^2  [\pf|K_{\mc\pf\ma}|\mb\ra^4\over  [\ma|K_{cfa}|\pd\ra^4}
{
[\mb\pe] \la \mc\pf\ra \la \mb\pd\ra   
[a|\K_{\mc\pf}|b\rangle
\over \la \mb\pe\ra [\mc\pf]
\la \pe\pd\ra  
[a|\K_{\mc\pf}|e\rangle  
}
{
  [\ma\pf]  [\pf|\K_{cfa}|\pd\ra \over [\mc\ma] [\mc|\K_{cfa}|\pd\ra   
}
\notag \\
\end{align}
and the other coefficients are obtained by relabelling. The expressions for the box coefficients have the appropriate symmetries under exchange of external legs.
Specifically the two-mass hard is invariant under the joint operation of $(a,b,c) \leftrightarrow (d,f,e)$ and conjugation. The one-mass coefficient
is invariant under $a\leftrightarrow b$ although this is not manifest (and similarly under $e\leftrightarrow f$).

\subsection{Canonical basis approach for triangle and bubble coefficients}

\begin{figure}
\includegraphics[scale=0.4]{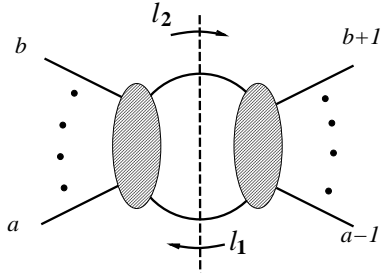}
\caption{A two-particle cut of a one-loop amplitude}
\label{cutfigure}
\end{figure}

The canonical basis approach~\cite{Dunbar:2009ax} is a systematic method to determine the coefficients of triangle and bubble integral functions from
the three and two-particle cuts.  
\noindent{The} two particle cut is shown in fig.~\ref{cutfigure} and is 
\begin{equation}
C_{2}  \equiv  i   \int d^4 \ell  \delta( \ell_1^2)\delta(\ell_2^2) 
A^{\rm tree}(-\ell_1, a, \cdots, b, \ell_2) \times A^{\rm tree}(-\ell_2,\cdots, \ell_1)\,.
\end{equation}
The product of tree amplitudes appearing in the two-particle cut can be decomposed 
in terms of  canonical forms ${\cal F}_i$, 
\begin{equation}
A^{\rm tree}(-\ell_1, \cdots, \ell_2) \times A^{\rm tree}(-\ell_2,\cdots, \ell_1)  = 
\sum e_i {\cal F}_i ({\ell_j}),
\end{equation} 
where the $e_i$ are coefficients independent of $\ell_j$.  
We then use
substitution rules to replace the ${\cal F}_i ({\ell_j})$ by the
rational functions $F_i(K)$ to obtain the coefficient of the bubble integral function as
\begin{equation}
\sum_i e_i F_i(K) \,.
\end{equation}
Similarly, we can obtain the coefficient of the triangle functions from 
the
 triple cut~\cite{Darren,BjerrumBohr:2007vu,Dunbar:2009ax} as shown in figure~\ref{threecut},
 \begin{align}
C_3\,=\,
-\int d^4 \ell \delta(\ell_0^2)  \delta(\ell_1^2) \delta(\ell_2^2) 
&  A^{\rm tree}(-\ell_0,  \cdots, \ell_1 )
\notag \\
&\hspace{-3cm}
\,\times\,
A^{\rm tree}(-\ell_1, \cdots,  \ell_2 )
\,\times\,
A^{\rm tree}(-\ell_2, \cdots,  \ell_0 )\, .
\end{align}

The product of tree amplitudes can, again, be expressed in terms of standard forms of $\ell_i$,  
\begin{equation}
A^{\rm tree}(-\ell_0, \cdots, \ell_1) \times A^{\rm tree}(-\ell_1,\cdots, \ell_2)  
\times A^{\rm tree}(-\ell_2,\cdots, \ell_0)  = 
\sum_i e_i {\cal E}_i ({\ell_j})
\end{equation}
and
substitution rules used to replace the ${\cal E}_i ({\ell_j})$ by the
functions $E_i(K_j)$ to obtain the triangle coefficient as 
\begin{equation}
\sum_i e_i E_i(K_j)\,. 
\end{equation}
\begin{center}
\begin{figure}[H]
\begin{picture}(40,100)(-240,0)
\Line(30,30)(70,40)
\Line(30,30)(70,20)
\SetWidth{2}
\Line(30,30)(60,50)
\Line(30,30)(60,10)
\SetWidth{1}
\Line(30,30)(-30,30)
\Line(-30,30)(0,75)
\Line(30,30)(0,75)
\Line(-30,30)(-70,40)
\Line(-30,30)(-70,20)
\SetWidth{2}
\Line(-30,30)(-60,50)
\Line(-30,30)(-60,10)
\SetWidth{1}
\Text(-60,30)[]{$\bullet$}
\Line(0,75)(-10,105)
\Line(0,75)(10,105)
\SetWidth{2}
\Line(0,75)(-20,95)
\Line(0,75)(20,95)
\Text(0,100)[]{$\bullet$}
\Text(57,41)[]{$\bullet$}
\Text(57,18)[]{$\bullet$}
 \SetWidth{1}
\DashCArc(45,20)(40,100,190){4}
\DashCArc(-50,20)(40,-10,80){4}
\DashCArc(00,100)(40,220,320){4}
\CCirc(30,30){8}{Black}{Purple}
\CCirc(-30,30){8}{Black}{Purple}
\CCirc(0,75){8}{Black}{Purple}
\Text(0,10)[]{$\ell_1$}
\Text(-30,65)[]{$\ell_2$}
\Text(30,65)[]{$\ell_3$}
\Text(80,30)[]{$K_3$}
\Text(-80,30)[]{$K_1$}
\Text(0,115)[]{$K_2$}
\end{picture}
\caption{Triple Cut}
    \label{threecut}
\end{figure}
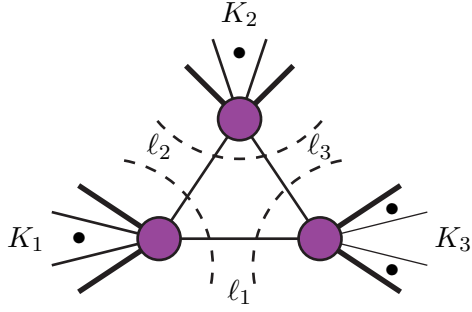
\end{center}

In general the integrands are rational functions of $\lambda(\ell_i)$ of degree $d=d_{num}-d_{denom}$. The 
simplest canonical forms have $d_{denom}=1$ or $2$.  More complex  denominators can be expressed in terms 
of the simplest forms by partial fractioning.  Terms in the integrand with $d<0$ only contribute to higher point integral functions.  The degree generally decreases with increasing
supersymmetry and for maximally supersymmetric Yang-Mills and supergravity there are no triangles. With increasing $d$ the canonical forms become increasingly complex. 

\subsection{Triangles}
Using the truncated box functions, we only need to compute the coefficient of the (IR finite) three-mass triangle:
$b^{3m}_{\{ \ma^-,\pd^+\}\{ \mb^-,\pe^+\}\{\mc^-,\pf^+\}}$. 
Using the Kawai-Lewellen-Tye (KLT) relations~\cite{Kawai:1985xq}, with a scalar circulating in the loop the cut integrand is
\begin{align}
C_3^0= 
M(\ell_1, & \ma^-,\pd^+,-\ell_2) 
M(\ell_2,\mb^-,\pe^+,-\ell_3)
M(\ell_3,\mc^-,\pf^+,-\ell_1)
\notag \\
&=
s_{\ma\pd}
A(\ell_1,\ma,\pd,\ell_2)
A(\ell_1,\pd,\ma,\ell_2)
s_{\mb\pe}
A(\ell_2,\mb,\pe,\ell_3)A(\ell_2,\pe,\mb,\ell_3)
\notag \\ &
\qquad
\times s_{\mc\pf}
A(\ell_3,\mc,\pf,\ell_1)A(\ell_3,\pf,\mc,\ell_1)
\notag \\ 
= &  s_{\ma\pd}s_{\mb\pe}s_{\mc\pf}   
C_3^{YM,0} ( \ma,\pd ;  \mb,\pe  ; \mc,\pf  ) 
C_3^{YM,0} ( \pd,\ma ;  \pe,\mb  ; \pf,\mc  ) \,.
 \end{align}
The four-point Yang-Mills amplitudes above are simultaneously MHV and ${\rm \overline{MHV}} $ amplitudes,
so there is a choice of two expressions for each. For algebraic convenience we take a mixed form for the first ordering:
\begin{align}
C_3^{YM,0} ( \ma,\pd ;  \mb,\pe  ; \mc,\pf  )& =
 {\spa{\ma}.{l_1} \spa{\ma}.{l_2}^2 \over \spa\ma.\pd\spa{\pd}.{l_2} \spa{l_1}.{l_2} }
 \times
   {\spb{\pe}.{l_3} \spb{\pe}.{l_2}^2 \over \spb\mb.\pe\spb{\mb}.{l_2} \spb{l_2}.{l_3} }
 \times
   {\spa{\mc}.{l_3} \spa{\mc}.{l_1}^2 \over \spa\mc.\pf\spa{\pf}.{l_1} \spa{l_3}.{l_1} } 
   \label{ord1}
\end{align}
and a purely MHV form for the second:
\begin{align}
C_3^{YM,0} ( \pd,\ma ;  \pe,\mb  ; \pf,\mc  )& =
  {\spa{\ma}.{l_2} \spa{\ma}.{l_1}^2 \over \spa\ma.\pd\spa{\pd}.{l_1} \spa{l_1}.{l_2} }
 \times
   {\spa{\mb}.{l_3} \spa{\mb}.{l_2}^2 \over \spa\mb.\pe\spa{\pe}.{l_2} \spa{l_2}.{l_3} }
 \times
   {\spa{\mc}.{l_1} \spa{\mc}.{l_3}^2 \over \spa\mc.\pf\spa{\pf}.{l_3} \spa{l_3}.{l_1} } \,.
   \label{ord2}
\end{align}
The contribution to the cut from a particle of helicity $h$ circulating in the loop is given by
\begin{equation}
C_3^0  \times X^{2h}\,,
\end{equation}
so that summing over the $\NeqFour$ matter multiplet  gives
\begin{equation}
C_3^0 \times \left ( X^{-2} -4X^{-1}+6 -4X+X^2 \right)
=C_3^0\times  {( 1-X)^4 \over X^2 }
=C_3^0 \times \rho^2 \,.
\end{equation}
The $X$-factor can be written in several ways. The natural choices given \eqref{ord1} and \eqref{ord2} are 
\begin{align}
 X_1=&
{ \la \ma \ell_1\ra \over \la \ma \ell_2\ra  }
{ \la \mb \ell_2\ra \over \la \mb \ell_3\ra  }
{ \la \mc \ell_3\ra \over \la \mc  \ell_1\ra  }
\;\;\; {\rm and} \;\;\; 
X_2= 
{ \la \ma l\ell_1\ra \over \la \ma \ell_2\ra  }
{ \spb{\pd}.{\ell_3} \over \spb{\pd}.{\ell_2}   }
{ \la \mc \ell_3\ra \over \la \mc \ell_1\ra  }\,,
\end{align}
with the  corresponding $\rho$-factors being
\begin{align}
\rho_1&=  {
\spa{\ell_1}.Y^2 \spa{\pe}.\mb^2 
\over 
 \la \ma \ell_1\ra \la \ma \ell_2\ra  
 \la \mb \ell_2\ra  \la \mb \ell_3\ra  
\la \mc \ell_3\ra  \la \mc \ell_1\ra  \spb{\ell_2}.{\ell_3}^2 
}
\;\;\; {\rm and} \;\;\; 
\rho_2
={\spa{\ell_1}.Y^2 
\over 
\la \ma \ell_1\ra  \la \ma \ell_2\ra  
\spb{\pe}.{\ell_3}  \spb{\pe}.{\ell_2}   
\la \mc \ell_3\ra  \la \mc \ell_1\ra  
}\,,
\end{align}
where
\begin{align}
|Y\ra= & |\ma \ra [\pe|\pf|\mc\ra   +|\mc\ra[\pe|\pd|\ma\ra \,.
\end{align}
The cut is then
\begin{equation}
C_3^{\NeqFour}=
s_{\ma\pd}s_{\mb\pe}s_{\mc\pf} 
C_3^{YM,0} ( \ma,\pd ;  \mb,\pe  ; \mc,\pf  )
\times 
C_3^{YM,0} ( \pd,\ma ;  \pe,\mb  ; \pf,\mc  )
\times \rho_1\times\rho_2\,.
\end{equation}
Using
\begin{equation}
{\spa{x}.{\ell_2} \over \spa{y}.{\ell_2} } =
{\spa{x}.{\ell_2} \spb{\ell_2}.{\ell_3}\spa{\ell_3}.{\ell_1}\over \spa{y}.{\ell_2} \spb{\ell_2}.{\ell_3}\spa{\ell_3}.{\ell_1}} =
{ \la x| K_2 K_3 | \ell_1\ra \over \la y| K_2 K_3 | \ell_1\ra  }
\;\;,\;\;
{\spa{x}.{\ell_3} \over \spa{y}.{\ell_3} } =
{\spa{x}.{\ell_3} \spb{\ell_3}.{\ell_2}\spa{\ell_2}.{\ell_1}\over \spa{y}.{\ell_3} \spb{\ell_3}.{\ell_2}\spa{\ell_2}.{\ell_1}} =
{ \la x| K_2 K_1 | \ell_1\ra \over \la y| K_2 K_1 | \ell_1\ra  }
\end{equation}
and 
\begin{equation}
\spa{\pe}.{\ell_3}   \spb{\mb}.{\ell_3} \spa{\ell_1}.{\ell_2}\spb{\ell_2}.{\ell_3} \spa{\ell_3}.{\ell_1}
=[\mb| \ell_3| \ell_1\ra  \la \pe| \ell_3  \ell_2 |\ell_1\ra
=[\mb| K_3| \ell_1\ra  \la \pe| K_2  K_1 |\ell_1\ra
\end{equation} 
we find
\begin{align}
C_3^{\NeqFour}=& {\spa\mb.\pe \spb\ma.\pd\spb\mc.\pf\over \spb{\mb}.{\pe} \spa\ma.\pd\spa\mc.\pf} 
 {\spa{1}.{\ell_1}  \la 1| K_2 K_3 | \ell_1\ra \over \spa{\pd}.{\ell_1} \la \pd | K_2 K_3 | \ell_1\ra   }
 \times
   { \spb{\pe}.{\ell_2} \spa{\mb}.{\ell_2}\over [\mb| K_3| \ell_1\ra  \la \pe| K_2  K_1 |\ell_1\ra }
 \times
   {\spa{\mc}.{\ell_1} \la \mc| K_2 K_1 | \ell_1\ra\ \over  \spa{\pf}.{\ell_1} \la \pf| K_2 K_1 | \ell_1\ra } 
  \times
   {  \spa{Y}.{\ell_1}^4   \over  \la \ell_1 | K_1K_2 |\ell_1 \ra  } 
\notag \\
=&{\spa\mb.\pe \spb\ma.\pd\spb\mc.\pf\over \spb\mb.\pe  \spa\ma.\pd\spa\mc.\pf} 
\prod_{i=1}^6  { \spa{\alpha_i}.{\ell_1} \over \spa{A_i}.{\ell_i} } 
 \times
{ \spa{Y}.{\ell_1}^2   \over  \la \ell_1 | K_1K_2 |\ell_1 \ra  } 
 \times
 \left(   { \spb{\pe}.{\ell_1} \spa{\mb}.{\ell_1} } +[\pe|K_1|\mb\ra 
\right)  \,,  
\end{align}
where
\begin{align}
\{ |\alpha_i\ra \} &=\{ |\pd\ra , |\pf\ra , K_3|\mb] , K_1K_2|\pe\ra , K_3K_2|\pd\ra, K_1K_2|\pf\ra \}\,,
\notag\\
\{ |B_i\ra \} &=\{ |\ma\ra , |\mc\ra, |Y\ra , |Y\ra , K_1K_2|\mc\ra, K_3K_2|\ma\ra \}\,.
\end{align}

Now we may {\it partial fraction} the expression, i.e. use the identity 
\begin{equation}
\prod_{i=1}^6  { \spa{\alpha_i}.{\ell_1} \over \spa{A_i}.{\ell_1} } 
= \sum_{i=1}^6  C_{A_i}  { \spa{\alpha_6}.{\ell_1} \over \spa{A_i}.{\ell_1} } 
\end{equation}
with
\begin{equation}
C_{A_i} = { \prod_{j=1}^5 \spa{\alpha_j}.{A_i} \over \prod_{j\neq i} \spa{A_j}.{A_i} }\,,
\end{equation}
so that the cut is written as a sum of canonical forms:
\begin{equation}
{\cal J}_1^0[ d; a,b,c ; \ell ]=  { \spa{\ell}.a \spa{\ell}.b \spa{\ell}.c \over \la \ell | K_1 K_2 | \ell \ra \spa{\ell}.d }
\;\;{\rm and}\;\;\
{\cal J}^1_1[ f; b,c,d, e; B ; \ell ]= { [B |\ell | b\ra \spa{\ell }.c  \spa{\ell}.d  \spa{\ell}.e \over \la \ell | K_1 K_2 | \ell \ra  \spa{\ell}.f }\,,
\end{equation}
which have corresponding canonical functions
\begin{align}
J_1^0[d;a,b,c; \{K_j\}]= & {  \la b | [K_1, K_2]  | d \ra \la c | [K_1 ,K_2 ] | a \ra +\Delta_3 \spa{b}.d\spa{c}.a 
  \over  2\Delta_3 \la d | K_1 K_2  | d \ra }
- { \spa{d}.b \spa{d}.c   \la a | [ K_1 ,K_2] | d \ra
\over 2 \la d | K_1 K_2  | d \ra^2  } 
\end{align}
and
\begin{align}
J^1_1[f; b,c,d,e; B ;\{K_j\} ]
=&\hskip6pt
\sum_{P_{12}} {  [B |a_0|b \ra \la f |[K_1,K_2]|c \ra \la d |[K_1,K_2]|e \ra \over  4 \Delta_3\la f | K_1K_2|f\ra } 
\notag \\
&
-\sum_{P_6} {  [ B |a_0|f \ra \la b |[K_1,K_2]|c \ra \la d |[K_1,K_2]|e \ra \over 24 \Delta_3\la f | K_1K_2|f\ra } 
\notag \\
&
-\sum_{P_6} {  [ B |a_0|f \ra \spa{f}.b  \spa{f}.c \la d |[K_1,K_2]|e \ra
\over 12 \la f | K_1K_2|f\ra^2 } 
\notag \\
&
-\sum_{P_4}
{  [B |a_0|b \ra\la f |[K_1,K_2]|c \ra\la f |[K_1,K_2]|d \ra\la f |[K_1,K_2]|e \ra
\over 4 \Delta_3\la f | K_1K_2|f\ra^2 } 
\notag \\
&-\sum_{P_6} {  [ B |a_0|f \ra \spa{f}.b  \spa{f}.c \la f |[K_1,K_2]|d \ra\la f |[K_1,K_2]|e \ra
\over 12 \la f | K_1K_2|f\ra^3 } \,,
\end{align}
where
\begin{align}
\Delta_3=4(K_1\cdot K_2)^2-4K_1^2K_2^2 = (K_1^2)^2+  (K_2^2)^2+  (K_3^2)^2-2 (  K_1^2K_2^2+  K_2^2K_3^2 +K_3^2K_1^2) 
\; , 
\end{align}
\begin{align}
a_0^\mu  = { -K_3^2 ( K_1^2+K_2^2-K_3^2 )\over \Delta_3  }   K_1^\mu 
+{ K_1^2 ( K_3^2+K_2^2-K_1^2 )\over \Delta_3  }  K_3^\mu 
\end{align}
and $P_n$ is the set of $n$ permutations of $\{b,c,d,e\}$ necessary to generate symmetry in these variables. 

We obtain the coefficient
\begin{align}
b_{\{a,d\},\{b,e\},\{c,f\}}=
{\spa\mb.\pe \spb\ma.\pd\spb\mc.\pf\over \spb{\mb}.{\pe} \spa\ma.\pd\spa\mc.\pf} 
\times \sum_{i=1}^6   C_{A_i} 
\Bigl( 
J_1^1[& A_i;  Y,Y, \alpha_6,\mb  ,\pe ; \{ K_i\} ] 
\notag \\ &
-[\pe |K_1| \mb \ra J_1^0 [ A_i ;Y,Y, \alpha_6; \{ K_i\} ]  \Bigr)\,.
\end{align}

The corresponding three mass triangle integral function is~\cite{Lu:1992ny,Bern:1993kr}

\begin{equation}
I_{3}^{3  m}
=\ {i\over \sqrt{-\Delta_3}}  \sum_{j=1}^3
  \Biggl[ \Li_2\biggl(-\Bigl({1+i\delta_j \over 1-i\delta_j}\Bigr)\biggr)
       -  \Li_2\biggl(-\Bigl({1-i\delta_j \over 1+i\delta_j}\Bigr)\biggr)
  \Biggr]\ + \ \Ord(\e)\,, 
\end{equation}
where
\begin{align}
\delta_1 & =
{ K_1^2-K_2^2-K_3^2
\over
\sqrt{-\Delta_3}} \; , 
\delta_2  = {K_2^2- K_1^2-K_3^2 \over
\sqrt{-\Delta_3}}  
\;\;{\rm and}\;\;\
\delta_3  = { K_3^2- K_1^2-K_2^2 \over
\sqrt{-\Delta_3}}\; .
\end{align}

\section{Bubbles}   
The final cut constructible pieces are the bubbles integral functions $I_2(K^2)$. 
There are two distinct types of bubbles coefficients depending upon whether  $K$ is a two particle momentum,
$K^2_{ij}=(k_i+k_j)^2$, or three particle, $K^2_{ijk}=(k_i+k_j+k_k)^2$. In the three particle case the cut only involves MHV
tree amplitudes whereas the two particle case requires the NMHV tree amplitude.

For the $\K^2_{\ma\pd\pe}$ bubble the cut integrand with a scalar circulating in the loop is

\begin{align}
C^{ 0}_{2}& = M_5^\tree(\ell_1,\pd^+,\pe^+,\ma^-,-\ell_2)\times M_5^\tree(\ell_2,\mb^-,\mc^-,\pf^+,-\ell_1)
\notag \\
&=
\biggl(  s_{\ell_1\pd}s_{\pe\ma}A_5^\tree(\ell_1,\pd,\pe,\ma,\ell_2)
A_5^\tree(\pd,\ell_1,\ma,\pe,\ell_2)
        +s_{\ell_1\pe}s_{\pd\ma}A_5^\tree(\ell_1,\pe,\pd,\ma,\ell_2)
        A_5^\tree(\pe,\ell_1,\ma,\pd,\ell_2)
\biggr)
\notag \\
&\times
\biggl(  s_{\ell_2\mb}s_{\mc\pf}A_5^\tree(\ell_2,\mb,\mc,\pf,\ell_1)A_5^\tree(\mb,\ell_2,\pf,\mc,\ell_1)
        +s_{\ell_2\mc}s_{\mb\pf}A_5^\tree(\ell_2,\mc,\mb,\pf,\ell_1)
        A_5^\tree(\mc,\ell_2,\pf,\mb,\ell_1)
\biggr)
\notag \\
&=   C^{p:0}_2(\ma,\mb,\mc,\pd,\pe,\pf)
+C^{p:0}_2(\ma,\mb,\mc,\pe,\pd,\pf)
+C^{p:0}_2(\ma,\mc,\mb,\pd,\pe,\pf)
+C^{p:0}_2(\ma,\mc,\mb,\pe,\pd,\pf)\,,
\end{align}
where
\begin{align}
C^{p:0}_2(\ma,\mb,\mc,\pd,\pe,\pf) 
=& 
s_{\ell_1\pd}s_{\pe\ma}  
A_5^\tree(\ell_1,\pd,\pe,\ma,\ell_2)
A_5^\tree(\pd,\ell_1,\ma,\pe,\ell_2)
\notag \\
& \times s_{\ell_2\mb}s_{\mc\pf}
A_5^\tree(\ell_2,\mb,\mc,\pf,\ell_1)
A_5^\tree(\mb,\ell_2,\pf,\mc,\ell_1)
\notag \\
=&
{\la \ma \ell_1 \ra^3\la \ma \ell_2 \ra^3 [\pf\ell_1]^3 [\pf\ell_2]^3\over \la \pd\pe\ra [\mb\mc] \K^2}
\times
\biggl( {[\ell_1\pd][\pe\ma]\over \la \pd\ell_1\ra \la \ma\pe\ra \la \pe\ell_2\ra \la \ell_2\pd\ra} 
\biggr)
\times
\biggl(  {\la \ell_2\mb\ra \la \mc\pf\ra\over [\mb\ell_2] [\pf\mc] [\mc\ell_1] [\ell_1\mb] }  
\biggr)\,.
\end{align}
As in the three-mass triangle, the contribution from a particle of helicity $h$ is $C_2^{0} \times X^{2h}$. 
Summing over the multiplet has 
the effect of multiplying by $\rho^2$. For this cut
\begin{equation}
X={\la \ma\ell_1\ra \over \la \ma\ell_2 \ra} {[f\ell_1]\over [\pf\ell_2]}
\quad \to \quad
\rho={[\pf|\K|\ma\ra^2\over \la \ma\ell_1\ra  \la \ma\ell_2 \ra [f\ell_1] [\pf\ell_2]}\,.
\end{equation}
Multiplying $C_2^{p:0}$ by $\rho^2$ gives
\begin{align}
C^{p:\NeqFour}_2(\ma,\mb,\mc,\pd,\pe,\pf) 
&=
{\la \ma \ell_1 \ra \la \ma \ell_2 \ra [\pf\ell_1] [\pf\ell_2] [\pf|\K|\ma\ra^4\over \la \pd\pe\ra [\mb\mc] \K^2}
\notag \\
&\hskip30pt
\times
\biggl( {[\ell_1\pd][\pe\ma]\over \la \pd\ell_1\ra \la \ma\pe\ra \la \pe\ell_2\ra \la \ell_2\pd\ra}
\biggr)
\biggl(  {\la \ell_2\mb\ra \la \mc\pf\ra\over [\mb\ell_2] [\pf\mc] [\mc\ell_1] [\ell_1\mb] }
\biggr)\,.
 \end{align}
 Using
 \begin{equation}
 { \spa{x}.{\ell_2} \over \spa{y}.{\ell_2} }={ \la x |K|\ell_1] \over \la y |K|\ell_1] }
 \;\;{\rm and}\;\;
 { \spb{x}.{\ell_2} \over \spb{y}.{\ell_2} }={ [ x |K|\ell_1\ra  \over [ y |K|\ell_1 \ra }
 \end{equation}
 we find 
 \begin{equation}
 C_2^{p:\NeqFour} 
 ={ [\pf|\K|\ma\ra^4\spb{\ma}.\pe \spa{\mc}.\pf \over \la \pd\pe\ra [\mb\mc] \spa{\ma}.\pe \spb{\mc}.\pf \K^2}
 \times { \prod_{i=1}^2 \spa{\alpha_i}.{\ell_1} \over \prod_{i=1}^2 \spa{A_i}.{ \ell_1} }
 \times { \prod_{j=1}^4 \spb{\beta_j}.{\ell_1} \over \prod_{j=1}^4 \spb{B_j}.{ \ell_1} }\,,
 \end{equation}
where
\begin{align}
\{ |\alpha_i\ra \}=\{ |\ma\ra, K|\pf] \}
\;\;   ,  \;
\{ |A_i\ra \}=\{|\pd\ra ,K|\mb] \}\,,
\;\;
\notag\\
\{ |\beta_j]=|\pd],|\pf], |\K|\ma\ra , |\K|\mb\ra \}
\;\;  , \; 
\{ |B_j] \}=\{ |\mb], |\mc], |\K|\pd\ra , |\K|\pe\ra \}\,.
\end{align}
Partial fractioning on $|\ell_1\ra$ and $|\ell_1]$  yields
\begin{equation}
C_2^{p:\NeqFour} 
 = { [\pf|\K|\ma\ra^4\spb{\ma}.\pe \spa{\mc}.\pf \over \la \pd\pe\ra [\mb\mc] \spa{\ma}.\pe \spb{\mc}.\pf \K^2}
 \times\sum_i \sum_j    C_{i,j} { [ \beta_4 \ell_1]\over [B_j \ell_1]}{\la \alpha_2 \ell_1 \ra \over \la  A_i \ell_1 \ra}\,,
 \label{bubbit}
 \end{equation}
where
\begin{equation}
C_{i,j}=  { \spa{\alpha_1}.{A_i} \over \prod_{k\neq i} \spa{A_k}.{A_i} }
 \times  {  \prod_{k=1}^3 \spb{\beta_k}.{B_j} \over \prod_{k\neq j} \spb{B_k}.{B_j}  }\,.
\end{equation}
The cut is now expressed in terms of the canonical form
\def\BB{B}
\def\BA{A}
\def\TB{b}
\def\TA{a}
\begin{equation}
{\cal H}^d_0={\la \TA \ell_1 \ra \over \la \BA \ell_1 \ra}
{[\TB \ell_1]\over [\BB \ell_1]}
\end{equation}
which has the  corresponding canonical function
\begin{align}
H^d_0\bigl[\BB,\BA,\TB,\TA \, ;\, \K \bigr] &=\begin{cases}
\displaystyle
{ \K^2[\TB\BB]\spa{\TA}.\BB \over [\BB|\K|\BA\ra [\BB|\K| \BB\ra}
  +{ [\TB|\K|\BA\ra \la \TA|\K|\BA] \over [\BB|\K|\BA\ra [\BA|\K|\BA\ra}
   &[\BB|\K|\BA\ra \neq 0  
\\[1em]
\displaystyle
 {[\TB\BA]\spa{\TA}.\BB \over [\BB\BA]\spa{\BA}.\BB}  & [\BB|\K|\BA\ra = 0\,. 
\end{cases}
\label{Hcanonform:eq}
\end{align}
The contribution of \eqref{bubbit} to the bubble coefficient is then
\begin{equation}
c_p(\ma,\mb,\mc,\pd,\pe,\pf)={ [\pf|\K|\ma\ra^4\spb{\ma}.\pe \spa{\mc}.\pf \over \la \pd\pe\ra [\mb\mc] \spa{\ma}.\pe \spb{\mc}.\pf \K^2}
\times \sum_i \sum_j    C_{i,j} H_0^d [ B_j,A_i,\beta_4,\alpha_2] \,,
\end{equation}
leading to the full bubble coefficient
\begin{align}
c_{\{\ma,\pd,\pe\},\{\mb,\mc,\pf\}}=& c_p(\ma,\mb,\mc,\pd,\pe,\pf)
                     +c_p(\ma,\mb,\mc,\pe,\pd,\pf)
                     +c_p(\ma,\mc,\mb,\pd,\pe,\pf)
                     +c_p(\ma,\mc,\mb,\pe,\pd,\pf)\,.
\end{align}

The coefficient of the bubble $I_2(K_{cd}^2)$ is obtained from the cut
\begin{equation}
C_2=  \sum_h  M_4^\tree(-\ell_1^h, \mc^-, \pd^+,-\ell_2^{-h} )\times
M_6^\tree(\ell_2^h, \ma^-,\mb^-,\pe^+,\pf^+,\ell_1^{-h} ).
\end{equation}
This cut can also be decomposed in canonical forms. However
the six-point NMHV tree amplitude in the cut 
is a sum of fourteen terms~\cite{BjerrumBohr:2006yw}
which leads to a lengthy expression for this bubble coefficient.  Full expressions for the bubble coefficients  of this type are given in appendix~\ref{bubapp}.
An explicit form of the bubble coefficients is available in {\tt Mathematica} format at \url{http://pyweb.swan.ac.uk/~dunbar/sixgraviton/R6.html}.

\section{Cancellation Webs and Rational Descendants}

Although we may split the amplitude into cut-constructible pieces 
and rational terms, when we examine the singularities in the amplitude there is a mixing between the two 
which is important when we reconstruct $R_n$ from its singularities. This has proven useful in the context of QCD amplitudes~\cite{Bern:2005cq,Berger:2006ci,Berger:2008sj}.
The cut-constructible pieces of the amplitude introduce a number of singularities that cannot be present in the full amplitude. These can be 
spurious singularities that occur at kinematic points where the full amplitude should be finite or higher order singularities occurring at points
where the amplitude has a simple pole. If these poles are of sufficiently high order, they generate rational descendants. 

\subsection{Higher Order Poles}

As an example of a higher order pole consider the 
behaviour of the one-mass box contribution $a^{1m[\NeqFour]}_{a,d,b,\{c,e,f\}}I_4^{a,d,b,\{c,e,f\}}$ as $[ab]\to 0$.
The box coefficient  $a^{1m[\NeqFour]}_{a,d,b,\{c,e,f\}}$ in eq.(\ref{onemassbox:eq})
contains a factor of $[ab]^{-4}$ and the expansion of the box integral function 
around $U (\equiv s_{ab})=0$ is~\cite{Dunbar:2011xw}
\begin{align}
 I_{}^{1{\rm m,trunc}}(S,T,U,M)=\fDelta_S \log(S)+\fDelta_T \log(T)+\fDelta_M \log(M^2)+\fDelta_R \,,
\end{align}
where
\begin{align}
\fDelta_S =& -2{U\over S T^2} + {U^2\over S T^3} -{2\over 3}{U^3\over S T^4} + {\cdots}
\notag \\
\fDelta_T =& -2{U\over S^2 T} + {U^2\over S^3 T} -{2\over 3}{U^3\over S^4 T} + {\cdots}
\notag \\
\fDelta_M =&  2{MU\over S^2 T^2} - {U^2M^2\over S^3 T^3} +{2\over 3}{U^3\over S^4 T^4} + {\cdots}
\notag \\
\fDelta_R^{1m}= &-{U^2\over S^2 T^2} +{1\over 3}{U^3M\over S^3 T^3} + {\cdots}
\end{align}

As the cut constructible terms contain all of the logarithms and dilogarithms in the amplitude, the logarithmic pieces  of this expansion
must combine with the bubbles to give an effective  coefficients that are linearly divergent as $U\to 0$.  We have confirmed numerically 
that the relevant cancellations between the one-mass box contributions and the bubble contributions occur. The quadratic divergence in the rational
descendant, $a^{1-m[\NeqFour]}_{a,d,b,\{c,e,f\}} \fDelta_R^{1m}$, must be cancelled by the rational piece of the amplitude. The full rational part of the 
amplitude will ultimately be obtained by recursion and one contribution to it will arise from this rational descendant if the shift {\it excites} the 
$[ab]=0$ pole.

There are corresponding $\la de\ra^{-4}$ poles in 
$a^{1-m[\NeqFour]}_{ d,a,e,\{b,c,f\}}I_4^{ d,a,e,\{b,c,f\}}$  which are obtained from those above by conjugation.

\subsection{$[a|\K_{be}|d\ra^{-4}$ Spurious Singularity}

The coefficients of the two mass hard boxes have singularities of the form $[a|\K_{be}|d\ra^{-4}$. These singularities also occur in the 
three mass triangle contributions: three powers of the pole are explicit in the leading term of the canonical form and a fourth arises
in the partial fractioning that splits the cut integrand into canonical forms.

In terms of kinematic variables, $[a|\K_{be}|d\ra \to 0$ corresponds to $UT-M_3^2M_4^2\to 0$. 
The two mass hard integral functions depend on $S$, $T$, $M_3^2$ and  $M_4^2$, while the three mass triangle depends on  $S$, $M_3^2$ and  $M_4^2$
and there is the kinematic constraint: $S+T+U=M_3^2+M_4^2$. For given $S$, $M_3^2$ and  $M_4^2$ it is possible find $T_0$ and 
$U_0$($=M_3^2+M_4^2-S-T_0$) so that $U_0T_0-M_3^2M_4^2= 0$. Close to the pole the two mass hard box integral functions can be expanded as a series
in terms of $(T-T_0)$. In this context it is convenient to work with dimensionless box integral functions $F^{2mh:trunc}$ defined by
\begin{align}
 I_4^{2mh:trunc} \equiv -2 \rg {F^{2mh:trunc} \over ST  }.
\end{align}
The expansion of this dimensionless box function is
\begin{align}
F^{2mh:trunc}(S,T,M_3^2,M_4^2) =&F^{2mh:trunc}(S,T_0,M_3^2,M_4^2)
                             +\fDelta^{2m}_{M_3}\log(M_3^2)
                             +\fDelta^{2m}_{M_4}\log(M_4^2)
                             \notag \\
                            & +\fDelta^{2m}_{S}\log(S)
                             +\fDelta^{2m}_{T}\log(T)
                             +\fDelta^{2m}_{R}\,,
\end{align}
where
\begin{align}
 \fDelta^{2m}_{M_3} &= 2\biggl[ {M_3^2\over T(T-M_3^2)}(T-T_0) +\Bigl( {(M_3^2)^2\over T^2(T-M_3^2)^2} +2{M_3^2\over T^2(T-M_3^2)} \Bigr){(T-T_0)^2\over 2}
 \notag \\
                          & \hskip30pt+\Bigl( 2{(M_3^2)^3\over T^3(T-M_3^2)^3}+6{(M_3^2)^2\over T^3(T-M_3^2)^2} +6{M_3^2\over T^3(T-M_3^2)} \Bigr){(T-T_0)^3\over 6} 
                          +\cdots\biggr]\,,
\notag \\[1em]
 \fDelta^{2m}_{M_4} &= 2\biggl[ {M_4^2\over T(T-M_4^2)}(T-T_0) +\Bigl( {(M_4^2)^2\over T^2(T-M_4^2)^2} +2{M_4^2\over T^2(T-M_4^2)} \Bigr){(T-T_0)^2\over 2}
 \notag \\
                          & \hskip30pt+\Bigl( 2{(M_4^2)^3\over T^3(T-M_4^2)^3}+6{(M_4^2)^2\over T^3(T-M_4^2)^2} +6{M_4^2\over T^3(T-M_4^2)} \Bigr){(T-T_0)^3\over 6} 
                          +\cdots\biggr] \,,
\notag \\[1em]
 \fDelta^{2m}_{S} &= 2\biggl[{(T-T_0)\over T} +{(T-T_0)^2\over 2T^2}+{(T-T_0)^3\over 3T^3} \biggr]\,,
\notag \\[1em]
  \fDelta^{2m}_{T} &=- \fDelta^{2m}_{S} -\fDelta^{2m}_{M_3}-\fDelta^{2m}_{M_4}
\end{align}
and
\begin{align}
 \fDelta^{2m}_{R} =&\Bigl( {M_3^2\over T^2(T-M_3^2)} + {M_4^2\over T^2(T-M_4^2)} +{1\over T^2}\Bigr)
 \notag \\
          &\hskip 3.0truecm \times
                                  {(M_3^2+M_4^2-S-2T)^2\bigl(UT-M_3^2M_4^2\bigr)^2\over \Delta_t^2}\Bigl( 1+6{\bigl(UT-M_3^2M_4^2\bigr)\over \Delta_t}\Bigr)
\notag \\                           
                 & +{1\over 3}\Bigl(  2{(M_3^2)^2\over (T-M_3^2)^2T^3} +5{M_3^2\over T^3(T-M_3^2)} 
                          + 2{(M_4^2)^2\over (T-M_4^2)^2T^3} +5{M_4^2\over T^3(T-M_4^2)} +{3\over T^3}\Bigr)
 \notag \\
          &\hskip 3.0truecm \times
          {(M_3^2+M_4^2-S-2T)^3\bigl(UT-M_3^2M_4^2\bigr)^3\over \Delta_t^3},
\end{align}
with
\begin{align}
 \Delta_t=S^2+M_3^2+M_4^2-2\bigl(S M_3^2+S M_4^2+M_3^2M_4^2\bigr).
\end{align}

The same pole appears in the two boxes with integral function: 
$I_4^{2mh, a,d,\{b,e\},\{c,f\}}$ and $I_4^{2mh,a,d,\{c,f\},\{b,e\}}$. 
On the pole the coefficients of these two boxes are not
equal and neither integral function vanishes. However, the sum of the dimensionless integral functions vanishes, i.e.
\begin{equation}
\left(F^{2mh}_{a,d,\{b,e\},\{c,f\}}+  F^{2mh}_{a,d,\{c,f\},\{b,e\}}\right)\biggr\vert_{[a|\K_{be}|d\ra=0}=0.
\label{fsumid}
\end{equation}

On this pole the dilogarithms in the individual  boxes and triangles survive,
but cancel between them. Setting
\begin{equation}
 F^{3mt} \equiv -i \sqrt{\Delta_3} I_{3}^{3  m}\,,
\end{equation}
the integral functions are related by
\begin{equation}
\left(F^{2mh}_{a,d,\{b,e\},\{c,f\}}-  F^{2mh}_{a,d,\{c,f\},\{b,e\}}\right)\biggr\vert_{[a|\K_{be}|d\ra=0}=
\pm F^{3mt}_{\{a,d\},\{b,e\},\{c,f\}}\biggr\vert_{[a|\K_{be}|d\ra=0}
\label{fcomboid}
\end{equation}
where the sign ambiguity is associated with the choice of sign for $\sqrt{\Delta_3}$. 
Schematically, expressing the box coefficients in terms of their sum and difference, $a_{\rm box\, 1}={\cal S}+{\cal D}$, 
$a_{\rm box\, 2}={\cal S}-{\cal D}$,
the box and triangle contributions are
\begin{equation}
\left({\cal S}+{\cal D}\right) F_{\rm box\,1} +\left( {\cal S}-{\cal D}\right) F_{\rm box\,2} +\tilde b_{tri} F^{3mt}
={\cal S}\left(F_{\rm box\,1}+F_{\rm box\,2}\right) + {\cal D}\left(F_{\rm box\,1}-F_{\rm box\,2}\right) +\tilde b_{tri} F^{3mt}\,.
\end{equation}
Expanding about $[a|\K_{be}|d\ra=0$, thanks to \eqref{fsumid} there is no dilogarithm component in the first term for any ${\cal S}$.
However, we can only use \eqref{fcomboid} if ${\cal D}$ and $\tilde b_{tri}$ are equal. In fact 
\begin{equation}
{\cal D} = \pm\tilde b_{tri} +{\cal O}( [a|\K_{be}|d\ra^0)\,.
\end{equation}
Hence the dilogarithms vanish from any term that is singular as $[a|\K_{be}|d\ra\to 0$, as required by the factorisation theorems~\cite{Bern:1995ix}.

As in the one-mass case, there are subleading singularities at cubic order multiplying logarithms. These combine with the bubble contributions
and cancel up to and including order $[a|\K_{be}|d\ra^{-1}$, leaving no  spurious singularity in the logarithms.

The rational descendant of this combination of boxes and triangle contain both $[a|\K_{be}|d\ra^{-2}$ and $[a|\K_{be}|d\ra^{-1}$ singularities.
Both of these singularities must be cancelled by the rational piece of the amplitude $R_n$. As the expansion has been performed about a singularity specified in 
terms of $S$, $M_3^2$ and $M_4^2$, there is no need to expand the three mass triangle integral function when determining this rational descendant.

\subsection{$\Delta_3 $ Spurious Singularity}

The three mass triangle contributions have $\Delta_3^{-2}$ poles which can be seen explicitly in the canonical forms. Around the $\Delta_3=0$ pole
the integral function has the expansion~\cite{Berger:2008sj},
\begin{align}
I_3(\mtria,\mtrib,\mtric)=\hskip 10pt & 
            \log(\mtria)\Bigl(-{2\over (\mtria-\mtrib-\mtric)}+{2\over 3}{\Delta_3\over ( \mtria-\mtrib-\mtric)^3}  +\cdots\Bigr) 
\notag \\ 
           +&\log(\mtrib)\Bigl(-{2\over(-\mtria+\mtrib-\mtric)}+{2\over 3}{\Delta_3\over (-\mtria+\mtrib-\mtric)^3}  +\cdots\Bigr)
\notag \\ 
           +&\log(\mtric)\Bigl(-{2\over(-\mtria-\mtrib+\mtric)}+{2\over 3}{\Delta_3\over (-\mtria-\mtrib+\mtric)^3}  +\cdots\Bigr) 
\notag \\  +& \fDelta_R^{3mt}\,,
\end{align}
where
\begin{align}
\fDelta_R^{3mt}           
  =         
          -{4\over 3}{\Delta_3\over ( \mtria-\mtrib-\mtric)(-\mtria+\mtrib-\mtric)(-\mtria-\mtrib+\mtric)} +\cdots\,.
\end{align}
The logarithmic terms in this expansion combine with the bubble contributions to yield a finite contribution on the pole. The rational piece in the expansion
must cancel with the rational part of the amplitude. 

\section{Obtaining ${\calR}_6$ By Recursion}

BCFW~\cite{Britto:2005fq} recursion applies
complex analysis to amplitudes.  Using Cauchy's theorem, if a complex function is
analytic except at simple poles $z_i$ (all non-zero) and $f(z)\longrightarrow
0$ as $|z|\longrightarrow\infty$ then
by considering the integral
\begin{equation}
 \oint_C  f(z) {dz\over z}\,,
\end{equation}
where the contour $C$ is the circle at infinity, we obtain  
\begin{equation}
f(0) =-\sum_i  { {\rm Residue}( f,z_i) \over z_i }\,.
\end{equation}
We wish to apply this with $f(z)={\calR}_6(z)$, where ${\calR}_6$ has been complexified by a BCFW shift of momenta.  Since
\begin{equation}
{\calM}_6 ={\cal C}_{\rm box} +{\cal C}_{\rm tri} +{\cal C}_{\rm bub} + {\calR}_6 
\to
{\calR}_6 ={\calM}_6 -{\cal C}_{\rm box} -{\cal C}_{\rm tri} -{\cal C}_{\rm bub} \,,
\end{equation}
the singularities and residues of ${\calR}_6$ are both those arising from the physical factorisations of ${\calM}_6$ and those 
induced by the necessity to cancel the spurious singularities of the cut-constructible pieces.

\subsection{Choice of Shift}

The rational part of an amplitude can be obtained recursively if the factorisation properties of the amplitude are understood at all of the
relevant poles. There are three main obstacles to this: quadratic poles in the amplitude, non-standard factorisations for complex momenta 
and contributions for large shifted momenta. Quadratic  poles in the amplitude lead to recursive contributions that depend on the off-shell behaviour
of the factorised currents. This can be addressed using augmented 
recursion~\cite{Dunbar:2010xk,Alston:2015gea}. 
For non-supersymmetric theories there are double poles of the form
\begin{equation}
V(a^+,b^+,K^+) \times { 1 \over \spb{a}.b^2 }\times M^{\tree}_{n-1}(K^-,\cdots , n)\,.
\end{equation}
For the six-point NMHV amplitude the tree amplitude vanishes since it has a single positive helicity leg.
(This is no  longer the case for seven and higher point NMHV amplitudes.)

Non-standard factorisations for complex momenta are unavoidable and are considered in detail below. The final obstacle is the possibility of contributions
from asymptotically large shifted momenta. The amplitude doesn't factorise in this limit, so the residue is undetermined. This issue may be avoided if the 
shift employed  causes the amplitude to vanish for asymptotically large shifted momenta. As the amplitude is as yet undetermined, its behaviour under any 
shift is unknown. However, if the cut constructible pieces don't vanish for asymptotically large shifted momenta there is little hope that the rational 
pieces would.

For example under a shift involving two negative helicity legs, 
\begin{equation}
\bar\lambda_a \to {\bar\lambda}_{\hat a}= \bar\lambda_a+z\bar\lambda_b
\qquad,
\qquad
\lambda_b\to \lambda_{\hat b}=\lambda_b -z \lambda_a \,,
\label{abshift}
\end{equation}
the cut constructible pieces of the amplitude are divergent for large $z$.

However, for a shift involving one negative helicity leg and one positive helicity leg,
\begin{equation}
\bar\lambda_a \to {\bar\lambda}_{\hat a}= \bar\lambda_a+z\bar\lambda_d,
\qquad,
\qquad
\lambda_d\to \lambda_{\hat d}=\lambda_d -z \lambda_a,
\label{adshift}
\end{equation}
the cut-constructible pieces all vanish at large $z$, at least suggesting that the rational piece is also well behaved there. The shift \eqref{adshift} 
will be used to obtain ${\calR}_6$.

The contributions to ${\calR}_6$ can be grouped into three classes: 
standard factorisations,
non-factorising contributions and
rational descendants of the cut-constructible pieces :
\begin{align}
 {\calR}_6={\cal R}_6^{\rm SF}+{\cal R}_6^{\rm NF}+{\cal R}_6^{\rm RD}\,.
\end{align}

\subsection{Standard Factorisations}

The standard factorisations of a six-point one-loop amplitude have the forms:
\begin{equation}
 {\calM}^{\rm tree}_3 {1\over \K^2} {\calM}^{\rm 1-loop}_5
 \qquad,
\qquad
 {\calM}^{\rm tree}_4 {1\over \K^2} {\calM}^{\rm 1-loop}_4
 \qquad,
\qquad
 {\calM}^{\rm tree}_5 {1\over \K^2} {\calM}^{\rm 1-loop}_3\,.
\end{equation}
In a supersymmetric theory the 3-point loop amplitudes vanish and so the third class are absent in this case.
With the shift \eqref{adshift} the factorisations of the first type are:
\begin{align}
{\calM}^{\rm tree}_3(\hat a^-,m_1^-, \hat \K^+) &
{1\over \K^2} 
{\calM}^{\rm 1-loop}_5 (-\hat \K^-,m_2^-,\hat d^+,p_1^+,p_2^+),\notag \\
{\calM}^{\rm tree}_3(\hat a^-,\hat\K^-, p^+_1) 
&{1\over \K^2} 
{\calM}^{\rm 1-loop}_5 (m_1^-,m_2^-,p_2^+,\hat d^+,-\hat\K^+),\notag \\
{\calM}^{\rm tree}_3(\hat\K^-,\hat d^+,p^+_1) 
&{1\over \K^2} 
{\calM}^{\rm 1-loop}_5 (\hat a^-,m_1^-,m_2^-,p_2^+,-\hat\K^+),\notag \\
{\calM}^{\rm tree}_3(m_1^-,\hat d^+,\hat\K^+) 
&{1\over \K^2} 
{\calM}^{\rm 1-loop}_5 (\hat a^-,-\hat\K^-,m_2^-,p_1^+,p_2^+).
\label{SF:tf}
\end{align}

While the factorisations of the second type are:
\begin{align}
 {\calM}^{\rm tree}_4 (\hat a^-,\hat\K^-,p_1^+,p_2^+) &{1\over \K^2}{\calM}^{\rm 1-loop}_4 (m_1^-,m_2^-,\hat d^+,-\hat\K^+),\notag \\
 {\calM}^{\rm tree}_4 (\hat a^-,m_1^-,p_1^+,\hat\K^+)&{1\over \K^2} {\calM}^{\rm 1-loop}_4 (-\hat\K^-,m_2^-,\hat d^+,p_2^+),\notag \\
 {\calM}^{\rm 1-loop}_4 (\hat a^-,\hat\K^-,p_1^+,p_2^+) &{1\over \K^2}{\calM}^{\rm tree}_4 (m_1^-,m_2^-,\hat d^+,-\hat\K^+),\notag \\
 {\calM}^{\rm 1-loop}_4 (\hat a^-,m_1^-,p_1^+,\K^+)&{1\over \K^2} {\calM}^{\rm tree}_4 (-\hat\K^-,m_2^-,\hat d^+,p_2^+).
 \label{SF:ff}
\end{align}

For generic six-point kinematics, the kinematic points at which the 4- and 5-point loop amplitudes appearing in these factorisations are evaluated  are in no 
way special, hence the rational contribution to the residue comes solely from the rational part of the 4- and 5-point loop amplitudes. Each factorisation therefore
gives a contribution to ${\calR}_6$ of
\begin{align}
{\calR}_6^{SF:i}={\rm Res}\biggl( { {\cal M}^{\rm tree:i}_{8-n} \;{\calR}_n^i \over z\K_i^2}\biggr)\biggr\vert_{z\to z_i}
= {\calM}^{\rm tree:i}_{8-n}(z_i) \times { 1\over \K_i^2}\times {\calR}_n^i (z_i)\,,
\end{align}
with~\cite{Dunbar:1994bn,Dunbar:2010fy}
\begin{align}
{\calR}_4(a^-,b^-,c^+,d^+)&={1\over 2}{\spa{a}.b^4 \spb{c}.d^2 \over \spa{c}.d^2}
\notag 
\shortintertext{and}
{\calR}_5(a^-,b^-,c^+,d^+,e^+] &={\cal R}^b_5(a^-,b^-,c^+,d^+,e^+) + {\cal R}^a_5(a^-,b^-,c^+,d^+,e^+)\,,
\end{align}
 where
\begin{align}
{\cal R}^a_5(a^-,b^-,c^+,d^+,e^+) &=-{\spa{a}.b^4\over 2}{\spb{c}.d^2\over\spa{c}.d^2}{\spb{b}.e\over\spa{b}.e}{\spa{b}.c\over \spa{c}.e}{\spa{b}.d\over\spa{d}.e}
+{\cal P}_6 (\{a,b\},\{c,d,e\})\,,
\notag \\
{\cal R}^b_5(a^-,b^-,c^+,d^+,e^+) &=-\spa{a}.b^4{\spb{c}.d\over\spa{c}.d}{\spb{c}.e\over\spa{c}.e}{\spb{d}.e\over\spa{d}.e}
\end{align}
and ${\cal P}_6$ denotes a sum over the six distinct permutations of $\{a,b\}$ and $\{c,d,e\}$ noting the symmetry of ${\cal R}^a_5$ under $c \leftrightarrow d$.
The full contribution of the standard factorisations is then 
\begin{align}
 {\calR}_6^{SF}=\sum_i{\calR}_6^{SF:i}\,,
\end{align}
where the sum is over all of the standard factorisation channels given in \eqref{SF:tf} and \eqref{SF:ff}.

\subsection{Contribution Of Rational Descendants}

As discussed above, higher order poles in the coefficients of the box and triangle contributions to the amplitude can generate rational descendants
when those poles are excited. The shift \eqref{adshift} excites some poles of each type. Specifically we have the various singularities listed in 
table~\ref{poles.table}
(with $p_i\neq d$, $m_i\neq a$).

\renewcommand{\arraystretch}{1.9}

\begin{table}[ht]
\begin{tabular}{|c|c|}
\hline
$I_4^{1m}( \hat a^-, p_1^+, m_2^-, \{p_2^+,p_3^+,m_3^-\})$
    &  $\spb{a}.{m_2}^{-4}$ 
    \\  \hline
$I_4^{1m}( \hat d^+, m_1^-, p_2^+, \{m_2^-,m_3^-,p_3^+\})$
    &  $\spa{d}.{p_2}^{-4}$   
     \\  \hline
$I_4^{2mh}(a^-,d^+,\{m_1,p_1\},\{m_2,p_2\} )$
    & $[\hat a|\K_{p_im_i}|\hat d\ra^{-4}$     
     \\  \hline
$I_4^{2mh}(a^-,p_2^+,\{m_1,d\},\{m_2,p_2\} )$
    & $[\hat a|\K_{p_im_i}|p_2\ra^{-4}$     

         \\  \hline
$I_4^{2mh}(m_1^-,d^+,\{a,p_1\},\{m_2,p_2\} )$
    & $[m_1|\K_{p_2m_2}|\hat d\ra^{-4}$     
    \\ \hline
$I_4^{2mh}(m_1^-,p_1^+,\{a,p_2\},\{m_2,\hat d \} )$
    & $[m_1|\K_{\hat d m_2}|p_2\ra^{-4}$     
    \\ \hline
$I_3^{3m}$
    & $\Delta_3^{-2}$      \\ \hline
    \end{tabular}
\caption{The various non-physical poles which induce terms in $R_6$}
\label{poles.table}
\end{table}

Denoting the rational descendant in each case by $\fDelta_R^i$, the corresponding coefficient by $c_i$ and the value of $z$ on the pole by $z_i$,
the contribution on each of these poles is
\begin{align}
{\cal R}_6^{\rm RD:i}=-{\rm Res}\biggl( {c_i(\hat a,b,c,\hat d,e,f) \fDelta_R^i(\hat a,b,c,\hat d,e,f)\over z}\biggr)\biggr\vert_{z\to z_i}
\end{align}
so that
\begin{align}
{\cal R}_6^{\rm RD}=\sum_i{\cal R}_6^{\rm RD:i} \,,
\end{align}
where the sum is over all of the poles listed above.

Individual terms in the bubble coefficients contain a range of other higher order poles. In principle these could also generate rational descendants, however
in the full bubble coefficients these are at most simple poles and so do not generate further rational descendants:
\begin{equation}
{\cal R}_6^{\rm RD: bub}=0 \,.
\end{equation}

\subsection{Non-standard Factorisations for Complex Momenta}

Factorisations of the amplitude occur when propagators go on shell. The standard factorisation channels arise when the on-shell propagator is
not in the loop and is explicit in, for example a Feynman diagram approach.  

The loop momentum integral may also generate poles in the amplitude~\cite{Bern:1995ix} particularly for complex momenta.  Since we are computing the amplitude by recursion in complex momenta we must determine these
complex factorisations.  
\vskip2.0truecm
\begin{figure}[H]
\centerline{
    \begin{picture}(-50,90)(50,-50)  
      \ArrowLine(0,-30)(0,30)
      \ArrowLine(0,30)(60,0) 
      \ArrowLine(60,0)(0,-30)
      \Line(0,-30)(-10,-40) 
      \Line(0, 30)(-10, 40)
      \Line(60,0)(50, 20) 
      \Line(60,0)(50,-20) 
      \Line(60,0)(78,15) 
      \Line(60,0)(78,-15)
      \BCirc(60,0){8} 
      \Text(60,0)[c]{$\tau$} 
      \Text(-13,-40)[r]{$\hat {a}^-$}
      \Text(-13, 40)[r]{$b^-$}
      \Text(50, 28)[l]{$c^-$} 
      \Text(77, 25)[l]{$d^+$}
      \Text(72,-22)[l]{$ e^+$}
      \Text(50,-30)[l]{$ f^+$}
      \Text(40,-15)[l]{$^{+h}$}
      \Text(40,+10)[l]{$^{-h}$}     
      \Text(25,-30)[l]{$A$}
      \Text(25,+30)[l]{$B$}     
      \Text(-10,0)[l]{$\ell$}     
    \end{picture}
    }
    \caption{Non-standard factorisations channel}
    \label{NStri}
\end{figure}
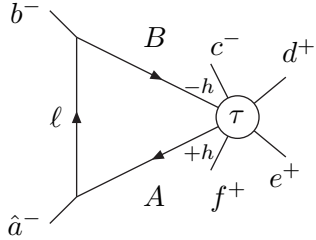

Poles can arise when two adjacent massless legs on a loop became collinear as illustrated in \figref{NStri}. This case 
has been discussed in the context of amplitudes with a single negative helicity leg~\cite{Dunbar:2010xk,Alston:2015gea}.
In the integration region $\ell\propto b$ the three propagators connected to $a$ and $b$ all become
on shell when $a$ and $b$ are collinear. The diagrams of 
interest can be grouped together to form a one mass triangle in the integral reduction sense (i.e. the massive corner represents a sum of all 
possible tree diagrams). The integration 
region of interest has all three propagators on shell and so the pole  may be determined by the triple cut of this triangle. 
This triple cut wouldn't normally exist, but {\it opens up} when $a$ and $b$ are collinear.

Using an axial gauge with reference spinor $q$~\cite{Schwinn:2005pi,Dunbar:2010xk,Alston:2015gea}, 
the contribution of \figref{NStri} with a scalar particle circulating in the loop is
\begin{align}
\int d^D\ell{1\over \ell^2 A^2 B^2}\left( {[q|a\ell|q]\over [aq]^2}\right)^2
\left( {[q|b\ell|q]\over [bq]^2}\right)^2{\tau}^{\rm tree}_{\rm MHV}( A, B,c^-,d^+,e^+,f^+)\,,
\label{NSint}
\end{align}
where ${\tau}^{\rm tree}_{\rm MHV}$ represents the sum of all possible tree diagrams.  $A$ and $B$ are given by
\begin{equation}
A=\ell+ a \;\;{\rm and} \;\; B=-\ell+b 
\end{equation}
and satisfy $A+B=a+b$. The integrands for other particle types in the loop are related
to the scalar contribution by an $X$-factor:
\begin{align}
X={\spa{\ell}.a\over  \spa{A}.a}{\spa{B}.b\over  \spa{\ell}.b}{\spa{A}.c\over  \spa{B}.c} = {[q|A|c\ra \over [q|B|c\ra}
\end{align}
and the $\NeqFour$ contribution is obtained by multiplying the integrand by a factor of $\rho^2$ where,
\begin{align}
\rho= {[q|a+b|c\ra^2 \over [q|A|c\ra[q|B|c\ra}\,.
\end{align}

\begin{figure}[H]
  \begin{center}
    {
      \begin{picture}(150,144)(0,0)
    \SetOffset(50,72)
        \ArrowLine(-20,-20)(-20, 20)
        \ArrowLine(-20, 20)( 20, 20)
        \ArrowLine( 20, 20)( 20,-20)
        \ArrowLine( 20,-20)(-20,-20)
        \Line(-20,-20)(-35,-35)
        \Line(-20, 20)(-35, 35)
        \Line( 20, -20)( 35, -35)
        \Line( 20, 20)( 20, 42)
        \Line( 20, 20)( 42, 20)
        \Text(0,25)[bc]{$B$}
        \Text(0,-25)[tc]{$A$}
        \Text(-25,0)[c]{$\ell$}
        \Text(-35, 40)[bc]{$b$}
        \Text(-35,-40)[tc]{$a$}
        \Text( 43,-40)[bc]{$x$}
        \Text( 37, 34)[bc]{$\bullet$}       
        \Text( 28, 39)[bc]{$\bullet$}       
        \Text( 42, 25)[bc]{$\bullet$}       
      \end{picture}
      \begin{picture}(150,144)(0,0)
    \SetOffset(50,72)
        \ArrowLine(-20,-20)(-20, 20)
        \ArrowLine(-20, 20)( 20, 20)
        \ArrowLine( 20, 20)( 20,-20)
        \ArrowLine( 20,-20)(-20,-20)
        \Line(-20,-20)(-35,-35)
        \Line(-20, 20)(-35, 35)
        \Line( 20, -20)( 35, -35)
        \Line( 20, 20)( 20, 42)
        \Line( 20, 20)( 42, 20)
        \Text(0,-25)[tc]{$\ell$}
        \Text(-28,0)[c]{$B$}
      \Text(28,0)[c]{$A$}
              \Text(-35, 40)[bc]{$y$}
        \Text(-35,-40)[tc]{$b$}
        \Text( 43,-40)[bc]{$a$}
        \Text( 37, 34)[bc]{$\bullet$}       
        \Text( 28, 39)[bc]{$\bullet$}       
        \Text( 42, 25)[bc]{$\bullet$}       
      \end{picture}    }
    \\
    \caption{The box integrals associated with \eqref{boxexample}  \label{sampbox} }
  \end{center}
\end{figure}
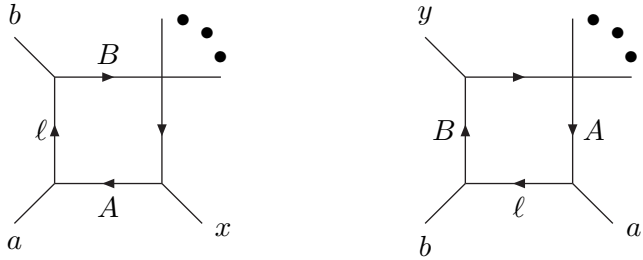

For $\ell\propto b$ the integrand of \eqref{NSint} contains $\spa{a}.b$ factors in its numerator,  leaving a pole in $[ab]$ in the integral. As ${\cal M}^{\rm tree}_{\rm MHV}$ is finite as
$[ab]\to0$, ${\tau}^{\rm tree}_{\rm MHV} \to {\cal M}^{\rm tree}_{\rm MHV}$ in the region of interest. For a scalar particle circulating in the loop
the KLT relations~\cite{Kawai:1985xq} give
\begin{align}
 {\cal M}^{\rm tree:0}_{\rm MHV}( A, B,e,f,c,d)=\Bigl[-i s_{AB}s_{cf}A^{YM:0}(A, B, e,f,c,d) \bigl[ s_{ec} & 
A^{YM}( B, A,c,e,f,d) \notag \\
                                                                       +(s_{ec}+s_{ef})A^{YM}(B ,A ,c,f,e,d)
                                                                       \bigr] \Bigr]&
                                                                       +{\cal P}(B ,e,f)\,.
 \label{KLTsix}\end{align}
Of the six terms in the permutation sum in \eqref{KLTsix}, the two which don't permute  $B$ can be neglected due to the explicit $s_{AB}$ factor.
The remaining four form two pairs with the members of each pair being related by interchange of legs $e$  and $f$. 
The $\NeqFour$ contributions of one member of each of these pairs are  
\begin{align}
C_1= \int d^D\ell{1\over \ell^2 A^2 B^2}{\spa{A}.a^2 \spa{B}.b^2 \spa{A}.c\spa{B}.c^2[Ae][Be]\over       
                                        \spa{A}.d  \spa{A}.e  \spa{B}.d \spa{B}.e \spa{B}.f}
                                        {[cf][q|a+b|c\ra^4 \over [aq]^2 [bq]^2\spa{c}.d\spa{c}.f\spa{d}.f}
\end{align}
and
\begin{align}
C_2= -\int d^D\ell{1\over \ell^2 A^2 B^2}{\spa{A}.a^2 \spa{B}.b^2 \spa{A}.c\spa{B}.c[Ae][Bc][f|B+c|d\ra\over       
                                        \spa{A}.d  \spa{A}.e  \spa{B}.d \spa{B}.f}
                                        {[cf][q|a+b|c\ra^4 \over [aq]^2 [bq]^2\spa{c}.d\spa{d}.e\spa{d}.f\spa{e}.f}\,.
\end{align}
Partial fractioning  the integrand  of $C_1$ using the $\spa{A}.c\spa{B}.c^2$ factor in the numerator 
yields six terms whose integrands have loop momentum dependence
\begin{align}
{1\over \ell^2 A^2 B^2}{\spa{A}.a^2 \spa{B}.b^2[Ae][Be]\over       
                                        \spa{A}.x  \spa{B}.y}\,,
\label{nicePF}
\end{align}
with $x\in\{d,e\}$ and $y\in\{d,e,f\}$. In the integration region of interest $A^2$ and $B^2$ are negligible allowing the integrands 
to be rewritten as quartic pentagon integrands
\begin{align}
{[e|A|a\ra[e|B|b\ra[x|A|a\ra[y|B|b\ra \over \ell^2 A^2 B^2(A+x)^2(B-y)^2}\,.
\end{align}
For $x\neq y$, using 
\begin{align}
\la b|ByxA|a\ra=2B.y\la b|xA|a\ra -2A.x \la b|yB|a\ra +\la b |y(a+b)Ax|a\ra 
\end{align}
splits each of these quartic pentagons into a pair of cubic one-mass boxes and a cubic pentagon which can be neglected. As a box with two adjacent corners attached to
single external legs of the same helicity has a vanishing quadruple cut, these cubic one-mass box integrals reduce to bubble and 
rational contributions only. The bubble coefficients can be  evaluated by direct parametrisation. For example the box integral \eqref{boxexample} which
is illustrated in \figref{sampbox} has bubbles associated with its $\{a,x\}$ and $\{b,a,x\}$ cuts. 
\begin{align}
\int d^D\ell {[e|A|a\ra[e|B|b\ra[x|A|a\ra \over \ell^2 A^2 B^2(A+x)^2}
\label{boxexample}
\end{align}
\noindent{The} $\{a,x\}$ cut of \eqref{boxexample}  gives the bubble coefficient
\begin{align}
C^{\rm bub}_{ax}={\spa{a}.b^2\over [ab]}{\spa{x}.a [ea]^2 \over \spa{b}.x^2 } +{\cal O}\bigl([ab]\bigr),
\end{align}
where terms of order $[ab]^1$ have been extracted from the leading term to simplify its denominator as far as possible. 
The remaining $\spa{b}.x$ singularity in this bubble coefficient is spurious and must cancel with the  $\{b,a,x\}$ bubble as
$\spa{b}.x\to 0$. So that this singularity is not present in the logarithmic part of the integral, the sum of the two bubble coefficients
 must be finite.  The sum of the two bubble contributions then involves the singular parts of the $\{a,x\}$ bubble coefficient multiplied by the difference
 of the integral functions of the two bubbles. With $s_{ab}$ and $s_{bx}$ both being small, the rational descendant of the bubbles on the $\spa{b}.x\to 0$
 pole is
\begin{align}
 {\spa{a}.b^2\over [ab]}{\spa{a}.x [ea]^2 \over \spa{b}.x^2 }\Bigl( {s_{bx}\over s_{ax}} -{1\over 2} {s_{bx}^2\over s_{ax}^2} +\cdots\Bigr) 
 +{\cal O}\bigl([ab]^0\bigr).
\end{align}
The leading term of the rational descendant has a $\spa{b}.x^{-1}$ spurious pole. This must be cancelled by the rational piece of the integral, allowing the
rational term to be identified as,
\begin{align}
{\spa{a}.b^2\over [ab]}{[ae][be] \over \spa{b}.x}+{\cal O}\bigl([ab]^0\bigr)\,.
\end{align}
For $x=y$ in \eqref{nicePF} the reduction to boxes uses the identity
\begin{align}
{1\over (A+x)^2(B-x)^2}= {1\over 2P.x}\Bigl( {1\over (A+x)^2} +{1\over (B-x)^2}\Bigr) +{\cal O}\bigl( A^2,B^2\bigr)
\end{align}
which yields a pair of quartic box integrals whose rational pieces are evaluated using the approach  described above.  
The full rational contribution of $C_1$ is
\begin{align}
 R^{C_1}= {\spb{f}.c [q|P_{ab}|c\ra^4 \over [aq]^2 [bq]^2 \spa{c}.d \spa{c}.f \spa{d}.e\spa{d}.f }
   \Biggl(                        & \hskip 5pt                        
                                       {\spa{c}.d\spa{c}.d^2\over\spa{e}.d\spa{f}.d} I^{5a}_{d,d,2}
                                      +{\spa{c}.d\spa{c}.e^2\over\spa{d}.e\spa{f}.e} I^{5a}_{d,e,1}
                         \notag \\ &
                                      +{\spa{c}.d\spa{c}.f^2\over\spa{d}.f\spa{e}.f} I^{5a}_{d,f,1}
                                      -{\spa{c}.e\spa{c}.d^2\over\spa{e}.d\spa{f}.d} I^{5a}_{e,d,1}
                          \notag \\ &
                                      -{\spa{c}.e\spa{c}.e^2\over\spa{d}.e\spa{f}.e} I^{5a}_{e,e,2}
                                      -{\spa{c}.e\spa{c}.f^2\over\spa{d}.f\spa{e}.f} I^{5a}_{e,f,1}
 \hskip 5pt \Biggr)\,,
\end{align}
where
\def\fsp{-}
\def\fdx{-}
\def\fdp{-}
\def\fsub{}
\def\fqa{}
\def\fqb{-}
\def\fqc{}
\def\fqd{-}
\begin{align}
I^{5a}_{x,y,1}=&\fsp{1\over 4}{\spa{a}.b^2 [ae][be]\over[ab]\spa{x}.y}
\;\;{\rm and}\;\;
 I^{5a}_{x,x,2}={\spa{a}.b^3  \over 12 [ab][x|P_{ab}|x\ra}\Bigl(  \fdx{ [eb]^2 [ax]\over\spa{a}.x } 
                                                             \fdp{ [ea]^2 [bx]\over\spa{b}.x } \Bigr) \,.
\end{align}

The $C_2$ contributions involve both quintic and quartic 
pentagon integrals, but their rational pieces can be obtained in a similar fashion to the $C_1$ contributions.
Separating the quintic and quartic 
pentagon integrals,
\begin{align}
 R^{C_2}=R^{C_2}_{\rm quin}+\fsub R^{C_2}_{\rm quar},
\end{align}
where
\begin{align}
R^{C_2}_{\rm quin}={ [q|P_{ab}|c\ra^4 \over [aq]^2 [bq]^2 \spa{c}.d \spa{d}.e^2 \spa{e}.f \spa{d}.f }
\Bigl(  \spa{c}.d I^{5b}_d 
       -\spa{c}.e I^{5b}_e \Bigr)
\end{align}
with
\begin{align}
I^{5b}_x= {1\over\spa{x}.f}\Biggl(&
                                  \fqd{1\over 6}\spa{a}.f 
                                     [eb][cb]\spa{a}.b^2\Bigl( {\spa{c}.b [af] \over \spa{f}.a}
                                                               +{\spa{f}.b\spa{a}.c [af]\over\spa{f}.a^2 }
                                                               -{\spa{a}.c [af]^2\over 2\spa{f}.a [bf]}
                                                         \Bigr)
 \notag \\ &                                                        
                                 +\fqa{\spa{a}.x\over 4\spa{x}.f\spa{a}.f}\la b|f P_{ab}|f\ra
                                       [ea]\spa{a}.c [cb]\spa{a}.b
  \notag \\ &                                                        
                                \fqb\spa{a}.x
                                       [ea] [fa] [ca] \spa{a}.x \spa{c}.b {\spa{a}.b^2\over 6 \spa{b}.x^2}{s_{bx}\over s_{ax}}
 \notag \\ &                                                        
                                 +\fqc{\spa{a}.x\over\spa{x}.f} \la b|f P_{ab}|x\ra
                                       [ea] [ca] \spa{a}.x \spa{c}.b {\spa{a}.b\over 4\spa{b}.x^2} {s_{bx}\over s_{ax}}
                                   \Biggr)
\end{align}
and
\begin{align}
R^{C_2}_{\rm quar}={[fc] [q|P_{ab}|c\ra^4 \over [aq]^2 [bq]^2 \spa{d}.e^2 \spa{d}.f^2 \spa{e}.f}
                                 \Biggl(
                                   \spa{c}.d^2        I^{5c}_{d,d,2}
                                  -\spa{c}.e\spa{c}.d I^{5c}_{e,d,1}
                                  -\spa{c}.d\spa{c}.f I^{5c}_{d,f,1}
                                  +\spa{c}.e\spa{c}.f I^{5c}_{e,f,1}
                                 \Biggr)
\end{align}
with
\begin{align}
I^{5c}_{x,y,1}=&\fsp{1\over 4}{\spa{a}.b^2 [ae][bc]\over[ab]\spa{x}.y}
\;\;{\rm and}\;\;
 I^{5c}_{x,y,2}={\spa{a}.b^3  \over 12 [ab][x|P_{ab}|x\ra}\Bigl(  \fdx{ [bc][be] [ax]\over\spa{a}.x } 
                                                             \fdp{ [ac][ae] [bx]\over\spa{b}.x } \Bigr) \,.
\end{align}
The contribution of these non-standard factorisations to the rational part of the 6-pt amplitude is obtained by recursion:
\begin{align}
{\cal R}_6^{C_i}(a,b,c,d,e,f)={\rm Res}\Biggl( {R^{C_i}(\hat a,b,c,\hat d,e,f)\over z}\Biggr)\Biggr\vert_{[\hat a b]\to 0} 
\end{align}
The contributions arising from the conjugate poles, e.g. $\spa{\hat d}.e\to 0$, can be obtained using the flip-conjugation symmetry of the amplitude.
Defining 
\begin{align}
{\cal R}_6^{C}(a,b,c,d,e,f)&={\cal R}_6^{C_1}(a,b,c,d,e,f)+{\cal R}_6^{C_2}(a,b,c,d,e,f)
\shortintertext{and} 
{\cal R}_6^{\tilde C}(a,b,c,d,e,f)&={\cal R}_6^C(d,e,f,a,b,c)\bigr\vert_{\spa{x}.y\leftrightarrow [xy]},
\end{align}
the full non-factorising contribution to ${\cal R}_6$ is
\begin{align}
 {\cal R}_6^{\rm n-f}(a,b,c,d,e,f)=&\hskip 8pt {\cal R}_6^C(a,b,c,d,e,f)+{\cal R}_6^C(a,b,c,d,f,e)+{\cal R}_6^C(a,c,b,d,e,f)+{\cal R}_6^C(a,c,b,d,f,e)
 \notag \\
& + {\cal R}_6^{\tilde C}(a,b,c,d,e,f)+{\cal R}_6^{\tilde C}(a,b,c,d,f,e)+{\cal R}_6^{\tilde C}(a,c,b,d,e,f)+{\cal R}_6^{\tilde C}(a,c,b,d,f,e)\,.
\end{align}

We have computed $R_6$ systematically using its pole structure.  Underlying this is the assumption that the amplitude
vanishes for large shifts. This is difficult to justify a priori. However the expression obtained has the correct symmetries and collinear limits (checked numerically). Generically a BCFW recursion produces terms which are not manifestly symmetric and the restoration of  symmetry is a good indicator that the amplitude has been correctly determined.

An explicit form of $R_6$ is  available in {\tt Mathematica} format at \url{http://pyweb.swan.ac.uk/~dunbar/sixgraviton/R6.html}.

\section{Conclusions}

Graviton scattering amplitudes have a rich structure.  In particular $\NeqEight$ supergravity has proven to have a much softer UV behaviour then previously expected with the underlying symmetry reason still unclear.  It is important to understand which structures of $\NeqEight$ survive in theories with lower
supersymmetry.  It is also important to study amplitudes beyond MHV since this can often have a misleadingly simple structure.  In this article we have constructed the six-point NMHV amplitude in $\NeqFour$ supergravity.   Of particular interest is the rational term since in the MHV case a particularly simple and 
suggestive structure appears~\cite{Dunbar:2011dw}.   The rational terms in the NMHV case do not appear to have any such simple structure although this may be hiding given the algebraic complexity of the amplitude.  

Computing the rational terms has required a blending of techniques including obtaining the rational descendants of the 
cut-constructible pieces. Amongst the cut-constructible pieces the coefficients of the bubble integral functions have been particularly cumbersome although, fortunately, 
these do not generate any rational descendants in this amplitude.

\section{Acknowledgements}

This work was supported by STFC grant ST/L000369/1.

\appendix

\section{Six-Point Tree Amplitude Expression}
\label{sixpointtreeapp}

The six-point tree amplitude needed for computing the bubble coefficients is 
\begin{equation}
M(  (l_1)^-_h, a^-, b^- , e^+, f^+,(l_2)^+_h )
=\sum_{i=1}^{14} T_i(h) = \sum_{i=1}^{14} A_i (X_i)^{2h}.
\label{treeexpA}
\end{equation}
The fourteen terms  in \eqref{treeexpA}  are
\def\t(#1,#2,#3){t_{#1#2#3}}
\def\s(#1,#2){s_{#1#2}}
\def\spaN(#1,#2){\left\langle#1#2\right\rangle}
\def\spbN(#1,#2){\left[#1#2\right]}

\begin{align}
T_1&=-i {\la ab \ra^7 \la el_2\ra [bl_1] [ef]^7
          \over
         \la al_1\ra  \la bl_1\ra  [e|\K_{abl_1}|a\ra
               [l_2|\K_{abl_1}|a\ra 
               [f|\K_{abl_1}|b\ra 
               [f|\K_{abl_1}|l_1\ra
               [el_2] [fl_2] t_{abl_1}
        }
        \Big[ \delta _{h,-2}\Big], 
\notag \\
T_2&=-i{ \la bl_1\ra [f|\K_{bfl_1}|a\ra^8  [el_2]
          \over
         \la ae\ra  \la al_2\ra \la el_2\ra [b|\K_{bfl_1}|a\ra [l_1|\K_{bfl_1}|a\ra 
         [f|\K_{bfl_1}|e\ra [f|\K_{bfl_1}|l_2\ra  [bf] [bl_1] [fl_1] t_{bfl_1}
        }  
        \bigg[-i {\la al_2\ra  [fl_1] \over [f|\K_{bfl_1}|a\ra }\bigg]^{A},
\notag \\
T_3&=-i{ \la ab\ra^7 \la el_1\ra [bl_2] [ef]^7
          \over
         \la al_2\ra \la bl_2\ra [e|\K_{efl_1}|a\ra [l_1|\K_{efl_1}|a\ra  [f|\K_{efl_1}|b\ra [f|\K_{efl_1}|l_2\ra
         [el_1] [fl_1]  t_{efl_1}
         }
         \big[\delta _{h,2}\big],  
         \cr
\notag \\
T_4&=i { \la al_1\ra^7  \la be\ra [ef]^7 [l_1l_2]
       \over
       \la al_2\ra \la l_1l_2\ra  [b|\K_{bef}|a\ra [e|\K_{bef}|a\ra [f|\K_{bef}|l_1\ra [f|\K_{bef}|l_2\ra 
       [be] [bf] t_{bef} 
       }
       \Biggl[ i{ \la al_2 \ra \over \la al_1\ra }\Biggr]^{A},  
\notag \\
T_5&=i{ \spa{a}.b^7 \spa{l_1}.{l_2} \spb{b}.e \spb{f}.{l_2}^7
       \over
      \spa{a}.e \spa{b}.e [l_1|\K_{abe}|a\ra [l_2|\K_{abe}|a\ra [f|\K_{abe}|b\ra [f|\K_{abe}|e\ra
       \spb{f}.{l_1} \spb{l_1}.{l_2} t_{abe}
       }
      \Biggl[i{ [fl_1]\over [fl_2] } \Biggr]^{A},
\notag \\
T_6&=-i{ \la al_1\ra^7 \la bl_2\ra [el_1] [fl_2]^7
        \over
        \la ae\ra \la el_1\ra [b|\K_{ael_1}|a\ra[l_2|\K_{ael_1}|a\ra[f|\K_{ael_1}|e\ra[f|\K_{ael_1}|l_1\ra
        [bf][bl_2]t_{ael_1}
        }
        \Biggl[{[f|\K_{ael_1}|a\ra \over \la al_1\ra [fl_2]} \Biggr]^{A},
\notag \\
T_7&=i { [ef][l_2|\K_{aef}|a\ra^7 \bigl( \la al_2\ra \la bl_1\ra   [l_1|\K_{aef}|e\ra [bl_2]
                                   -\la al_1\ra \la bl_2\ra   [l_2|\K_{aef}|e\ra [bl_1]\bigr)
       \over
       \la af\ra^2 \la ef\ra [b|\K_{aef}|a\ra [b|\K_{aef}|e\ra [l_1|\K_{aef}|a\ra [l_1|\K_{aef}|e\ra
       [l_2|\K_{aef}|e\ra  [bl_1] [bl_2] [l_1l_2] t_{aef}
       }
       \Biggl[i {[l_1|\K_{aef}|a\ra \over  [l_2|\K_{aef}|a\ra}\Biggr]^{A},
\notag \\
T_8&= i{ [e|\K_{bel_1}|a\ra^7 [fl_2] ( \la ae\ra \la bl_1\ra [l_1|\K_{bel_1}|l_2\ra [be]
                                     -\la be\ra \la al_1\ra [e|\K_{bel_1}|l_2\ra [bl_1] )
       \over
        \la af\ra^2  \la fl_2\ra
         [b|\K_{bel_1}|a\ra [l_1|\K_{bel_1}|a\ra [b|\K_{bel_1}|l_2\ra [e|\K_{bel_1}|l_2\ra [l_1|\K_{bel_1}|l_2\ra
         [be] [bl_1] [el_1] t_{bel_1}
       }
       \Biggl [i { \la l_2a\ra [el_1] \over [e|\K_{bel_1}|a\ra} \Biggr]^{A},
\notag \\
T_9&=i{
   \la al_1\ra^8 [el_2]^7 [fl_1]\bigl( \la al_2\ra \la be\ra [b|\K_{afl_1}|l_1\ra [el_2]
                                      -\la ab\ra \la el_2\ra [l_2|\K_{afl_1}|l_1\ra [be]\bigr)
    \over
    \la af\ra^2 \la fl_1\ra [b|\K_{afl_1}|a\ra [e|\K_{afl_1}|a\ra [l_2|\K_{afl_1}|a\ra [b|\K_{afl_1}|l_1\ra
                            [e|\K_{afl_1}|l_1\ra [l_2|\K_{afl_1}|l_1\ra
     [be] [bl_2] t_{afl_1}
     }
    \Biggl[ i { [e|\K_{afl_1}|a\ra \over \la l_1a\ra [el_2]}\Biggr]^{A},
\notag \\
T_{10}&=i{\la ab\ra^8  [bf][el_2]^7\bigl( \la al_2\ra \la el_1\ra [l_1|\K_{abf}|b\ra [el_2]
                                         -\la al_1\ra \la el_2\ra [l_2|\K_{abf}|b\ra [el_1]\bigr)
         \over
         \la af\ra^2 \la bf\ra 
         [e|\K_{abf}|a\ra [l_1|\K_{abf}|a\ra [l_2|\K_{abf}|a\ra[e|\K_{abf}|b\ra[l_1|\K_{abf}|b\ra[l_2|\K_{abf}|b\ra
         [el_1] [l_1l_2] t_{abf}
         }
       \Biggl[i {[el_1]\over [el_2]}\Biggr]^{A},
\notag \\
T_{11}&=i{\la ae\ra \la bl_1\ra^7 [ef]^8 \bigl( \la bl_2\ra [e|\K_{aef}|l_1\ra [bl_1][fl_2]
                                               -\la bl_1\ra [e|\K_{aef}|l_2\ra [bl_2][fl_1]\bigr)
         \over
         \la bl_2\ra \la l_1l_2\ra [e|\K_{aef}|b\ra [f|\K_{aef}|b\ra [e|\K_{aef}|l_1\ra
                                   [f|\K_{aef}|l_1\ra [e|\K_{aef}|l_2\ra [f|\K_{aef}|l_2\ra
        [ae] [af]^2t_{aef}
        }
        \Biggl[i{\la bl_2\ra \over \la bl_1\ra} \Biggr]^{A},
\notag \\
T_{12}&=-i{\la al_2\ra \la bl_1\ra^7 [fl_2]^8\bigr( \la be\ra [l_2|\K_{bel_1}|l_1\ra [bl_1][ef]
                                                   +\la bl_1\ra [l_2|\K_{bel_1}|e\ra [be][fl_1]\bigr)
          \over
           \la be\ra  \la el_1\ra [f|\K_{bel_1}|b\ra [l_2|\K_{bel_1}|b\ra [f|\K_{bel_1}|l_1\ra 
                                  [f|\K_{bel_1}|e\ra [l_2|\K_{bel_1}|e\ra [l_2|\K_{bel_1}|l_1\ra 
           [al_2] [af]^2  t_{bel_1} 
          }
          \Biggl[i{[f|\K_{bel_1}|b\ra  \over \la l_1b\ra [fl_2]}\Biggr]^{A},
\notag \\
T_{13}&=-i {\la al_1\ra [f|\K_{afl_1}|b\ra^7 (\la be\ra  [l_1|\K_{afl_1}|l_2\ra [bf] [el_2]
                                            +\la el_2\ra  [l_1|\K_{afl_1}|b\ra [be] [fl_2]\bigr)
           \over
           \la be\ra \la bl_2\ra \la el_2\ra
           [l_1|\K_{afl_1}|b\ra [f|\K_{afl_1}|e\ra [l_1|\K_{afl_1}|e\ra 
           [f|\K_{afl_1}|l_2\ra [l_1|\K_{afl_1}|l_2\ra  
           [af]^2 [al_1] t_{afl_1}
           }
           \Biggl[i { \la l_2b\ra [fl_1] \over  [f|\K_{afl_1}|b\ra} \Biggr]^{A}, 
\notag \\
T_{14}&=i{ \la ab\ra [f|\K_{abf}|l_1\ra^7\bigl(\la el_2\ra [b|\K_{abf}|l_1\ra [el_1] [fl_2]
                                             -\la el_1\ra [b|\K_{abf}|l_2\ra [el_2] [fl_1]\bigr)
         \over
         \la el_1\ra \la el_2\ra \la l_1l_2\ra 
         [b|\K_{abf}|e\ra [f|\K_{abf}|e\ra [b|\K_{abf}|l_1\ra [b|\K_{abf}|l_2\ra [f|\K_{abf}|l_2\ra 
         [ab] [af]^2 t_{abf}
         }
         \Biggl[i { [f|\K_{abf}|l_2\ra \over [f|\K_{abf}|l_1\ra}\Biggr]^{A},
\label{treeexp}
\end{align}
with $A=4-2h$.

\section{Bubble coefficient}
\label{bubapp}
\def\spacycomA{ , \; }
\def\spacycomB{\, ,\,}
The 2:4 bubbles involve the 6-pt NMHV tree amplitude. This has fourteen terms and consequently, the bubble coefficient has fourteen sources. Each of these generates a collection of terms leading
to an algebraically complicated expression,
\begin{equation}
c = \sum_i C_{Ti}\,.
\end{equation}

Of these fourteen terms two ($T_1$ and $T_3$) don't enter the $\NeqFour$ matter multiple calculation and the rest split evenly into {\it massless}
and {\it massive} types. Massive terms involve a factor of $((\ell+Q)^2)^{-1}$ where $Q^2\neq 0$. 
Terms $T_4$, $T_5$, $T_7$,  $T_{10}$, $T_{11}$ and $T_{14}$ are of the massless type and their $\ell$ dependent factors
in the denominator
are of the from $\la x\ell\ra$ or $[x\ell]$. 
The bubble coefficients for these massless type terms can  be evaluated using the $H_0^d$   
canonical forms presented above (see \eqref{Hcanonform:eq}). The overall result for these terms is then
\begin{equation}
C_{T4}+C_{T5}+C_{T7}+C_{T10}+C_{T11}+C_{T14} =  \sum_{j=1}^{65}  D_j H_0^d [ B_j,A_j,b_j,a_j ] \,.
\end{equation} 

The explicit results are
\begin{align}
C_{T4} =&  
{ 
\spb{c}.d \spa{a}.c^4 \spa{b}.e \spb{e}.f^7 s_{cd}  
\over 
[b|\K_{bef}|a\ra [e|\K_{bef}|a\ra \spa{c}.d \spb{b}.e  \spb{b}.f  \K^2_{bef}  
} \sum_{i=1}^2\sum_{j=1}^2 \beta_i \alpha_j   H^d_0[ B_i,A_j,b_2,a_2 ]
\end{align}
\begin{align}
\{ |a_j\ra \} =& \{ |c \ra ,  |a\ra \} \;\;\ \{ |A_j\ra \} = \{  |d\ra ,  \K_{be}|f]  \} 
\\
\{ |b_i ]  \} =& \{ |d] , K_{cd}|a\ra  \}  \;\;\;\ \{ |B_i]\}=\{ |c] , K_{cd}  K_{be}|f]   \} 
\end{align}
\begin{align}
\alpha_j = { \prod_{i=1}^{n_A-1} \spa{a_i}.{A_j} \over \prod_{k\neq j} \spa{A_k}.{A_j} }
\;\;\;
\beta_i=  {\prod_{i=1}^{n_B-1} \spb{b_i}.{B_i} \over \prod_{k\neq i} \spb{B_k}.{B_i} }
\label{alphabetaeqn}
\end{align}
For $C_{T4}$, $n_A=n_B=2$ and the numerator products simplify to $\spa{a_1}.{A_j}$ and $\spb{b_1}.{B_i}$.

\begin{align}
C_{T5} =&  
-{
[f|\K_{cd}|c\ra^4\spa{a}.b^7\spb{b}.e
\over
\spa{c}.d^2\spa{a}.e \spa{b}.e [f|\K_{abe}|b\ra [f|\K_{abe}|e\ra \K^2_{abe} }
\sum_{i=1}^2\sum_{j=1}^2 \beta_i \alpha_j   H^d_0[ B_i,A_j,b_2,a_2 ]
\end{align}
where 
\begin{align}
\{ |a_j\ra \} =& \{ |c \ra , \K_{cd}|f ] \} \;\;\ \{ |A_j\ra \} = \{  |d\ra ,  \K_{cd}\K_{abe}|a\ra  \} 
\\
\{ |b_i ]  \} =& \{ |f] , |d] , \}  \;\;\;\ \{ |B_i]\}=\{ |c] , K_{be}|a\ra   \} 
\end{align}
and $\alpha_i$ and $\beta_j$ are given by \ref{alphabetaeqn} with $n_A=n_B=2$. 
\begin{align}
C_{T7} =&  %
{\spb{e}.f \spa{c}.d^2 [d|\K_{aef}|a\ra^4
\over
\spa{a}.f^2 \spa{e}.f [b| \K_{aef}|a\ra  [b|K_{aef}|e\ra\K^2_{aef};
}
\Biggl( 
\notag \\
&
[b|\K_{aef}|a\ra
\sum_{i=1}^2\sum_{j=1}^3 \beta_i \alpha_j   H^d_0[ B_i,A_j,b_2,a_3 ] 
-\K^2_{aef} \spa{b}.e
\sum_{i=1}^2\sum_{j=1}^3 \delta_i \gamma_j  H^d_0[ D_i,C_j,d_2,c_3 ]; 
\Biggr) 
\end{align}
\begin{align}
\{ |a_j\ra \} =& \{ |b \ra , |c\ra , \K_{cd}\K_{aef} | a\ra \}  \;\;\;\  \{ |A_j\ra \} = \{  |d\ra ,  \K_{cd}|b] ,  \K_{cd}\K_{aef}|e \ra  \} 
\notag \\
\{ |b_i ]  \} =& \{ |d] , K_{aef}|a\ra , \}  \;\;\;\ \{ |B_i]\}=\{ |b] , |c]    \} 
\notag \\
\{ |c_j\ra \} =& \{ |a\ra , |c\ra , \K_{cd}\K_{aef}|a\ra \} \;\;\;\ \{ |C_j\ra = \{ |d\ra , \K_{cd}|b] ,  \K_{cd}\K_{aef}|e\ra  \}
\notag \\
\{ |d_i] \} =& \{ |f] , \K_{aef}|a\ra  \} \;\;\; \{ |D_i] = \{ |c] , K_{aef}|e]  \} 
\end{align}
where $\alpha_j$ and $\beta_i$ are given in eq.(\ref{alphabetaeqn}) with $n_A=3,n_B=2$ 
and
\begin{align}
\gamma_j = { \prod_{k=1}^{n_C-1} \spa{c_k}.{C_j} \over \prod_{k\neq j} \spa{C_k}.{C_j} }
\;\;\;
\delta_i={ \prod_{k=1}^{n_D-1}  \spb{d_k}.{D_i} \over \prod_{k\neq i} \spb{D_k}.{D_i} }
\label{gammadeltaeq}
\end{align}
with $n_C=3,n_D=2$.

\begin{align}
C_{T10} =&  
-{ \spa{a}.b^8 \spb{b}.f [e|\K_{cd}|c\ra^4 \spb{c}.d
\over 
\spa{c}.d \spa{a}.f^2 \spa{b}.f [e|\K_{abf}|a \ra  \K^2_{abf} [e|\K_{abf}|b\ra  s_{cd}    }
\Biggl( 
\notag  \\
&
-[e|\K_{cd}|a\ra 
\sum_{i=1}^3\sum_{j=1}^3 \beta_i \alpha_j   H^d_0[ B_i,A_j,b_3,a_3 ]
+  
\la e|\K_{cd}|\K_{abf}|b \ra
\sum_{i=1}^3\sum_{j=1}^3 \delta_i \gamma_j  H^d_0[ D_i,C_j,d_3,c_3 ]; 
\Biggr) 
\end{align}
where
\begin{align}
\{ |a_j\ra \} =& \{ |e \ra , |c\ra , \K_{cd} | e] \}  \;\;\;\  \{ |A_j\ra \} = \{  |d\ra ,  \K_{cd}\K_{abf}|a\ra ,  \K_{cd}\K_{abf}|b \ra  \} 
\notag \\
\{ |b_i ]  \} =& \{ |d] , |e] , \K_{abf}|b\ra , \}  \;\;\;\ \{ |B_i]\}=\{ |c] , \K_{abf}|b\ra ,\K_{abf}|a\ra ,   \} 
\notag \\
\{ |c_j\ra \} =& \{ |a\ra , |c\ra , \K_{cd}|e] \} \;\;\;\ \{ |C_j\ra = \{ |d\ra , \K_{cd}\K_{abf}|a\ra  , \K_{cd}\K_{abf}|b \ra    \}
\notag \\
\{ |d_i] \} =& \{ |e] , |e] , |d]   \} \;\;\; \{ |D_i] = \{ |c] , \K_{abf}|a\ra , \K_{abf}|b\ra \} 
\end{align}
and the $\alpha_i,\beta_j,\gamma_i,\delta_j$ are given in eqns.(\ref{alphabetaeqn}) and (\ref{gammadeltaeq}) with
$n_A=n_B=n_C=n_D=3$.

\begin{align}
C_{T14} =&  
{ \spa{a}.b [f| \K_{abf}| \K_{cd}| d]^4
\over 
\spb{a}.b\spb{a}.f^2\K^2_{abf} [b|\K_{abf}|e\ra [f|\K_{abf}|e\ra \spa{c}.d^2 }
\Biggl( 
\notag \\
&
-[f|\K_{cd}|e\ra 
\sum_{i=1}^3\sum_{j=1}^2 \beta_i \alpha_j   H^d_0[ B_i,A_j,b_3,a_2 ]
-[b|\K_{abf}|\K_{c,d}|e]
  \sum_{i=1}^3\sum_{j=1}^2 \delta_i \gamma_j  H^d_0[ D_i,C_j,d_3,c_3 ]; 
\Biggr) 
\end{align}
\begin{align}
\{ |a_j\ra \} =& \{ |c \ra ,  \K_{abf} | f] \}  \;\;\;\  \{ |A_j\ra \} = \{  |e\ra ,  |d\ra  \} 
\notag \\
\{ |b_i ]  \} =& \{ |d] , |e] , \K_{cd}\K_{abf}|f] , \}  \;\;\;\ \{ |B_i]\}=\{ |c] , \K_{cd}|e\ra ,\K_{cd}\K_{abf}|b\ra ,   \} 
\notag \\
\{ |c_j\ra \} =& \{ |c\ra  , \K_{abf}|f] \} \;\;\;\ \{ |C_j\ra = \{ |d\ra , \K_{abf}|b]    \}
\notag \\
\{ |d_i] \} =& \{ |d] , |f] , \K_{cd}\K_{abf}|f]   \} \;\;\; \{ |D_i] = \{ |c] , \K_{cd}|e\ra , \K_{cd}\K_{abf}|b] \} 
\end{align}
and the $\alpha_i,\beta_j,\gamma_i,\delta_j$ are given in eqns.(\ref{alphabetaeqn}) and (\ref{gammadeltaeq}) with
$n_A=2, n_B=3, n_C=2,n_D=3$.

\begin{align}
C_{T11} =&  
{ \spb{c}.d^2\spa{a}.e \spb{e}.f^8\spa{b}.c^4
\over 
K^2_{aef}\spb{a}.e \spb{a}.f^2 [e|\K_{aef}|b\ra [f|\K_{aef}|b\ra      
} 
\Biggl( 
\\
&
[f|\K_{abef}|b\ra \sum_{i=1}^3\sum_{j=1}^2 \beta_i \alpha_j   H^d_0[ B_i,A_j,b_3,a_2 ]
-K^2_{aef}\spb{b}.e  \sum_{i=1}^3\sum_{j=1}^2 \delta_i \gamma_j  H^d_0[ D_i,C_j,d_3,c_3 ]; 
\Biggr) 
\end{align}
where
\begin{align}
\{ |a_j\ra \} =& \{ |b\ra , |c\ra \} \;\;\ \{ |A_j] \} = \{  |d] , K_{ae}  |f  \ra \} 
\notag \\
\{ |b_i ]  \} =& \{ |b] , |d] , K_{cd} |b\ra \}  \;\;\;\ \{ |B_i]\}=\{ |c] , K_{cd}K_{aef}|f] , K_{cd}K_{aef}|f] \} 
\notag \\
\{ |c_j\ra \} =& \{ |b\ra , |b\ra , |c\ra \} \;\;\;\ \{ |C_j\ra = \{ |d\ra , K_{aef}|e] ,  K_{aef}|f]  \}
\notag \\
\{ |d_i] \} =& \{ |f] , |d] , K_{cd}|b\ra \} \;\;\; \{ |D_i] = \{ |c] , K_{cd}K_{aef}|e] , K_{cd}K_{aef}|e] \} 
\end{align}
and the $\alpha_i,\beta_j,\gamma_i,\delta_j$ are given in eqns.(\ref{alphabetaeqn}) and (\ref{gammadeltaeq}) with
$n_A=n_B=n_C=n_D=3$.


The remaining pieces come from $T_2$, $T_6$, $T_8$, $T_9$, $T_{12}$ and $T_{13}$ and all involve  massive propagators.
These terms  generically take the form
\begin{equation}
\sim { f(\ell) \over  (\ell+Q)^2 }\,.
\end{equation}
Other denominators such as $[\ell|Q|\ell\ra^{-1}$ and $[\alpha|K_{\ell+Q}|\beta\ra^{-1}$ appear but these can be manipulated 
into a common $[\ell|Q|\ell\ra^{-1}$  form using
\begin{equation}
[\alpha |(\ell+Q)|\beta\rangle
=[\ell|\alpha] \langle \beta \ell \ra + [ \alpha |Q|\beta \ra 
=[\ell|\left( \alpha] \langle \beta +{ [ \alpha |Q|\beta \ra \over P^2 } P  \right)  \ell \ra
={ [ \alpha |Q|\beta \ra \over P^2 } [\ell|  \tilde{Q} | \ell \ra 
\end{equation}
with
\begin{equation}
\tilde{Q}= P + {P^2 \over [ \alpha |Q|\beta \ra }\bar\lambda_{\alpha}\lambda_{\beta}
\end{equation}
where we have used that, on the cut $(l_1-P)^2=0$, so that $[l_1|P|l_1\ra=P^2$. 
Also 
\begin{equation}
{ (\ell+Q)^2 }={  [\ell|Q|\ell\rangle +Q^2 }
\equiv
{  [\ell|Q|\ell\rangle +Q^2 { [\ell|P|\ell\rangle\over P^2 } }
= [\ell|\tilde Q|\ell\rangle 
\end{equation}
where 
\begin{equation}
\tilde Q^\mu =Q^\mu + {Q^2\over P^2} P^\mu  \,.
\end{equation}

The previous six terms were of overall order $\ell^4$. When combined with the other  tree amplitude and multiplying by 
the $\rho$-factor the resulting cut was of order $\ell^0$.  
The $\ell$ count of these terms is $+6$ and there is no straightforward way of implementing a reduction. 
As in the massless case the $\rho$-factors lower the overall power count of the $\NeqFour$ contribution by $8$,  
leading to an overall power count of $+2$. This significantly increases the complexity of the expressions. 
The leading large $\ell$ contributions cancel between the terms at large $\ell$ indicating that there is probably an underlying  
simpler version of the bubble coefficient.   
The form of the six-point NMHV amplitude was obtained using a BCFW shift on legs $a^-$ and $f^+$. We have evaluated alternative
forms using alternative shifts e.g. $a^-$ and $b^-$ but the resultant expressions
still include terms which would be $\ell^6$ or have equivalent problems. 

Fortunately only three of these contributions are required since
\begin{align}
 C_{T6}+C_{T9}+C_{T12}=C_{T2}+C_{T8}+C_{T13}\,.
\end{align}
We present the analysis of  term $C_{T2}$ in detail below and the results 
for  $C_{T8}$ and $C_{T13}$ after.

Term $T_2$ is
\begin{align}
T_2&=-i{ \la bl_1\ra [f|P_{bfl_1}|a\ra^8  [el_2]
          \over
         \la ae\ra  \la al_2\ra \la el_2\ra [b|P_{bfl_1}|a\ra [l_1|P_{bfl_1}|a\ra 
         [f|P_{bfl_1}|e\ra [f|P_{bfl_1}|l_2\ra  [bf] [bl_1] [fl_1] t_{bfl_1}
        }  
        \bigg[-i {\la al_2\ra  [fl_1] \over [f|P_{bfl_1}|a\ra }\bigg]^{A}, 
\label{termt2}
\end{align}
where $A=4-2h$.

The denominator of \eqref{termt2}  includes a  products of three massive factors:
\begin{align}
{1\over    [b|P_{bfl_1}|a\ra  [f|P_{bfl_1}|e\ra t_{bfl_1} } \,.
\end{align}
These denominator factors can be rewritten as

\begin{align}
[b|P_{bfl_1}|a\ra &=[b|f|a\ra+[l_1b]\la al_1\ra 
= [l_1\biggl|\biggl({[b|f|a\ra\over P^2}P+\bar\lambda_b \lambda_a\biggr)\biggr|l_1\ra
\equiv {[b|f|a\ra\over P^2}[l_1|Q_{2:1}|l_2\ra
\notag \\
[f|P_{bfl_1}|a\ra &=[f|b|a\ra+[l_1f]\la al_1\ra 
= [l_1\biggl|\biggl({[f|b|a\ra\over P^2}P+\bar\lambda_f \lambda_a\biggr)\biggr|l_1\ra
\equiv {[f|b|a\ra\over P^2}[l_1|Q_{2:2}|l_1\ra 
\notag \\
t_{bfl_1}&= [l_1|P_{bf}|l_1\ra +s_{bf}
= [l_1\biggl|\biggl({s_{bf}\over P^2}P+ P_{bf}\biggr)\biggr|l_1\ra
\equiv [l_1|Q_{2:3}|l_1\ra
\end{align}
\def\qs{
Q21= [b|f|a>/P^2 P +|b]<a| scaled to Q=P+...
Q22= [f|b|e>/P^2 P +|f]<e| scaled to Q=P+...
Q23= s[b,f]/P^2  P +b +f 
}
Drawing in two $l_1$ dependent factors from the numerator, the massive factor can be separated using 
\begin{align}
{\la xl_1\ra \la yl_1\ra\over [l_1|Q_{2:1}|l_1\ra[l_1|Q_{2:2}|l_1\ra[l_1|Q_{2:3}|l_1\ra}
= &
{ \la yl_1\ra\over [l_1|Q_{2:1}|l_1\ra [l_1|Q_{2:2}Q_{2:3}|l_1]}
\biggl( {[l_1|Q_{2:2}|x\ra\over [l_1|Q_{2:2}|l_1\ra}-{[l_1|Q_{2:3}|x\ra\over [l_1|Q_{2:3}|l_1\ra}
\biggr)
\notag \\
= &
{1\over  [l_1|Q_{2:2}Q_{2:3}|l_1]}
\Biggl(
{ [l_1|Q_{2:2}|x\ra\la yl_1\ra\over [l_1|Q_{2:1}Q_{2:2}|l_1] }
\biggl( {[l_1|Q_{2:1}|y\ra\over [l_1|Q_{2:1}|l_1\ra}-{[l_1|Q_{2:2}|y\ra\over [l_1|Q_{2:2}|l_1\ra} \biggr)
\notag \\ &
-{ [l_1|Q_{2:3}|x\ra \over [l_1|Q_{2:1}Q_{2:3}|l_1] }
\biggl( {[l_1|Q_{2:1}|y\ra\over [l_1|Q_{2:1}|l_1\ra}-{[l_1|Q_{2:3}|y\ra\over [l_1|Q_{2:3}|l_1\ra}\biggr)
\Biggr)\,.
\end{align}
The $[l_1|Q_{2:i}Q_{2:j}|l_1]$ factors can be split by  defining $\hat Q_2^{ij}=Q_{2:i}-\alpha Q_{2:j}$  where $(\hat Q_2^{ij})^2=0$, so that, 
\begin{equation}
[l_1|Q_{2:i}Q_{2:j}|l_1]=[l_1|(Q_{2:i}-\alpha Q_{2:j})Q_{2:j}|l_1]=[l_1|\hat Q_2^{ij}Q_{2:j}|l_1]=[l_1 \hat Q_2^{ij}]\la \hat Q_2^{ij}|Q_{2:j}|l_1]\,.
\end{equation}
These factors can then be treated in the same way as the massless factors.  
As the full denominator may contain factors of the form $\spa{x}.{l_1}$, partial fractioning on both $|l_1\ra$ and $|l_1]$ yields
terms with loop momentum dependent factors of the form:
\begin{equation}
 {[A|l_1|B\ra [C|l_1|D\ra+\gamma [E|l_1|F\ra\over [l_1|Q|l_1\ra}{[Yl_1]\over [Xl_1]}{\la yl_1\ra \over \la x l_1\ra }\,.
 \label{calG111}
\end{equation}
The canonical form arising from the terms in  \eqref{calG111} are the $G_{111}$ functions as defined in appendix \ref{Canonformsection}. 

The cut momentum $P_{cd}$ is specified by the sum of the two null momenta $c$ and $d$, however it is often convenient to express $P_{cd}$ as the sum of two alternative null momenta. For any null momentum
$x$, setting 
\def\hash{*}
\begin{align}
 x^\hash={P_{cd}^2\over [x|P_{cd}|x\ra } x,
\end{align}
gives
\begin{align}
P= P^\flat_x +x^\hash 
\end{align}
with $(P^\flat_x)^2=0$.

Defining
\begin{align}
\lambda_{X}&=   [f|P_{bc}|a\ra \lambda_c +[fd]\spa{c}.a \lambda_d
\notag \\
\lambda_{Q_{2:1}^x} &=   {s_{cd}\over [b|f|a\ra}\lambda_a \qquad \bar\lambda_{Q_{2:1}^x}=\lambda_{Q_{2:1}^x}^*
\notag \\
\lambda_{Q_{2:2}^x} &=   {s_{cd}\over [f|b|e\ra}\lambda_e \qquad \bar\lambda_{Q_{2:2}^x}=\lambda_{Q_{2:2}^x}^*   
\notag \\
\{ |b_i] \} &=\{ \la X_2|P_{cd}|  \spacycomA  \la X_2|P_{cd}|\spacycomA  \la X_2|P_{cd}|\spacycomA  \la X_2|P_{cd}| \}
\notag \\
\{ |B_i] \}&=\{ [b| \spacycomA [c| \spacycomA  \la e|P_{cd}| \spacycomA  \la a|P_{bf}|\spacycomA [f|P_{ae}P_{cd}| \}
\notag \\
\{ |a_i^1 \rangle \}&=\{ \la b| \spacycomA \la c| \spacycomA [f|Q_{2:1}| \spacycomA \la a|P_{cd}Q_{2:1}| \}
\notag \\
\{ |a_i^2 \rangle \}&=\{ \la b| \spacycomA \la c| \spacycomA [f|Q_{2:2}| \spacycomA \la a|P_{cd}Q_{2:2}| \}
\notag \\
\{ |a_i^3 \rangle \}&=\{ \la b| \spacycomA \la c| \spacycomA [f|Q_{2:3}| \spacycomA \la a|P_{cd}Q_{2:3}| \}
\notag \\
\{ |A_i^1 \rangle \}&=\{ \la d| \spacycomA \la f| \spacycomA \la \hat Q_2^{1,2}| \spacycomA [ \hat Q_2^{1,2}|Q_{2:2}| \spacycomA \la a|P_{bf}Q_{2:3}| \}
\notag \\
\{ |A_i^2 \rangle \}&=\{ \la b| \spacycomA \la d| \spacycomA \la \hat Q_2^{1,2}| \spacycomA [ \hat Q_2^{1,2}|Q_{2:2}| \spacycomA \la e|P_{bf}Q_{2:3}| \}
\notag \\
\{ |A_i^3 \rangle \}&=\{ \la b| \spacycomA \la d| \spacycomA \la f| \spacycomA  \la a|P_{bf}Q_{2:3}| \spacycomA \la e|P_{bf}Q_{2:3}| \}
\end{align}

\begin{align}
\alpha^k_j=&
{\prod_{i=1}^4 \spa{a^k_i}.{A^k_j} \over \prod_{j\neq i} \spa{A^k_i}.{A^k_j} }
\;\;\; 
\beta_j=
{\prod_{i=1}^4 \spb{b_i}.{B_j} \over \prod_{i\neq j}^5 \spb{B_i}.{B_j} }
\end{align}
the contribution of term $T_2$ to the bubble coefficient is            
\begin{align}
C_{T2}=
{\spb{c}.d  \spa{f}.b \over \spa{a}.e \spb{b}.f s_{cd}^2\spa{c}.d 
}
& \sum_{j=1}^5\sum_{i=1}^5\sum_{k=1}^3 
\rho^k
\alpha_j^k\beta_i
\notag \\
                   \times \biggl( & \hskip 52pt  G_{111}^s\bigr[ [e|P_{cd}| ,A_i^k, [d|, B_j, [f| , \la a|, Q_{2:1} ,\{ b^\hash, P^\flat_b \}, P_{cd}, Q_{2:1}^x \bigr]
\notag \\
        &-2[f|b|a\ra           G_{011}  \bigr[ [e|P_{cd}| ,A_i^k, [d|, B_j, [f| , \la a|, Q_{2:1} ,\{  b^\hash, P^\flat_b\}, P_{cd}, Q_{2:1}^x \bigr]
\notag \\
        &+[f|b|a\ra^2          G_{s11}  \bigr[ [e|P_{cd}| ,A_i^k, [d|, B_j, [f| , \la a|, Q_{2:1} ,\{  b^\hash, P^\flat_b\}, P_{cd}, Q_{2:1}^x \bigr]
        \biggr)
\end{align}
where
\begin{equation}
\rho^1= { 1 \over [f|b|e\ra} , \;\; \rho^2={1 \over [b|f|a\rangle } , \;\;
\rho^3= -{ \spa{f}.b \over s_{cd} } \,.
\end{equation}

In the above $x$ and $w$ represent arbitrary null momenta.


\vfill\eject


\vfill\eject

{\bf $C_{T8}$ and $C_{T13}$} 

For $C_{T8}$ we have 
\begin{align}
C_{T8}= & 
 {\spa{a}.e \spb{c}.d\over \spa{a}.f^3 \spa{c}.d s_{cd}^2 \spb{e}.b }      \sum_{i=1}^6\sum_{j=1}^3  
          \beta_i \Biggl( 
\notag \\ &  \alpha_j^1
\begin{pmatrix}          
               G_{111}^s[\la c|,A_j, [d|,B_i, [e|, \la a|,Q_{8:1},\{ b^*,P_b^\flat\},P_{cd},{{Q^x_{8:1}}}]       
\\
                 -[e|b|a\ra G_{011}[\la c|,A_j, [d|,B_i, [e|, \la a|,Q_{8:1},\{ b^*,P_b^\flat\},P_{cd},{{Q^x_{8:1}}}] 
\end{pmatrix}
\notag \\ &  -\alpha_j^2
\begin{pmatrix}
                           G_{111}^s[\la c|,A_j, [d|,B_i, [e|, \la a|,Q_{8:2},\{ x^*,P_x^\flat\},P_{cd},{{w}}]       
\\
                  -[e|b|a\ra G_{011}[\la c|,A_j, [d|,B_i, [e|, \la a|,Q_{8:2},\{ x^*,P_x^\flat\},P_{cd},{{w}}] 
\end{pmatrix} \Biggr)
\notag \\ 
    +   &       {\spa{b}.e [cd]\over \spa{a}.f^3\spa{c}.d s_{cd}^2 [be]^2} \sum_{i=1}^6\sum_{j=1}^2 
\gamma_i \Biggl(
\notag \\ & \delta_j^1
\begin{pmatrix} 
                              G_{111}^s[\la c|,C_j, [d|,D_i, [e|, \la a|,Q_{8:1},\{ b^*,P_b^\flat\},P_{cd},{{Q^x_{8:1}}}]       
\\
                     -[e|{b}|a\rangle G_{011}[\la c|,C_j, [d|,D_i, [e|, \la a|,Q_{8:1},\{ b^*,P_b^\flat\},P_{cd},{{Q^x_{8:1}}}] 
\end{pmatrix}
\notag \\ & -\delta_j^2
\begin{pmatrix}                       
                       G_{111}^s[\la c|,C_j, [d|,D_i, [e|, \la a|,Q_{8:2},\{ x^*,P_x^\flat\},{{w}}]       
\\
            -[e|b|a\ra   G_{011}[\la c|,C_j, [d|,D_i, [e|, \la a|,Q_{8:2},\{ x^*,P_x^\flat\},{{w}}] 
\end{pmatrix} \Biggr)
\end{align}
where we define 
\begin{align}
\lambda_{X}=   -\lambda_{c}[ e|{b}|a \ra 
                   + \lambda_{c} \spb{e}.c \spa{a}.c  
                   + \lambda_{d} \spb{e}.d \spa{a}.c 
\end{align}
\begin{align}
   \lambda_{Q^a_{8:1}}&=   \lambda_{a}{ s_{cd}\over [b|e|a \ra}
   \,\, ,\,\,
\bar\lambda_{Q^a_{8:1}}=\bar\lambda_{b}
 \,\, ,\,\,
\lambda_{Q^x_{8:1}}=          \lambda_{Q^a_{8:1}}
 \,\, ,\,\,
\bar\lambda_{Q^x_{8:1}}=         \lambda_{Q^a_{8:1}}^*
\end{align}
\begin{align}
 Q_{8:1}= {Q^a_{8:1}}+k_c+k_d
\end{align}

\begin{align}
   \lambda_{Q^a_{8:2} }=   \lambda_{c} {s_{be}\over s_{cd} }
  \,\, ,\,\,   
\bar\lambda_{Q^a_{8:2} }=\bar\lambda_{c }
  \,\, ,\,\,
   \lambda_{Q^b_{8:2} }=   \lambda_{d }
     \,\, ,\,\,
\bar\lambda_{Q^b_{8:2} }=\bar\lambda_{d }{s_{be}\over s_{cd}} 
\end{align}
\begin{align}
 Q_{8:2}=Q^a_{8:2}+Q^b_{8:2}+k_b+k_e
\end{align}
\begin{align}
 \hat P_8=P_{cd}+\alpha P_{af} \,\, {\rm s.t.} \,\, \hat P_8^2=0
\end{align}

\begin{align}
\{ | A_j \ra \} & = \{ \la a| \spacycomA  \la d| \spacycomA \la e|      \}
\; , \;\;
\{ | a_j^1 \ra \}=\{ \la b| \spacycomA  \la a|P_{cd}Q_{8:1}|          \}
\; , \;\; 
\{ | a_j^2 \ra \}=\{ \la b| \spacycomA  \la a|P_{cd}Q_{8:2}|          \}
\notag \\
\{ | B_i ]  \} &=  \{ [b| \spacycomA  \la a|P_{be}|  \spacycomA  [{b}|P_{af}P_{cd}| \spacycomA
                 [c| \spacycomA \la{f}|P_{cd}|  \spacycomA  [{e}|P_{af}P_{cd}|    \}
\notag \\
\{ | b_i ]  \}&=\{ \la{a}|P_{cd}| \spacycomA  \la X|P_{cd}| \spacycomA \la X|P_{cd}| \spacycomA \la X|P_{cd}| \spacycomA \la X|P_{cd}|     \}
\notag \\
\{ | C_j \ra \} &= \{ \la{d}| \spacycomA \la{e}|      \}
\; , \;\;
\{ | c_j^1 \ra \}=\{ \la{a}|P_{cd}Q_{8:1}|           \}
\; , \;\;
\{ | c_j^2 \ra \}=\{ \la{a}|P_{cd}Q_{8:2}|            \}
\notag \\
\{ | D_i ]  \}  &=\{ [c| \spacycomA  \la{a}|P_{be}|  \spacycomA  [{b}|P_{af}P_{cd}| \spacycomA
                \la{f}|P_{cd}|  \spacycomA  [{\hat P_8}|  \spacycomA  \la{\hat P_8}|P_{af}|            \}
\notag \\
\{ | d_i ]  \} & =\{ \la{a}|P_{cd}| \spacycomA  \la X|P_{cd}| \spacycomA \la X|P_{cd}| \spacycomA \la X|P_{cd}| \spacycomA \la X|P_{cd}|     \}
\end{align}

\begin{align}
\alpha_j^k =&
{\prod_{i=1}^2 \spa{a^k_i}.{A^k_j} \over \prod_{i\neq j} \spa{A^k_i}.{A^k_j} }
\,\, ,\,\,
\beta_j=
{\prod_{i=1}^5 \spb{b_i}.{B_j} \over \prod_{i\neq i} \spb{B_i}.{B_j} }
\end{align}
\begin{align}
\gamma_j^k=
{ \spa{c^k_1}.{C^k_j} \over \prod_{i\neq j} \spa{C^k_i}.{C^k_j} }
\,\, ,\,\,
\delta_j=&
{\prod_{i=1}^5 \spb{d_i}.{D_j} \over \prod_{i\neq j} \spb{D_i}.{D_j} }
\end{align}

Finally, 
\begin{align}
C_{T13}=
 -& {[fb][cd]\spb{e}.b \over \spa{c}.d \spa{e}.b^2 \spb{f}.a^3 s_{cd}^2}     \sum_{i=1}^6 \sum_{j=1}^3 
  \beta_i  \Biggl(    
\notag \\ \alpha_j^1&
\begin{pmatrix}
                             G_{111}^s[\la c|,A_j,[d|,B_i,[f| , \la b|,Q_{13:1},\{ f^*,P_f^\flat\},P_{cd},{{Q_{13:1}^x}}] 
\\
                   -[f|a|b\ra  G_{011}[\la c|,A_j,[d|,B_i,[f| , \la b|,Q_{13:1},\{ f^*,P_f^\flat\},P_{cd},{{Q_{13:1}^x}}] 
\end{pmatrix}
\notag \\ -\alpha_j^2&
\begin{pmatrix}
                             G_{111}^s[\la c|,A_j,[d|,B_i,[f| , \la b|,Q_{13:2},\{ x^*,P_x^\flat\},P_{cd},{{w}}] 
\\
                   -[f|a|b\ra  G_{011}[\la c|,A_j,[d|,B_i,[f| , \la b|,Q_{13:2},\{ x^*,P_x^\flat\},P_{cd},{{w}}] 
\end{pmatrix}
\Biggr)
\notag \\ +  & {\spb{c}.d\spb{e}.b \over \spa{c}.d\spa{e}.b^2 \spb{f}.a^3 s_{cd}^2}     \sum_{i=1}^6 \sum_{j=1}^3
       \delta_i\Biggl(
\notag \\ & \gamma_j^1
\begin{pmatrix}                 
                            G_{111}^s[\la c|,D_j,[d|,C_i,[f| , \la b|,Q_{13:1},\{ f^*,P_f^\flat\},P_{cd},{{Q_{13:1}^x}}] 
\\
                   -[f|a|b\ra G_{011}[\la c|,D_j,[d|,C_i,[f| , \la b|,Q_{13:1},\{ f^*,P_f^\flat\},P_{cd},{{Q_{13:1}^x}}] 
\end{pmatrix}
\notag \\ &-\gamma_j^2
\begin{pmatrix}                 
                             G_{111}^s[\la c|,D_j,[d|,C_i,[f| , \la b|,Q_{13:2},\{ x^*,P_x^\flat\},P_{cd},{{w}}] 
\\
                   -[f|a|b\ra  G_{011}[\la c|,D_j,[d|,C_i,[f| , \la b|,Q_{13:2},\{ x^*,P_x^\flat\},P_{cd},{{w}}] 
\end{pmatrix}  \Biggr)                                
\end{align}
where
\begin{align}
\lambda_{X}=   \lambda_{c} [ f|a|b \ra 
                   + \lambda_{c} \spa{c}.b [fc]
                   + \lambda_{d} \spa{c}.b [fd ]
\end{align}
\begin{align}
   \lambda_{Q_{13:1}^a}=   \lambda_{e} { s_{cd}\over [f|a|e \ra}
\,\, ,\,\,
   \bar\lambda_{Q_{13:1}^a}=\bar\lambda_{f}
\,\, ,\,\,
   \lambda_{Q_{13:1}^x }=          \lambda_{Q_{13:1}^a}
\,\, ,\,\,
\bar\lambda_{Q_{13:1}^x}=          \lambda_{Q_{13:1}^a}^*
\end{align}
\begin{align}
Q_{13:1}= Q_{13:1}^a+k_c+k_d
\end{align}
\begin{align}
   \lambda_{Q_{13:2}^a}=   \lambda_{c}{[a|f|a\ra\over s_{cd}}
\,\, ,\,\,
\bar\lambda_{Q_{13:2}^a }=\bar\lambda_{c}
\,\, ,\,\,
   \lambda_{Q_{13:2}^b }=   \lambda_{d}
\,\, ,\,\,
\bar\lambda_{Q_{13:2}^b}=\bar\lambda_{d}{ [a|f|a\ra\over s_{cd}}
\end{align}

\begin{align}
Q_{13:2}=Q_{13:2}^a+Q_{13:2}^b+k_a+k_f
\end{align}

\begin{align}
\{ |A_j\rangle \} &= \{ \la b| , \; \la d|  , \; [{b}|P_{cd}| \}
\; , \;\;
\{ |a_j^1\rangle \}=\{ [{e}|P_{cd}| ,  [{f}|Q_{13:1}|        \}
\; , \;\;
\{ |a_j^2\rangle \}=\{ [{e}|P_{cd}| , [{f}|Q_{13:2}|        \}
\notag \\
\{ |B_i] \}&=\{ [a| ,   \la{e}|P_{af}|  ,   [{f}|P_{be}P_{cd}| , 
                [c| ,  \la{b}|P_{af}|  ,  \la{e}|P_{cd}|                    \}
\notag \\
\{ |b_i] \}&=\{ \la{b}|P_{cd}| \spacycomA  \la{X}|P_{cd}|  \spacycomA \la{X}|P_{cd}|  \spacycomA 
                                               \la{X}|P_{cd}|  \spacycomA \la{X}|P_{cd}|            \}
\notag \\
\{ |C_j\rangle \} &= \{ \la{{b}}| \spacycomA \la{{d}}|     \spacycomA [{b}|P_{cd}| \}
\; , \;\;
\{ |c_j^1\rangle \}=\{ [{f}|P_{cd}|  \spacycomA [{f}|Q_{13:1}|        \}
\; , \;\;
\{ |c_j^2\rangle \}=\{ [{f}|P_{cd}|  \spacycomA [{f}|Q_{13:2}|        \}
\notag \\
\{ |D_i] \}&=\{  [{{a}}| \spacycomA  \la {e}|P_{af}|  \spacycomA  [{f}|P_{be}P_{cd}| \spacycomA
                 [{{c}}| \spacycomA  [{\hat P_{13}}|  \spacycomA  \la{\hat P_{13}}|P_{af}|                    \}
\notag \\
\{ |d_i] \}&=\{ \la{b}|P_{cd}| \spacycomA  \la{X}|P_{cd}|  \spacycomA \la{X}|P_{cd}|  \spacycomA 
                                               \la{X}|P_{cd}|   , \;  \la{X}|P_{cd}|            \}
\end{align}

\begin{align}
\alpha_j^k =&
{\prod_{i=1}^2 \spa{a^k_i}.{A^k_j} \over \prod_{i\neq j} \spa{A^k_i}.{A^k_j} }
\,\, ,\,\,
\beta_j=
{\prod_{i=1}^5 \spb{b_i}.{B_j} \over \prod_{i\neq i} \spb{B_i}.{B_j} }
\end{align}
\begin{align}
\gamma_j^k=
{\prod_{i=1}^2 \spa{c^k_i}.{C^k_j} \over \prod_{i\neq j} \spa{C^k_i}.{C^k_j} }
\,\, ,\,\,
\delta_j=&
{\prod_{i=1}^5 \spb{d_i}.{D_j} \over \prod_{i\neq j} \spb{D_i}.{D_j} }
\end{align}

\vfill\eject

\section{${G}_{111}$ Canonical Form}
\label{Canonformsection}

\def\showcase#1{#1}
\def\g{\gamma}
\def\tQ{\tilde Q}

The massive canonical form required for the bubble coefficients is
\begin{align}
 {\cal G}_{111}[  x,y ,a,A,b,B,f,e,\tQ,\ell] =[\ell x] \spa{\ell}.y {\spa{\ell}.a \over \spa{\ell}.A}{[\ell b] \over [\ell B]}{ [\ell f]\spa{\ell}.e \over [\ell|\tQ|\ell\ra}
\end{align}
with corresponding canonical function 
\begin{equation}
G_{111}[x,y,a,A,b,B,f,e,\tQ,c,d,P]
\end{equation}

Specifically the 2:4 bubble coefficients involve cut integrands of the form
\begin{align}
{\cal G}_{111}^s[     a,A,b,B,f,e,\tQ,\ell] &=                 { \spa{\ell}.a \over \spa{\ell}.A}{[\ell b]\over [\ell B] }{ [\ell f]^2 \spa{\ell}.e^2 \over [\ell|\tQ|\ell\ra}
\end{align}
and
\begin{align}
{\cal G}_{011}[      a,A,b,B,f,e,\tQ,\ell] &=                  { \spa{\ell}.a \over \spa{\ell}.A}{[\ell b]\over [\ell B] }{ [\ell f]   \spa{\ell}.e   \over [\ell|\tQ|\ell\ra}
\end{align}
which are special cases of ${\cal G}_{111}$ with corresponding functions 
given by 
\begin{align}
G_{111}^s[a,A,b,B,f,e,\tQ,c,d,P]=&G_{111}[f,e,a,A,b,B,f,e,\tQ,c,d,P]
\notag \\
G_{011}[a,A,b,B,f,e,\tQ,c,d,P]=& 
 -( G_{111}[c,c,a,A,b,B,f,e,\tQ,c,d,P]
\notag \\ & + G_{111}[d,d,a,A,b,B,f,e,\tQ,c,d,P])/s_{cd} 
\end{align}
The function of $G_{111}$ is derived below.

\subsubsection{Real $\tQ$}

For $\tQ$ real the canonical form $G_{111}$ can be evaluated using a basis for the loop momentum based 
on any pair of real null momenta $c$ and $d$ since,  if the momentum crossing the cut is the sum of two null momenta, $P=c+d$, the  cut loop momentum can be parametrised by 
\begin{align}
\lambda_{\ell}=\cos{\theta\over 2}\lambda_c +\sin{\theta\over 2}\e^{-i\phi}\lambda_d \quad ,\qquad
\bar\lambda_{\ell}=\cos{\theta\over 2}\bar\lambda_c +\sin{\theta\over 2}\e^{i\phi}\bar\lambda_d \,.
\end{align}

The expression for the canonical form has a range of special cases 
if certain combinations of $A$, $B$, $P$ and $\tQ$ vanish:
\begin{align}
G_{111}=\begin{cases}
         G_{111}^g    & [B|P|A\ra \neq 0, \la A|P\tQ|A\ra \neq 0, [B|P\tQ|B] \neq 0 \\
         G_{111}^x    & [B|P|A\ra   =  0, \la A|P\tQ|A\ra \neq 0, [B|P\tQ|B] \neq 0 \\
         G_{111}^{y1} & [B|P|A\ra \neq 0, \la A|P\tQ|A\ra  =   0, [B|P\tQ|B] \neq 0 \\
         G_{111}^{y2} & [B|P|A\ra \neq 0, \la A|P\tQ|A\ra \neq 0, [B|P\tQ|B]   =  0 \\
         G_{111}^{y12}& [B|P|A\ra \neq 0, \la A|P\tQ|A\ra  =   0, [B|P\tQ|B]   =  0 \\
         G_{111}^{xy} & [B|P|A\ra  =   0, \la A|P\tQ|A\ra  =   0, [B|P\tQ|B]   =  0 \\         
        \end{cases}
\end{align}
The full canonical form can be split into terms involving just the $[\ell|\tQ|\ell\ra^{-1}$ pole
and terms involving one or both of $ {\spa{\ell}.A}^{-1}$ and $[\ell B]^{-1}$. The contributions from terms involving no extra pole (np), an extra angle pole (ap), an extra square pole (sp) 
and both extra poles (dp) are given explicitly below. The decomposition of each of the special cases for $G_{111}$ into these pieces is:

\def\Cnp0{C_{np:0}}
\def\CnpgR{C_{np:\g R}}
\def\CnpgI{C_{np:\g I}}
\def\Cap0{C_{ap:0}}
\def\Capa{C_{ap:a}}
\def\CapgR{C_{ap:\g R}}
\def\CapgI{C_{ap:\g I}}
\def\Csp0{C_{sp:0}}
\def\Csps{C_{sp:s}}
\def\CspgR{C_{sp:\g R}}
\def\CspgI{C_{sp:\g I}}
\def\Cdp0{C_{dp:0}}
\def\Cdpa{C_{dp:a}}
\def\Cdps{C_{dp:s}}
\def\CdpgR{C_{dp:\g R}}
\def\CdpgI{C_{dp:\g I}}

\begin{align}
G_{111}^g= ( & \hskip 5pt \Cnp0
                                                 +\CnpgR
                                                 +\CnpgI
                                                 + \Cap0
                                                 + \Capa
                                                 +\CapgR
                                                 +\CapgI
                                                  \notag \\&
                                                + \Csp0
                                                 + \Csps
                                                 +\CspgR
                                                 +\CspgI
                                                 + \Cdp0
                                                 + \Cdpa
                                                 + \Cdps
                                                 +\CdpgR
                                                 +\CdpgI
                                            )
\end{align}                                            
\def\Cdpax{C_{dp:ax}}
\def\Cdpsx{C_{dp:sx}}
\def\CdpgRx{C_{dp:\g Rx}}
\def\CdpgIx{C_{dp:\g Ix}}      

\begin{align}
G_{111}^x= ( & \hskip 5pt \Cnp0
                                                 +\CnpgR
                                                 +\CnpgI
                                                 + \Cap0
                                                 + \Capa
                                                 +\CapgR
                                                 +\CapgI
                                                  \notag \\&
                                                + \Csp0
                                                 + \Csps
                                                 +\CspgR
                                                 +\CspgI
                                                 + \Cdp0
                                                 + \Cdpax
                                                 + \Cdpsx
                                                 +\CdpgRx
                                                 +\CdpgIx
                                            )
\end{align}                                            
\def\Capax{C_{ap:ax}}
\def\CapgRx{C_{ap:\g Rx}}
\def\CapgIx{C_{ap:\g Ix}}
\def\Cdpay{C_{dp:ay}}
\def\CdpgRy1{C_{dp:\g Ry1}}
\def\CdpgIy1{C_{dp:\g Iy1}}

\begin{align}
G_{111}^{y1}= ( & \hskip 5pt \Cnp0
                                                 +\CnpgR
                                                 +\CnpgI
                                                 + \Cap0
                                                 + \Capax
                                                 +\CapgRx
                                                 +\CapgIx
                                                  \notag \\&
                                                + \Csp0
                                                 + \Csps
                                                 +\CspgR
                                                 +\CspgI
                                                 + \Cdp0
                                                 + \Cdpay
                                                 + \Cdps
                                                 +\CdpgRy1
                                                 +\CdpgIy1
                                            )
\end{align}                                            
  
\def\Cspsx{C_{sp:sx}}
\def\CspgRx{C_{sp:\g Rx}}
\def\CspgIx{C_{sp:\g Ix}}
\def\Cdpsy{C_{dp:sy}}
\def\CdpgRy2{C_{dp:\g Ry2}}
\def\CdpgIy2{C_{dp:\g Iy2}}

\begin{align}
G_{111}^{y2}= ( & \hskip 5pt \Cnp0
                                                 +\CnpgR
                                                 +\CnpgI
                                                 + \Cap0
                                                 + \Capa
                                                 +\CapgR
                                                 +\CapgI
                                                  \notag \\&
                                                + \Csp0
                                                 + \Cspsx
                                                 +\CspgRx
                                                 +\CspgIx
                                                 + \Cdp0
                                                 + \Cdpa
                                                 + \Cdpsy
                                                 +\CdpgRy2
                                                 +\CdpgIy2
                                            )
\end{align}                                            

\def\Cdpaxy{C_{dp:axy}}
\def\Cdpsxy{C_{dp:sxy}}
\def\CdpgRxy{C_{dp:\g Rxy}}
\def\CdpgIxy{C_{dp:\g Ixy}}

\begin{align}
G_{111}^{xy}= ( & \hskip 5pt \Cnp0
                                                 +\CnpgR
                                                 +\CnpgI
                                                 + \Cap0
                                                 + \Capax
                                                 +\CapgRx
                                                 +\CapgIx
                                                  \notag \\&
                                                + \Csp0
                                                 + \Cspsx
                                                 +\CspgRx
                                                 +\CspgIx
                                                 + \Cdp0
                                                 + \Cdpaxy
                                                 + \Cdpsxy
                                                 +\CdpgRxy
                                                 +\CdpgIxy
                                            )
\end{align}

Using the definitions,
\begin{align}
{\cal A}_1&=   \frac{\spa{d}.{a} \spa{d}.{e}}{\spa{d}.{A}^2 } \left(\spa{c}.{d}
    \left(\frac{\spa{a}.{A}
    \spa{d}.{y}}{\spa{d}.{a}}+\frac{\spa{e}.{A}
    \spa{d}.{y}}{\spa{d}.{e}}+\spa{y}.{A}\right)
    +2 \spa{c}.{A}
    \spa{d}.{y}\right)
\notag \\
{\cal A}_0&=   \frac{\spa{d}.{a} \spa{d}.{e}}{\spa{d}.{A}^3 } \biggl(\spa{c}.{A}
    \spa{c}.{d} \left(\frac{\spa{a}.{A}
    \spa{d}.{y}}{\spa{d}.{a}}+\frac{\spa{e}.{A}
    \spa{d}.{y}}{\spa{d}.{e}}+\spa{y}.{A}\right)
\notag \\ & \hskip 70pt    +\spa{c}.{d}^2 \left(\frac{\spa{a}.{A} \spa{e}.{A}
    \spa{d}.{y}}{\spa{d}.{a} \spa{d}.{e}}+\frac{\spa{a}.{A}
    \spa{y}.{A}}{\spa{d}.{a}}+\frac{\spa{e}.{A}
    \spa{y}.{A}}{\spa{d}.{e}}\right)
    +\spa{c}.{A}^2 \spa{d}.{y}\biggr)
\notag \\
{\cal A}_p&=   \frac{\spa{a}.{A} \spa{e}.{A} \spa{y}.{A}
    \spa{c}.{d}^3}{\spa{d}.{A}^4}
\end{align}
\begin{align}
{\cal S}_1&= \frac{\spb{c}.{b} \spb{c}.{f}}{\spb{c}.{B}^2 } \left(\spb{d}.{c}
    \left(\frac{\spb{b}.{B}
    \spb{c}.{x}}{\spb{c}.{b}}+\frac{\spb{f}.{B}
    \spb{c}.{x}}{\spb{c}.{f}}+\spb{x}.{B}\right)+2 \spb{d}.{B}
    \spb{c}.{x}\right)
\notag \\
{\cal S}_0&= \frac{\spb{c}.{b} \spb{c}.{f} }{\spb{c}.{B}^3}\biggl(\spb{d}.{B}
    \spb{d}.{c} \left(\frac{\spb{b}.{B}
    \spb{c}.{x}}{\spb{c}.{b}}+\frac{\spb{f}.{B}
    \spb{c}.{x}}{\spb{c}.{f}}+\spb{x}.{B}\right)
\notag \\ &\hskip 100pt        +\spb{d}.{c}^2 \left(\frac{\spb{b}.{B} \spb{f}.{B}
    \spb{c}.{x}}{\spb{c}.{b} \spb{c}.{f}}+\frac{\spb{b}.{B}
    \spb{x}.{B}}{\spb{c}.{b}}+\frac{\spb{f}.{B}
    \spb{x}.{B}}{\spb{c}.{f}}\right)  
    +\spb{d}.{B}^2 \spb{c}.{x}\biggr)
\notag \\
{\cal S}_p&=\frac{\spb{b}.{B} \spb{f}.{B} \spb{x}.{B}
    \spb{d}.{c}^3}{\spb{c}.{B}^4}
\end{align}
and
\begin{align}
C_{pre}={\spa{d}.{a}\over \spa{d}.{A}  }{\spb{c}.{b}\over \spb{c}.{B}  }{\spa{d}.{e}\spb{c}.{f}\over [c|\tQ|d\ra}\spa{d}.{y}\spb{c}.{x} 
\end{align}
the contributions of the no-extra-pole piece are
\begin{align}
C_{np:0}=-{C_{pre}\over 2}[c|\tQ|d\ra
\Biggl[ { {\cal A}_0 {\cal S}_1 + {\cal S}_0 {\cal A}_1 \over [d|\tQ|c\ra }
-{ {\cal A}_0{\cal S}_0 \bigl( [c|\tQ|c\ra +  [d|\tQ|d\ra \bigr)
                                          \over [d|\tQ|c\ra^2 }
 \Biggr],
\end{align}

\begin{align}
C_{np:\g R}=-{C_{pre}\over 4}
\Bigl( -{\cal P}_Q(1+2r_Q) + {\cal A}_1+{\cal S}_1
-{1\over  Q_C}\bigl( {\cal A}_0 {\cal S}_1 + {\cal S}_0 {\cal A}_1 \bigr)
+{{\cal P}_Q(1+2r_Q)\over  Q_C^2} {\cal A}_0{\cal S}_0 \Bigr),                                     
\end{align}
and                                    
\begin{align}
C_{np:\g I}=-{C_{pre}\over 16{\cal P}_\delta^2}
\Biggl[&
 {\cal P}_Q^2\bigl[1 +{1\over Q_C^2}{\cal A}_0{\cal S}_0 \bigr]
                      \biggl( (2+8r_Q-3d_\delta){[d|\tQ|d\ra\over [c|\tQ|d\ra} -(8r_Q-3d_\delta){[c|\tQ|c\ra\over [c|\tQ|d\ra}\biggr)
\notag \\ &
+{\cal P}_Q    \Biggl[ 
               \bigl[{\cal A}_1+{ {\cal A}_0 {\cal S}_1 \over  Q_C}\bigr]
                         \biggl( (-2+4r_Q-3d_\delta){[d|\tQ|d\ra\over [c|\tQ|d\ra} -(-4+4r_Q-3d_\delta){[c|\tQ|c\ra\over [c|\tQ|d\ra}\biggr)
\notag \\ &
-\bigl[{\cal S}_1+{ {\cal S}_0 {\cal A}_1 \over  Q_C}\bigr] 
                         \biggl( (2+4r_Q-3d_\delta){[d|\tQ|d\ra\over [c|\tQ|d\ra} -(4r_Q-3d_\delta){[c|\tQ|c\ra\over [c|\tQ|d\ra}\biggr)
               \Biggr] 
\notag \\ &
         -2\bigl[{\cal A}_1{\cal S}_1 -Q_C -{{\cal A}_0{\cal S}_0\over Q_C} \bigr]
                         \biggl( (-2-3d_\delta){[d|\tQ|d\ra\over [c|\tQ|d\ra} -(-4-3d_\delta){[c|\tQ|c\ra\over [c|\tQ|d\ra}\biggr)
\notag \\ &                        
         +2 {\cal A}_0   \biggl( (-6-3d_\delta){[d|\tQ|d\ra\over [c|\tQ|d\ra} -(-8-3d_\delta){[c|\tQ|c\ra\over [c|\tQ|d\ra}\biggr)
\notag \\ &                                 
         +2 {\cal S}_0   \biggl( ( 2-3d_\delta){[d|\tQ|d\ra\over [c|\tQ|d\ra} -(  -3d_\delta){[c|\tQ|c\ra\over [c|\tQ|d\ra}\biggr) 
\Biggr]
\end{align}
with
\begin{align}
{\cal P}_Q&=\Bigl({[d|\tQ|d\ra - [c|\tQ|c\ra\over [c|\tQ|d\ra}\Bigr) 
\,\, ,\,\, 
r_Q= {[c|\tQ|c\ra\over [d|\tQ|d\ra-[c|\tQ|c\ra}
\,\, ,\,\, 
Q_C={[d|\tQ|c\ra\over [c|\tQ|d\ra}
\notag \\
{\cal P}_\delta&=\sqrt{{\bigl( [d|\tQ|d\ra -[c|\tQ|c\ra  \bigr)^2 +4[d|\tQ|c\ra[c|\tQ|d\ra\over [c|\tQ|d\ra^2} }
\notag \\
d_\delta&=-2\Biggl[{[c|\tQ|c\ra\bigl([c|\tQ|c\ra-[d|\tQ|d\ra  \bigr) +2[c|\tQ|d\ra[d|\tQ|c\ra\over [c|\tQ|d\ra^2}\Biggr]{1\over {\cal P}_\delta^2}
\end{align}

The contributions of the bonus angle pole pieces are

\begin{align}
 C_{ap:0}=-{C_{pre}{\cal A}_p\over 2}{\spa{d}.{A}\over \spa{c}.{A}}{[c|\tQ|d\ra\over [d|\tQ|c\ra}
\Biggl[{\cal S}_1  
-{\cal S}_0 
\Bigl(  {[c|\tQ|c\ra\over [d|\tQ|c\ra} + {[d|\tQ|d\ra\over [d|\tQ|c\ra}                                  
      +{\spa{d}.{A}\over \spa{c}.{A} }
\Bigr)
\Biggr]                                        
\end{align}

\begin{align}
C_{ap:a}=-{C_{pre}{\cal A}_p\over {\cal P}_{\cal A} }{\spa{d}.{A}^2\over \spa{c}.{A}^2}
\biggl[&
 \Bigl({\spa{c}.{A}^2\over \spa{d}.{A}^2}+{\spa{c}.{A}\over \spa{d}.{A}}{\cal S}_1 +{\cal S}_0 \Bigr)\Bigl( {[A|d|A\ra^2\over 2[A|P|A\ra^2}
                                                                                                    -b_{\cal A}{[A|d|A\ra\over[A|P|A\ra}\Bigr)
\notag \\ &
-\Bigl({\spa{c}.{A}\over \spa{d}.{A}}{\cal S}_1 +2{\cal S}_0\Bigr){[A|d|A\ra\over[A|P|A\ra}
\biggr]         
\end{align}

\begin{align}
C_{ap:ax}=&{C_{pre}{\cal A}_p }{\spa{d}.{A}^3 [c|\tQ|d\ra\over \spa{c}.{A}^3 [d|\tQ|A\ra}
\notag \\ &
\times
\biggl[ {1\over 3}{[A|d|A\ra^3\over[A|P|A\ra^3}\Bigl({\spa{c}.{A}^2\over \spa{d}.{A}^2}+{\spa{c}.{A}\over \spa{d}.{A}}{\cal S}_1 +{\cal S}_0 \Bigr)
-{1\over 2}{[A|d|A\ra^2\over[A|P|A\ra^2}\Bigl( {\spa{c}.{A}\over \spa{d}.{A}}{\cal S}_1 +2{\cal S}_0\Bigr)
+{[A|d|A\ra\over[A|P|A\ra} {\cal S}_0
\biggr]      
\end{align}

\begin{align}
C_{ap:\g R}=-{C_{pre}{\cal A}_p}
\biggl[ &-{ 1 \over 2{\cal P}_{\cal A}}
         \Bigl( \bigl({3\over 2}+b_1\bigr)
         +\bigl({1\over 2}+b_1\bigr){\cal S}_1{\spa{d}.{A}\over \spa{c}.{A}} 
         -\bigl({1\over 2}-b_1\bigr){\cal S}_0{\spa{d}.{A}^2\over \spa{c}.{A}^2} \Bigr)
\notag \\&
-{1\over 4 Q_C} \Bigl(  {\cal S}_1{\spa{d}.{A}\over \spa{c}.{A}}-{\cal S}_0{\spa{d}.{A}^2\over \spa{c}.{A}^2} \Bigr)
+{{\cal P}_Q(1+ 2r_Q)\over 4Q_C^2}{\cal S}_0{\spa{d}.{A}\over \spa{c}.{A}}
\biggr]                                                   
\end{align}

\begin{align}
C_{ap:\g Rx}=-{C_{pre}{\cal A}_p}
\biggl[& -{ 1 \over 2}{\spa{d}.{A}\over \spa{c}.{A}}{1\over   {\spa{c}.{A}\over \spa{d}.{A}} -{\cal P}_Q r_Q }
         \Bigl( {1\over 3}- {1\over 6}{\cal S}_1{\spa{d}.{A}\over \spa{c}.{A}} +{1\over 3}{\cal S}_0{\spa{d}.{A}^2\over \spa{c}.{A}^2} \Bigr)
\notag \\ &
-{1\over 4 Q_C} \Bigl(  {\cal S}_1{\spa{d}.{A}\over \spa{c}.{A}}-{\cal S}_0{\spa{d}.{A}^2\over \spa{c}.{A}^2} \Bigr)
+{{\cal P}_Q(1+ 2r_Q)\over 4Q_C^2}{\cal S}_0{\spa{d}.{A}\over \spa{c}.{A}}
\biggr]                                                   
\end{align}

\begin{align}
C_{ap:\g I}=-{C_{pre}{\cal A}_p\over 4{\cal P}_\delta^2}
\Biggl[ &-{{\cal P}_{r1}\over {\cal P}_{\cal A}}\biggl[1+ {\cal S}_1{\spa{d}.{A}\over \spa{c}.{A}}+{\cal S}_0{\spa{d}.{A}^2\over \spa{c}.{A}^2}\biggr]
                          \biggl[ 2 {[d|\tQ|d\ra\over [c|\tQ|d\ra} +(4r_1-4b_1-3d_\delta)
                                \Bigl( {[d|\tQ|d\ra\over [c|\tQ|d\ra} -{[c|\tQ|c\ra\over [c|\tQ|d\ra} \Bigr) \biggr] 
\notag \\ &
+4{{\cal P}_{r1}\over {\cal P}_{\cal A}}\biggl[ 2 + {\cal S}_1{\spa{d}.{A}\over \spa{c}.{A}} \biggr]
                                \Bigl( {[d|\tQ|d\ra\over [c|\tQ|d\ra} -{[c|\tQ|c\ra\over [c|\tQ|d\ra} \Bigr) 
\notag \\ &
                                +{{\cal P}_Q\over 2 Q_C} {\cal S}_1{\spa{d}.{A}\over \spa{c}.{A}}\biggl[ 2 {[d|\tQ|d\ra\over [c|\tQ|d\ra} 
            +(4r_Q-4-3d_\delta)\Bigl( {[d|\tQ|d\ra\over [c|\tQ|d\ra} -{[c|\tQ|c\ra\over [c|\tQ|d\ra} \Bigr)\biggr]  
\notag \\ &
            +{{\cal P}_Q\over 2 Q_C} {\cal S}_0{\spa{d}.{A}^2\over \spa{c}.{A}^2} \biggl[ 2 {[d|\tQ|d\ra\over [c|\tQ|d\ra} 
            +(4r_Q-3d_\delta)\Bigl( {[d|\tQ|d\ra\over [c|\tQ|d\ra} -{[c|\tQ|c\ra\over [c|\tQ|d\ra} \Bigr) \biggr]
\notag \\ &
+{\cal S}_0{\spa{d}.{A}\over \spa{c}.{A}}{{\cal P}_Q^2\over 2Q_C^2}\biggl[ 2 {[d|\tQ|d\ra\over [c|\tQ|d\ra} 
            +(8r_Q-3d_\delta)\Bigl( {[d|\tQ|d\ra\over [c|\tQ|d\ra} -{[c|\tQ|c\ra\over [c|\tQ|d\ra} \Bigr)\biggr] 
\notag \\ &
+{{\cal S}_0\over Q_C}{\spa{d}.{A}\over \spa{c}.{A}} \biggl[ 2 {[d|\tQ|d\ra\over [c|\tQ|d\ra} 
            +(-4-3d_\delta)\Bigl( {[d|\tQ|d\ra\over [c|\tQ|d\ra} -{[c|\tQ|c\ra\over [c|\tQ|d\ra} \Bigr) \biggr]          
\Biggr]                                                   
\end{align}

\begin{align}
C_{ap:\g Ix}=-{C_{pre}{\cal A}_p\over 4{\cal P}_\delta^2}
\Biggl[&{1\over 6} {\spa{d}.{A}\over \spa{c}.{A}}{{\cal P}_{r1}\over {\spa{c}.{A}\over \spa{d}.{A}} -{\cal P}_Q r_Q
                   }\biggl[1+ {\cal S}_1{\spa{d}.{A}\over \spa{c}.{A}}+{\cal S}_0{\spa{d}.{A}^2\over \spa{c}.{A}^2}\biggr]
\notag \\ &
\times
                          \biggl[ (8+12r_1-10d_\delta) {[d|\tQ|d\ra\over [c|\tQ|d\ra} +(15d_\delta^2-18d_\delta r_1 -16 c_\delta)
                                \Bigl( {[d|\tQ|d\ra\over [c|\tQ|d\ra} -{[c|\tQ|c\ra\over [c|\tQ|d\ra} \Bigr) \biggr] 
\notag \\ &
-{\spa{d}.{A}\over \spa{c}.{A}}{{\cal P}_{r1}\over {\spa{c}.{A}\over \spa{d}.{A}} -{\cal P}_Q r_Q}\biggl[ 2 + {\cal S}_1{\spa{d}.{A}\over \spa{c}.{A}} \biggr]
\biggl[ 2{[d|\tQ|d\ra\over [c|\tQ|d\ra} +(4 r_1 -3 d_\delta) \Bigl( {[d|\tQ|d\ra\over [c|\tQ|d\ra} -{[c|\tQ|c\ra\over [c|\tQ|d\ra} \Bigr) \biggr]
\notag \\ &
+4{\spa{d}.{A}\over \spa{c}.{A}}{{\cal P}_{r1}\over {\spa{c}.{A}\over \spa{d}.{A}} -{\cal P}_Q r_Q}
\Bigl( {[d|\tQ|d\ra\over [c|\tQ|d\ra} -{[c|\tQ|c\ra\over [c|\tQ|d\ra} \Bigr) 
\notag \\ &
+{{\cal P}_Q\over 2 Q_C} {\cal S}_1{\spa{d}.{A}\over \spa{c}.{A}}\biggl[ 2 {[d|\tQ|d\ra\over [c|\tQ|d\ra} 
            +(4r_Q-4-3d_\delta)\Bigl( {[d|\tQ|d\ra\over [c|\tQ|d\ra} -{[c|\tQ|c\ra\over [c|\tQ|d\ra} \Bigr)\biggr]  
\notag \\ &
            +{{\cal P}_Q\over 2 Q_C} {\cal S}_0{\spa{d}.{A}^2\over \spa{c}.{A}^2} \biggl[ 2 {[d|\tQ|d\ra\over [c|\tQ|d\ra} 
            +(4r_Q-3d_\delta)\Bigl( {[d|\tQ|d\ra\over [c|\tQ|d\ra} -{[c|\tQ|c\ra\over [c|\tQ|d\ra} \Bigr) \biggr]
\notag \\ &
+{\cal S}_0{\spa{d}.{A}\over \spa{c}.{A}}{{\cal P}_Q^2\over 2Q_C^2}\biggl[ 2 {[d|\tQ|d\ra\over [c|\tQ|d\ra} 
            +(8r_Q-3d_\delta)\Bigl( {[d|\tQ|d\ra\over [c|\tQ|d\ra} -{[c|\tQ|c\ra\over [c|\tQ|d\ra} \Bigr)\biggr] 
\notag \\ &
+{{\cal S}_0\over Q_C}{\spa{d}.{A}\over \spa{c}.{A}} \biggl[ 2 {[d|\tQ|d\ra\over [c|\tQ|d\ra} 
            +(-4-3d_\delta)\Bigl( {[d|\tQ|d\ra\over [c|\tQ|d\ra} -{[c|\tQ|c\ra\over [c|\tQ|d\ra} \Bigr) \biggr]          
\Biggr]                                                   
\end{align}
with
\begin{align}
{\cal P}_{\cal A}&={\spa{c}.{A}^2\over \spa{d}.{A}^2}-{\spa{c}.{A}\over \spa{d}.{A}}\biggl( {[c|\tQ|c\ra\over [c|\tQ|d\ra}
                                                                                   -{[d|\tQ|d\ra\over [c|\tQ|d\ra}\biggr) -{[d|\tQ|c\ra\over [c|\tQ|d\ra}
=-{\spa{c}.{d} \la A|P\tQ|A\ra \over \spa{d}.{A}^2 [c|\tQ|d\ra},
\notag \\
b_{\cal A}&= -\Bigl({\spa{c}.{A}\over \spa{d}.{A}}{[d|\tQ|d\ra\over [c|\tQ|d\ra} -{[d|\tQ|c\ra\over [c|\tQ|d\ra}\Bigr){1\over {\cal P}_{\cal A}}
=-{\spa{c}.{d} [d|\tQ|A\ra \over \spa{d}.{A} [c|\tQ|d\ra}{1\over {\cal P}_{\cal A}}, 
\end{align}
\begin{align}
c_\delta&={[c|\tQ|c\ra^2\over 4 [c|\tQ|d\ra^2}{1\over {\cal P}_\delta^2}.
\,\, , \,\, 
b_1=\Bigl({\cal P}_Q r_Q{\spa{c}.{A}\over \spa{d}.{A}}-{\spa{c}.{A}^2\over \spa{d}.{A}^2}\Bigr) {1\over {\cal P}_{A}}
\notag \\
{\cal P}_{r1} &=-{{\cal P}_Q\over 2}-{\spa{c}.{A}\over \spa{d}.{A}}
\,\, , \,\, 
r_1= \Bigl( -{{\cal P}_Qr_Q\over 2}+ {\spa{c}.{A}\over \spa{d}.{A}}\Bigr){1\over {\cal P}_{r1}}
\end{align}

Similarly the bonus square pole pieces give

\begin{align}
C_{sp:0}=-{C_{pre}{\cal S}_p\over 2}{[c|\tQ|d\ra\over [d|\tQ|c\ra} {\spb{c}.{B}\over \spb{d}.{B}}
\Biggl[
{\cal A}_1
-{\cal A}_0    
\biggl({\bigl(  [c|\tQ|c\ra + [d|\tQ|d\ra \bigr)\over [d|\tQ|c\ra }
                                      +{\spb{c}.{B}\over \spb{d}.{B}}
\biggr)\Biggr]                                        
\end{align}
\begin{align}
C_{sp:s}=&-{C_{pre}{\cal S}_p\over {\cal P}_{\cal S}}{\spb{c}.{B}^2\over \spb{d}.{B}^2}
\notag \\
&\times\Bigl(\bigl( {\spb{d}.{B}^2\over \spb{c}.{B}^2} +  {\spb{d}.{B}\over \spb{c}.{B}} {\cal A}_1 + {\cal A}_0 \bigr)
\bigl[{1\over 2}{[B|c|B\ra^2\over [B|P|B\ra^2} -b_{\cal S}{[B|c|B\ra\over [B|P|B\ra}\bigr]
       -\bigl({\spb{d}.{B}\over \spb{c}.{B}} {\cal A}_1 +2{\cal A}_0 \bigr){[B|c|B\ra\over [B|P|B\ra}  \Bigr) 
\end{align}
\begin{align}
C_{sp:sx}=&-C_{pre}{\cal S}_p{\spb{c}.{B}^3[c|\tQ|d\ra\over \spb{d}.{B}^2\spb{c}.{d} [B|\tQ|c\ra}
\notag \\
&\times
\biggl[
{1\over 3} {[B|c|B\ra^3\over[B|P|B\ra^3}\Bigl( {\spb{d}.{B}^2\over \spb{c}.{B}^2} +  {\spb{d}.{B}\over \spb{c}.{B}} {\cal A}_1 + {\cal A}_0 \Bigr)
       -{1\over 2} {[B|c|B\ra^2\over[B|P|B\ra^2}\Bigl({\spb{d}.{B}\over \spb{c}.{B}} {\cal A}_1 +2{\cal A}_0 \Bigr) 
       +{[B|c|B\ra\over[B|P|B\ra}{\cal A}_0 
\biggr]                          
\end{align}

\begin{align}
C_{sp:\g R}=-C_{pre}{\cal S}_p 
\Biggl[-&{ 1
\over 2{\cal P}_{\cal S} }
           \Bigl({1\over 2}-b_{\bar 2} 
          -\bigl( {1\over 2}+b_{\bar 2}\bigr) {\cal A}_1 {\spb{c}.{B}\over \spb{d}.{B}}
          -\bigl( {3\over 2}+b_{\bar 2}\bigr) {\cal A}_0 {\spb{c}.{B}^2\over \spb{d}.{B}^2} \Bigr)
\notag \\ &
-{1\over 4Q_C}\Bigl( {\cal A}_1 {\spb{c}.{B}\over \spb{d}.{B}} -{\cal A}_0 {\spb{c}.{B}^2\over \spb{d}.{B}^2} \Bigr)
+{{\cal P}_Q(1+2r_Q)\over 4Q_C^2} {\cal A}_0{\spb{c}.{B}\over \spb{d}.{B}}
\Biggr]
\end{align}

\begin{align}
C_{sp:\g Rx}=-C_{pre}{\cal S}_p 
\Biggl[&{ 1\over 2}
{1\over {\cal P}_Q r_Q {\spb{d}.{B}\over \spb{c}.{B}}-Q_C}
           \Bigl({1\over 3}-{1\over 6} {\cal A}_1 {\spb{c}.{B}\over \spb{d}.{B}}+{1\over 3} {\cal A}_0 {\spb{c}.{B}^2\over \spb{d}.{B}^2} \Bigr)
\notag \\ &
-{1\over 4Q_C}\Bigl( {\cal A}_1 {\spb{c}.{B}\over \spb{d}.{B}} -{\cal A}_0 {\spb{c}.{B}^2\over \spb{d}.{B}^2} \Bigr)
+{{\cal P}_Q(1+2r_Q)\over 4Q_C^2} {\cal A}_0{\spb{c}.{B}\over \spb{d}.{B}}
\Biggr]
\end{align}

\begin{align}
\hskip-100pt
C_{sp:\g I}={C_{pre}{\cal S}_p\over {\cal P}_\delta^2}
\Biggl[{ {\cal P}_{r2}\over 4{\cal P}_{\cal S}}
           \Bigl(\bigl[1+ {\cal A}_1{\spb{c}.{B}\over \spb{d}.{B}}+ {\cal A}_0 {\spb{c}.{B}^2\over \spb{d}.{B}^2}\bigr]
           \bigl[ (2-4b_{\bar 2}+4r_{\bar 2}-3d_\delta){[d|\tQ|c\ra\over [d|\tQ|d\ra} -(-4b_{\bar 2}+4r_{\bar 2}-3d_\delta){[c|\tQ|c\ra\over [c|\tQ|d\ra}\bigr]&
\notag \\
-4\bigl[ {\cal A}_1 {\spb{c}.{B}\over \spb{d}.{B}}+2 {\cal A}_0 {\spb{c}.{B}^2\over \spb{d}.{B}^2} \bigr]
                  \bigl[ {[d|\tQ|d\ra\over [c|\tQ|d\ra} -{[c|\tQ|c\ra\over [c|\tQ|d\ra}\bigr]&
                  \Bigr)
\notag \\
\hskip-20pt
+{{\cal P}_Q\over 8Q_C }\Bigl(  \bigl[{\cal A}_1 {\spb{c}.{B}\over \spb{d}.{B}}+ {\cal A}_0 {\spb{c}.{B}^2\over \spb{d}.{B}^2}\bigr]
                                              \bigl[(2+4r_Q-3d_\delta){[d|\tQ|d\ra\over [c|\tQ|d\ra} -(4r_Q-3d_\delta){[c|\tQ|c\ra\over [c|\tQ|d\ra}\bigr]
                                      -4{\cal A}_0 {\spb{c}.{B}^2\over \spb{d}.{B}^2}
                                      \bigl[ {[d|\tQ|d\ra\over [c|\tQ|d\ra} -{[c|\tQ|c\ra\over [c|\tQ|d\ra}\bigr]&
\Bigr)
\notag \\
-{\cal A}_0{{\cal P}_Q^2\bigl[(2+8r_Q-3d_\delta){[d|\tQ|d\ra} -(8r_Q-3d_\delta){[c|\tQ|c\ra}\bigr]
   -2Q_C[(2+3d_\delta){[d|\tQ|d\ra} -(4+3d_\delta){[c|\tQ|c\ra}\bigr]
   \over 8Q_C^2 [c|\tQ|d\ra} {\spb{c}.{B}\over \spb{d}.{B}}&
\Biggr]
\end{align}

\begin{align}
C_{sp:\g Ix}&=-{ C_{pre}{\cal S}_p \over [c|\tQ|d\ra }
\Biggl[{ {\cal P}_{r2} \spb{c}.{B}
\over  {\cal P}_Q r_Q {\spb{d}.{B}}-Q_C  \spb{c}.{B}}
           \Biggl(
{\cal A}_0 {\spb{c}.{B}^2\over \spb{d}.{B}^2}\Bigl( {[d|\tQ|d\ra} -{[c|\tQ|c\ra}          
\Bigr) %
\notag \\
&
          +{1\over 24}\biggl[1+ {\cal A}_1{\spb{c}.{B}\over \spb{d}.{B}}+ {\cal A}_0 {\spb{c}.{B}^2\over \spb{d}.{B}^2}\biggr]                         \biggl[ (8+12r_{\bar 2}-10d_\delta) {[d|\tQ|d\ra} +(15d_\delta^2-18d_\delta r_{\bar 2} -16 c_\delta)
                                \Bigl( {[d|\tQ|d\ra } -{[c|\tQ|c\ra} \Bigr) \biggr] 
\notag \\  &               
-{1\over 4}\biggl[ {\cal A}_1 {\spb{c}.{B}\over \spb{d}.{B}}+2 {\cal A}_0 {\spb{c}.{B}^2\over \spb{d}.{B}^2} \biggr]  
\biggl[ 2{[d|\tQ|d\ra} +(4 r_{\bar 2} -3 d_\delta) \Bigl( {[d|\tQ|d\ra} -{[c|\tQ|c\ra} \Bigr) \biggr]
                                                           \,\,\,     \Biggr)
\notag \\
-{{\cal P}_Q\over 8Q_C }&\Bigl(  \bigl[{\cal A}_1 {\spb{c}.{B}\over \spb{d}.{B}}+ {\cal A}_0 {\spb{c}.{B}^2
\over \spb{d}.{B}^2}\bigr]
                                              \bigl[(2+4r_Q-3d_\delta){[d|\tQ|d\ra} -(4r_Q-3d_\delta){[c|\tQ|c\ra}\bigr]
                                      -4{\cal A}_0 {\spb{c}.{B}^2\over \spb{d}.{B}^2}
                                      \bigl[ {[d|\tQ|d\ra} -{[c|\tQ|c\ra}\bigr]
\Bigr)
\notag \\
+ {\cal A}_0&{{\cal P}_Q^2\bigl[(2+8r_Q-3d_\delta){[d|\tQ|d\ra} -(8r_Q-3d_\delta){[c|\tQ|c\ra}\bigr]
   -2Q_C[(2+3d_\delta){[d|\tQ|d\ra} -(4+3d_\delta){[c|\tQ|c\ra}\bigr]
   \over 8Q_C^2}{\spb{c}.{B}\over \spb{d}.{B}}
\Biggr]
\end{align}
with
\begin{align}
{\cal P}_{\cal S}&=  {\spb{d}.{B}^2\over \spb{c}.{B}^2} 
          + {\spb{d}.{B}\over \spb{c}.{B}}(  {[c|\tQ|c\ra\over [c|\tQ|d\ra} -  {[d|\tQ|d\ra\over [c|\tQ|d\ra}) -{[d|\tQ|c\ra\over [c|\tQ|d\ra},
\notag \\
b_{\cal S}&=-\bigl ({\spb{d}.{B}\over \spb{c}.{B}}{[c|\tQ|c\ra\over [c|\tQ|d\ra} - {[d|\tQ|c\ra\over [c|\tQ|d\ra} \bigr) {1\over {\cal P}_{\cal S}}
\end{align}
\begin{align}
{\cal P}_{r2}&= {{\cal P}_Q\over 2} -{\spb{d}.{B}\over \spb{c}.{B}}
\,\, , \,\, 
r_{\bar 2}=\Bigl( {{\cal P}_Qr_Q\over 2} \Bigr) {1\over {\cal P}_{r2}}
\,\, , \,\, 
b_{\bar 2}=\Bigl(Q_C- {\cal P}_Q r_Q{\spb{d}.{B}\over \spb{c}.{B}}\Bigr){1\over {\cal P}_{\cal S}}
\end{align}

Finally  the double bonus  pole pieces are

\begin{align}             
 C_{dp:0}={C_{pre}{\cal A}_p{\cal S}_p\over 2 }
  {[c|\tQ|d\ra \spa{d}.{A}  \spb{c}.{B} \over [d|\tQ|c\ra \spa{c}.{A} \spb{d}.{B} }
 \Biggl[  {[c|\tQ|c\ra\over [d|\tQ|c\ra} 
              +  {[d|\tQ|d\ra\over [d|\tQ|c\ra}
       +{\spa{d}.{A}\over \spa{c}.{A}} + {\spb{c}.{B}\over \spb{d}.{B}} \Biggr] 
\end{align}

\def\calCdpa{ {C_{pre}{\cal A}_p{\cal S}_p\over {\cal D}_Q}\Bigl({\spa{d}.A\over \spa{c}.A}\Bigr)^4}

\begin{align}
 C_{dp:ab}
=\calCdpa
\biggl[ { \bigl(1-{[A|c|A\ra^2\over [A|P|A\ra^2} \bigr)\over 2 ({\cal E}_Q-1)({\cal E}_S-1) } - {\bigl(1-{[A|c|A\ra\over [A|P|A\ra}\bigr) ({\cal E}_Q+{\cal E}_S-2) \over  ({\cal E}_Q-1)^2({\cal E}_S-1)^2 }
\biggr] 
\end{align}

\begin{align}
C_{dp:ax}
={\calCdpa}
\biggl[ { \bigl(1 -{[A|c|A\ra^3\over [A|P|A\ra^3} \bigr) \over 3 ({\cal E}_Q-1)}
       -{ \bigl(1- {[A|c|A\ra^2\over [A|P|A\ra^2} \bigr) \over 2 ({\cal E}_Q-1)^2 } 
       +{\bigl(1-  {[A|c|A\ra\over [A|P|A\ra}     \bigr) \over   ({\cal E}_Q-1)^3 }
\biggr] 
\end{align}
\begin{align}
C_{dp:ay}
=\calCdpa
\biggl[ { \bigl(1 -{[A|c|A\ra^3\over [A|P|A\ra^3} \bigr) \over 3 ({\cal E}_S-1)}
       -{ \bigl(1- {[A|c|A\ra^2\over [A|P|A\ra^2} \bigr) \over 2 ({\cal E}_S-1)^2 } 
       +{\bigl(1-  {[A|c|A\ra\over [A|P|A\ra}     \bigr) \over   ({\cal E}_S-1)^3 }
\biggr] 
\end{align}
\def\calCdpa4{ {C_{pre}{\cal A}_p{\cal S}_p\over 4{\cal D}_Q}\Bigl({\spa{d}.A\over \spa{c}.A}\Bigr)^4}
\begin{align}
C_{dp:axy}
={\calCdpa4}
 { \bigl(1 -{[A|c|A\ra^4\over [A|P|A\ra^4} \bigr)}
\end{align}
with
\begin{align}
{\cal D}_Q=\Bigl[{\spa{c}.{A}\over \spa{d}.{A}}-{[c|\tQ|c\ra\over [c|\tQ|d\ra}\Bigr] 
\,\, , \,\,
{\cal E}_Q=-\Biggl[{ [d|\tQ|d\ra\over [c|\tQ|d\ra }+{[d|\tQ|c\ra\over [c|\tQ|d\ra}{\spa{d}.{A}\over \spa{c}.{A} } \Biggr]{1\over {\cal D}_Q }
 \,\, , \,\,
{\cal E}_S=-{\spb{d}.{B}\over \spb{c}.{B}}{\spa{d}.{A}\over \spa{c}.{A} }
\end{align}

\begin{align}
C_{dp:sb}={C_{pre}{\cal S}_p {\cal A}_p\over 2 {\cal P}_{\cal S}{\cal D}_A}{\spb{c}.{B}^2\over \spb{d}.{B}^2}
\biggl(1- {[B|d|B\ra^2\over [B|P|B\ra^2}-2({\cal F}_Q+ {\cal E}_A)\Bigl[1-{[B|d|B\ra\over [B|P|B\ra}\Bigr]\biggr)
\end{align}

\begin{align}
C_{dp:sx}=-{C_{pre}{\cal S}_p {\cal A}_p\over {\cal P}_{\cal S}}{\spb{c}.{B}^3\over \spb{d}.{B}^3}
\Biggl[ {1         \over 3}\Bigl(1 - {[B|d|B\ra^3\over [B|P|B\ra^3}\Bigr)
       -{{\cal F}_Q\over 2}\Bigl(1 - {[B|d|B\ra^2\over [B|P|B\ra^2}\Bigr)
       +{\cal F}_Q^2\Bigl(1 - {[B|d|B\ra\over [B|P|B\ra}\Bigr)
\Biggr]
\end{align}

\begin{align}
C_{dp:sy}=-{C_{pre}{\cal S}_p {\cal A}_p\over {\cal D}_A}{\spb{c}.{B}^3\over \spb{d}.{B}^3} 
{1\over {\spb{d}.{B}\over \spb{c}.{B}} -{[d|\tQ|d\ra\over [c|\tQ|d\ra}}
\biggl[ {1         \over 3}\Bigl(1 - {[B|d|B\ra^3\over [B|P|B\ra^3}\Bigr)
       -{{\cal E}_A\over 2}\Bigl(1 - {[B|d|B\ra^2\over [B|P|B\ra^2}\Bigr)
       +{\cal E}_A^2       \Bigl(1 - {[B|d|B\ra\over [B|P|B\ra}\Bigr)
\Biggr]
\end{align}

\begin{align}
C_{dp:sxy}=C_{pre}{\cal S}_p {\cal A}_p{\spb{c}.{B}^4\over \spb{d}.{B}^4}
{1\over {\spb{d}.{B}\over \spb{c}.{B}} -{[d|\tQ|d\ra\over [c|\tQ|d\ra}}
 {1         \over 4}\Bigl(1 - {[B|d|B\ra^4\over [B|P|B\ra^4}\Bigr)
\end{align}
with
\begin{align}
{\cal F}_Q= \Bigl( {\spb{d}.{B}^2\over \spb{c}.{B}^2} -{\spb{d}.{B}\over \spb{c}.{B}}{[d|\tQ|d\ra\over [c|\tQ|d\ra}  \Bigr){1\over {\cal G}_Q}
\,\, , \,\,
{\cal D}_A={\spb{d}.{B}\over \spb{c}.{B}} +{\spa{c}.{A}\over \spa{d}.{A}} 
 \,\, , \,\,
{\cal E}_A=-{\spb{d}.{B}\over \spb{c}.{B}}{1\over {\cal D}_A}
\end{align}

\def\mupx{{[d|\tQ|c\ra\over [c|\tQ|d\ra} }
\def\cupx{{[c|\tQ|c\ra\over [c|\tQ|d\ra} }
\def\dupx{{[d|\tQ|d\ra\over [c|\tQ|d\ra} }
\def\calPdsq{{\cal P}_d^s}

\begin{align} 
C_{dp \g:R}= -\frac{1}{4} {\cal A}_{p} {\cal S}_{p}
    C_{pre} \biggl(&\frac{\spa{d}.{A}^2 (1-2
    ({a_1}+{b_1}))}{\spa{c}.{A}^2
    {\cal P}_{a1} {\cal P}_{b1}}-\frac{\spb{c}.{B}^2 (-2
    ({a_1}+b_{\bar 2})-5)}{\spb{d}.{B}^2
    {\cal P}_{a1} {\cal P}_{b2}}
\notag \\ &     
    +\frac{\spa{d}.{A}
    \spb{c}.{B} {\cal P}_{Q} (2
    {r_Q}+1)}{\spa{c}.{A} \spb{d}.{B}
    {Q_C}^2}+\frac{\spa{d}.{A}^2 \spb{c}.{B}^2
    \left(\frac{\spa{c}.{A}}{\spa{d}.{A}}+\frac{\spb{d}.{B}}
    {\spb{c}.{B}}\right)}{\spa{c}.{A}^2 \spb{d}.{B}^2
    {Q_C}}\biggr)
\end{align}

\begin{align}
 C_{dp\g:Rx}=  -\frac{1}{2} {\cal A}_{p} {\cal S}_{p}         
    C_{pre} \biggl(&\frac{\spa{d}.{A}              
    \spb{c}.{B} {\cal P}_{Q} (2 {r_Q}+1)}{2            
    \spa{c}.{A} \spb{d}.{B}                                             
    {Q_C}^2}+\frac{\spa{d}.{A}^3
    \left({b_1}^2-\frac{1}{2}
    {b_1}+\frac{1}{3}\right)}{\spa{c}.{A}^3
    {\cal P}_{b1} }
\notag \\ &    
    +\frac{\spb{c}.{B}^3
    \left((b_{\bar 2}+1)^2+\frac{1}{2}
    (b_{\bar 2}+1)+\frac{1}{3}\right)}{\spb{d}.{B}^3
    {\cal P}_{b2}}\biggr)
\end{align}

\begin{align}
  C_{dp\g:Ry1}= -\frac{1}{2} {\cal A}_{p} {\cal S}_{p}         
    C_{pre} \biggl(&-\frac{\spb{c}.{B}^2 (-2       
    ({a_1}+b_{\bar 2})-5)}{2 \spb{d}.{B}^2      
    {\cal P}_{a1} {\cal P}_{b2}}+\frac{\spa{d}.{A}^3      
    \left({a_1}^2-\frac{1}{2}                                   
    {a_1}
    +\frac{1}{3}\right)}{\spa{c}.{A}^3                 
    {\cal P}_{a1}                                                      
    \left(\frac{\spa{c}.{A}}{\spa{d}.{A}}-{\cal P}_{Q}
    {r_Q}\right)}
\notag \\ &    
    +\frac{\spa{d}.{A} \spb{c}.{B}
    {\cal P}_{Q} (2 {r_Q}+1)}{2 \spa{c}.{A}
    \spb{d}.{B} {Q_C}^2}+\frac{\spa{d}.{A}^2
    \spb{c}.{B}^2
    \left(\frac{\spa{c}.{A}}{\spa{d}.{A}}+\frac{\spb{d}.{B}}
    {\spb{c}.{B}}\right)}{2 \spa{c}.{A}^2 \spb{d}.{B}^2
    {Q_C}}\biggr)
\end{align}

\begin{align}
 C_{dp\g:Ry2}=  -\frac{1}{2} {\cal A}_{p} {\cal S}_{p}         
    C_{pre} \biggl(&\frac{\spa{d}.{A}^2 (1-2       
    ({a_1}+{b_1}))}{2 \spa{c}.{A}^2         
    {\cal P}_{a1} {\cal P}_{b1}}-\frac{\spb{c}.{B}^2      
    \left({a_1}^2+\frac{5}{2}                                   
    {a_1}+\frac{11}{6}\right)}{\spb{d}.{B}^2                
    {\cal P}_{a1} \left(\frac{\spb{d}.{B} {\cal P}_{Q}      
    {r_Q}}{\spb{c}.{B}}-{Q_C}\right)}
    \notag \\ &    
+\frac{\spa{d}.{A} \spb{c}.{B} {\cal P}_{Q} (2 {r_Q}+1)}{2
    \spa{c}.{A} \spb{d}.{B}
    {Q_C}^2}+\frac{\spa{d}.{A}^2 \spb{c}.{B}^2
    \left(\frac{\spa{c}.{A}}{\spa{d}.{A}}
+\frac{\spb{d}.{B}}{\spb{c}.{B}}\right)}{2 \spa{c}.{A}^2 \spb{d}.{B}^2
    {Q_C}}\biggr)
\end{align}

\begin{align}
C_{dp\g:Ry12}=  -\frac{1}{2} {\cal A}_{p} {\cal S}_{p}
    C_{pre} \biggl(&-\frac{\spb{c}.{B}^2
    \left({a_1}^2+\frac{5}{2}
    {a_1}+\frac{11}{6}\right)}{\spb{d}.{B}^2
    {\cal P}_{a1} \left(\frac{\spb{d}.{B} {\cal P}_{Q}
    {r_Q}}{\spb{c}.{B}}-{Q_C}\right)}
+\frac{\spa{d}.{A}^3 \left({a_1}^2-\frac{1}{2}
    {a_1}+\frac{1}{3}\right)}{\spa{c}.{A}^3
    {\cal P}_{a1}
    \left(\frac{\spa{c}.{A}}{\spa{d}.{A}}-{\cal P}_{Q}
    {r_Q}\right)}
\notag \\ &       
    +\frac{\spa{d}.{A} \spb{c}.{B}
    {\cal P}_{Q} (2 {r_Q}+1)}{2 \spa{c}.{A}
    \spb{d}.{B} {Q_C}^2}+\frac{\spa{d}.{A}^2
    \spb{c}.{B}^2
    \left(\frac{\spa{c}.{A}}{\spa{d}.{A}}+\frac{\spb{d}.{B}}
{\spb{c}.{B}}\right)}{2 \spa{c}.{A}^2 \spb{d}.{B}^2
    {Q_C}}\biggr)
\end{align}

\begin{align}
C_{dp\g:Rxy}=  -\frac{1}{2} {\cal A}_{p} {\cal S}_{p}C_{pre}
    \biggl(&\frac{\spa{d}.{A} \spb{c}.{B} {\cal P}_{Q} (2
    {r_Q}+1)}{2 \spa{c}.{A} \spb{d}.{B}
    {Q_C}^2}+\frac{\spa{d}.{A}^2 \spb{c}.{B}^2
    \left(\frac{\spa{c}.{A}}{\spa{d}.{A}}+\frac{\spb{d}.{B}}
{\spb{c}.{B}}\right)}{2 \spa{c}.{A}^2 \spb{d}.{B}^2
    {Q_C}}
\notag \\ &        
+\frac{\spa{d}.{A}^4}{4 \spa{c}.{A}^4
    \left(\frac{\spa{c}.{A}}{\spa{d}.{A}}-{\cal P}_{Q}
    {r_Q}\right)}-\frac{\spb{c}.{B}^3}{4 \spb{d}.{B}^3
    \left(\frac{\spb{d}.{B} {\cal P}_{Q}
    {r_Q}}{\spb{c}.{B}}-{Q_C}\right)}\biggr)    
\end{align}

\begin{align}
 C_{dp\g:I}=  -\frac{{\cal A}_{p} {\cal S}_{p}               
    C_{pre} }{4 {\calPdsq}}  \biggl(&-\frac{\spa{d}.{A}^2           
    {\cal P}_{r1} (({\dupx}-{\cupx}) (-4           
    {a_1}-4 {b_1}+4 {r_1}-3     
    {d_\delta})+2 {\dupx})}{\spa{c}.{A}^2                
    {\cal P}_{a1} {\cal P}_{b1}}
\notag \\ &        
    -\frac{\spb{c}.{B}^2      
    {\cal P}_{r2} (({\dupx}-{\cupx}) (-4           
    {a_1}-4 b_{\bar 2}+4 r_{\bar 2}-3   
    {d_\delta}-12)+2 {\dupx})}{\spb{d}.{B}^2             
    {\cal P}_{a1} {\cal P}_{b2}}
\notag \\ &    
    +\frac{\spa{d}.{A}        
    \spb{c}.{B} {\cal P}_{Q}^2 (({\dupx}-{\cupx})
    (8 {r_Q}-3 {d_\delta})+2 {\dupx})}{2       
    \spa{c}.{A} \spb{d}.{B}                                             
    {Q_C}^2}
\notag \\ &        
    +\frac{\spa{d}.{A}^2 \spb{c}.{B}              
    {\cal P}_{Q} (({\dupx}-{\cupx}) (4               
    {r_Q}-3 {d_\delta})+2 {\dupx})}{2          
    \spa{c}.{A}^2 \spb{d}.{B} {Q_C}}
\notag \\ &        
    -\frac{\spa{d}.{A}
    \spb{c}.{B}^2 {\cal P}_{Q} (({\dupx}-{\cupx})
    (-3 {d_\delta}+4 {r_Q}-4)+2 {\dupx})}{2    
    \spa{c}.{A} \spb{d}.{B}^2 {Q_C}}
\notag \\ &        
    +\frac{\spa{d}.{A}
    \spb{c}.{B} ((-3 {d_\delta}-4)                                 
    ({\dupx}-{\cupx})+2 {\dupx})}{\spa{c}.{A} 
    \spb{d}.{B} {Q_C}}\biggr)
\end{align}

\begin{align}
 C_{dp\g:Ix}=   -\frac{{\cal A}_{p} {\cal S}_{p}
    C_{pre}}{4    {\calPdsq}} 
    \biggl(&\frac{\spa{d}.{A}^2
    \spb{c}.{B} {\cal P}_{Q} (({\dupx}-{\cupx})
    (4 {r_Q}-3 {d_\delta})+2 {\dupx})}{2
    \spa{c}.{A}^2 \spb{d}.{B} {Q_C}}
\notag \\ &    
    +\frac{\spa{d}.{A}
    \spb{c}.{B} {\cal P}_{Q}^2 (({\dupx}-{\cupx})
    (8 {r_Q}-3 {d_\delta})+2 {\dupx})}{2
    \spa{c}.{A} \spb{d}.{B} {Q_C}^2}
\notag \\ &    
    -\frac{\spa{d}.{A}
    \spb{c}.{B}^2 {\cal P}_{Q} (({\dupx}-{\cupx})
    (-3 {d_\delta}+4 {r_Q}-4)+2 {\dupx})}{2
    \spa{c}.{A} \spb{d}.{B}^2 {Q_C}}
\notag \\ &    
    +\frac{\spa{d}.{A}
    \spb{c}.{B} ((-3 {d_\delta}-4)
    ({\dupx}-{\cupx})+2 {\dupx})}{\spa{c}.{A}
    \spb{d}.{B} {Q_C}}
\notag \\ &      
    -\frac{\spa{d}.{A}^3
    {\cal P}_{r1}
    \begin{pmatrix}
    ({\dupx}-{\cupx})
    \left({d_\delta} (18 {b_1}-18
    {r_1}+15 {d_\delta})-24 {b_1}
    {r_1}+24 {b_1}^2-16
    {c_\delta}\right)
    \\
    +{\dupx} (-12 {b_1}+12
    {r_1}-10 {d_\delta}+8)
    \end{pmatrix}
    }{6 \spa{c}.{A}^3
    {\cal P}_{b1}}
\notag \\ &      
    +\frac{\spb{c}.{B}^3 {\cal P}_{r2}
    \begin{pmatrix}
    ({\dupx}-{\cupx}) \left({d_\delta} (18
    b_{\bar 2}-18 r_{\bar 2}+15 {d_\delta})-24
    b_{\bar 2} r_{\bar 2}+24 b_{\bar 2}^2-16
    {c_\delta}\right)
    \\
    +{\dupx} (-12 b_{\bar 2}+12
    r_{\bar 2}-10 {d_\delta}+8)
    \end{pmatrix}
    }{6 \spb{d}.{B}^3
    {\cal P}_{b2}}
\notag \\ &      
    -\frac{3 \spb{c}.{B}^3 {\cal P}_{r2}
    (({\dupx}-{\cupx}) (-4 b_{\bar 2}+4
    r_{\bar 2}-3 {d_\delta}-4)+2
    {\dupx})}{\spb{d}.{B}^3 {\cal P}_{b2}}\biggr)
\end{align}

\begin{align}
C_{dp\g:Iy1}=   -\frac{{\cal A}_{p} {\cal S}_{p}
    C_{pre}}{4 {\calPdsq}}
    \biggl(&-\frac{\spb{c}.{B}^2
    {\cal P}_{r2} (({\dupx}-{\cupx}) (-4
    {a_1}-4 b_{\bar 2}+4 r_{\bar 2}-3
    {d_\delta}-12)+2 {\dupx})}{\spb{d}.{B}^2
    {\cal P}_{a1} {\cal P}_{b2}}
\notag \\ &     
    -\frac{\spa{d}.{A}^3
    {\cal P}_{r1}
    \begin{pmatrix}
    ({\dupx}-{\cupx}) \left((24
    {a_1}+18 {d_\delta})
    ({a_1}-{r_1})-16 {c_\delta}+15
    {d_\delta}^2\right)
    \\
    +{\dupx} (-12 {a_1}+12
    {r_1}-10 {d_\delta}+8)
    \end{pmatrix}
    }{6 \spa{c}.{A}^3
    {\cal P}_{a1}
    \left(\frac{\spa{c}.{A}}{\spa{d}.{A}}-{\cal P}_{Q}
    {r_Q}\right)}
\notag \\ &     
    +\frac{\spa{d}.{A}^2 \spb{c}.{B}
    {\cal P}_{Q} (({\dupx}-{\cupx}) (4
    {r_Q}-3 {d_\delta})+2 {\dupx})}{2
    \spa{c}.{A}^2 \spb{d}.{B} {Q_C}}
\notag \\ &     
    +\frac{\spa{d}.{A}
    \spb{c}.{B} {\cal P}_{Q}^2 (({\dupx}-{\cupx})
    (8 {r_Q}-3 {d_\delta})+2 {\dupx})}{2
    \spa{c}.{A} \spb{d}.{B} {Q_C}^2}
\notag \\ &   
    -\frac{\spa{d}.{A}
    \spb{c}.{B}^2 {\cal P}_{Q} (({\dupx}-{\cupx})
    (-3 {d_\delta}+4 {r_Q}-4)+2 {\dupx})}{2
    \spa{c}.{A} \spb{d}.{B}^2 {Q_C}}
\notag \\ &     
    +\frac{\spa{d}.{A}
    \spb{c}.{B} ((-3 {d_\delta}-4)
    ({\dupx}-{\cupx})+2 {\dupx})}{\spa{c}.{A}
    \spb{d}.{B} {Q_C}}\biggr)
\end{align}

\begin{align}
C_{dp\g:Iy2}=  -\frac{{\cal A}_{p} {\cal S}_{p}
    C_{pre} }{4 {\calPdsq}}
    \biggl(&-\frac{\spa{d}.{A}^2
    {\cal P}_{r1} (({\dupx}-{\cupx}) (-4
    {a_1}-4 {b_1}+4 {r_1}-3
    {d_\delta})+2 {\dupx})}{\spa{c}.{A}^2
    {\cal P}_{a1} {\cal P}_{b1}}
\notag \\ &    
    -\frac{\spb{c}.{B}^2
    {\cal P}_{r2} 
    \begin{pmatrix}
    ({\dupx}-{\cupx}) \left((24
    {a_1}+18 {d_\delta})
    ({a_1}-r_{\bar 2})-16 {c_\delta}+15
    {d_\delta}^2\right)
    \\
    -18 ({\dupx}-{\cupx}) (-4
    {a_1}+4 r_{\bar 2}-3
    {d_\delta})
    \\
    +{\dupx} (-12 {a_1}+12
    r_{\bar 2}-10 {d_\delta}+8)
    \\
    +72
    ({\dupx}-{\cupx})-36 {\dupx}
    \end{pmatrix}
    }{6
    \spb{d}.{B}^2 {\cal P}_{a1} \left(\frac{\spb{d}.{B}
    {\cal P}_{Q}
    {r_Q}}{\spb{c}.{B}}-{Q_C}\right)}
\notag \\ & 
    +\frac{\spa{d}.{A} \spb{c}.{B} {\cal P}_{Q}^2
    (({\dupx}-{\cupx}) (8 {r_Q}-3
    {d_\delta})+2 {\dupx})}{2 \spa{c}.{A} \spb{d}.{B}
    {Q_C}^2}
\notag \\ &     
    +\frac{\spa{d}.{A}^2 \spb{c}.{B}
    {\cal P}_{Q} (({\dupx}-{\cupx}) (4
    {r_Q}-3 {d_\delta})+2 {\dupx})}{2
    \spa{c}.{A}^2 \spb{d}.{B} {Q_C}}
\notag \\ &     
    -\frac{\spa{d}.{A}
    \spb{c}.{B}^2 {\cal P}_{Q} (({\dupx}-{\cupx})
    (-3 {d_\delta}+4 {r_Q}-4)+2 {\dupx})}{2
    \spa{c}.{A} \spb{d}.{B}^2 {Q_C}}
\notag \\ &     
    +\frac{\spa{d}.{A}
    \spb{c}.{B} ((-3 {d_\delta}-4)
    ({\dupx}-{\cupx})+2 {\dupx})}{\spa{c}.{A}
    \spb{d}.{B} {Q_C}}\biggr)
\end{align}

\begin{align}
C_{dp\g:Iy12}=  -\frac{{\cal A}_{p} {\cal S}_{p}
    C_{pre}}{4 {\calPdsq}}
    \biggl(&-\frac{\spa{d}.{A}^3
    {\cal P}_{r1} 
    \begin{pmatrix}
    ({\dupx}-{\cupx}) \left((24
    {a_1}+18 {d_\delta})
    ({a_1}-{r_1})-16 {c_\delta}+15
    {d_\delta}^2\right)
    \\
    +{\dupx} (-12 {a_1}+12
    {r_1}-10 {d_\delta}+8)
    \end{pmatrix}
    }{6 \spa{c}.{A}^3
    {\cal P}_{a1}
    \left(\frac{\spa{c}.{A}}{\spa{d}.{A}}-{\cal P}_{Q}
    {r_Q}\right)}
\notag \\ &    
    -\frac{\spb{c}.{B}^2 {\cal P}_{r2}
    \begin{pmatrix}
    ({\dupx}-{\cupx}) \left((24
    {a_1}+18 {d_\delta})
    ({a_1}-r_{\bar 2})-16 {c_\delta}+15
    {d_\delta}^2\right)
    \\
    -18 ({\dupx}-{\cupx}) (-4
    {a_1}+4 r_{\bar 2}-3
    {d_\delta})
    \\
    +{\dupx} (-12 {a_1}+12
    r_{\bar 2}-10 {d_\delta}+8)
    \\
    +72
    ({\dupx}-{\cupx})-36 {\dupx}
    \end{pmatrix}
    }{6
    \spb{d}.{B}^2 {\cal P}_{a1} \left(\frac{\spb{d}.{B}
    {\cal P}_{Q}
    {r_Q}}{\spb{c}.{B}}-{Q_C}\right)}
\notag \\ &    
    +\frac{\spa{d}.{A}^2 \spb{c}.{B} {\cal P}_{Q}
    (({\dupx}-{\cupx}) (4 {r_Q}-3
    {d_\delta})+2 {\dupx})}{2 \spa{c}.{A}^2
    \spb{d}.{B} {Q_C}}
\notag \\ &    
    +\frac{\spa{d}.{A} \spb{c}.{B}
    {\cal P}_{Q}^2 (({\dupx}-{\cupx}) (8
    {r_Q}-3 {d_\delta})+2 {\dupx})}{2
    \spa{c}.{A} \spb{d}.{B} {Q_C}^2}
\notag \\ &    
    -\frac{\spa{d}.{A}
    \spb{c}.{B}^2 {\cal P}_{Q} (({\dupx}-{\cupx})
    (-3 {d_\delta}+4 {r_Q}-4)+2 {\dupx})}{2
    \spa{c}.{A} \spb{d}.{B}^2 {Q_C}}
\notag \\ &    
    +\frac{\spa{d}.{A}
    \spb{c}.{B} ((-3 {d_\delta}-4)
    ({\dupx}-{\cupx})+2 {\dupx})}{\spa{c}.{A}
    \spb{d}.{B} {Q_C}}\biggr)
\end{align}

\begin{align}
C_{dp\g:Ixy}= -\frac{{\cal A}_{p} {\cal S}_{p}
    C_{pre} }{4{\calPdsq}}
    \biggl(&\frac{\spa{d}.{A}^2
    \spb{c}.{B} {\cal P}_{Q} (({\dupx}-{\cupx})
    (4 {r_Q}-3 {d_\delta})+2 {\dupx})}{2
    \spa{c}.{A}^2 \spb{d}.{B} {Q_C}}
\notag \\ &    
    +\frac{\spa{d}.{A}
    \spb{c}.{B} {\cal P}_{Q}^2 (({\dupx}-{\cupx})
    (8 {r_Q}-3 {d_\delta})+2 {\dupx})}{2
    \spa{c}.{A} \spb{d}.{B} {Q_C}^2}
\notag \\ &    
    -\frac{\spa{d}.{A}
    \spb{c}.{B}^2 {\cal P}_{Q} (({\dupx}-{\cupx})
    (-3 {d_\delta}+4 {r_Q}-4)+2 {\dupx})}{2
    \spa{c}.{A} \spb{d}.{B}^2 {Q_C}}
\notag \\ &    
    +\frac{\spa{d}.{A}
    \spb{c}.{B} ((-3 {d_\delta}-4)
    ({\dupx}-{\cupx})+2 {\dupx})}{\spa{c}.{A}
    \spb{d}.{B} {Q_C}}
\notag \\ &    
    -\frac{\spa{d}.{A}^4
    {\cal P}_{r1} 
    \begin{pmatrix}
    ({\dupx}-{\cupx}) \left(4
    {c_\delta} (55 {d_\delta}-32 {r_1})+120
    {d_\delta}^2 {r_1}-105
    {d_\delta}^3\right)
    \\
    +{\dupx} \left(-80 {d_\delta}
    {r_1}+64 {r_1}-72 {c_\delta}+70
    {d_\delta}^2-56 {d_\delta}+48\right)
    \end{pmatrix}
    }{48
    \spa{c}.{A}^4
    \left(\frac{\spa{c}.{A}}{\spa{d}.{A}}-{\cal P}_{Q}
    {r_Q}\right)}
\notag \\ &    
    +\frac{\spb{c}.{B}^3 {\cal P}_{r2}
    \begin{pmatrix}
    -\frac{1}{2} ({\dupx}-{\cupx}) \left(-18
    {d_\delta} r_{\bar 2}-16 {c_\delta}+15
    {d_\delta}^2\right)
    \\
    +\frac{1}{48}
    ({\dupx}-{\cupx}) \left(4 {c_\delta} (55
    {d_\delta}-32 r_{\bar 2})+120 {d_\delta}^2
    r_{\bar 2}-105 {d_\delta}^3\right) 
    \\
    +\frac{1}{48}
    {\dupx} \left(-80 {d_\delta} r_{\bar 2}+64
    r_{\bar 2}-72 {c_\delta}+70 {d_\delta}^2-56
    {d_\delta}+48\right)
    \\
    +3 ({\dupx}-{\cupx}) (4
    r_{\bar 2}-3 {d_\delta}
    )-\frac{1}{2} {\dupx}
    (12 r_{\bar 2}-10 {d_\delta}+8)
    \\
    -4
    ({\dupx}-{\cupx})+6
    {\dupx}
    \end{pmatrix}
    }{\spb{d}.{B}^3 \left(\frac{\spb{d}.{B}
    {\cal P}_{Q}
    {r_Q}}{\spb{c}.{B}}-{Q_C}\right)}\biggr)
\end{align}


\begin{align}
\calPdsq= (\dupx-\cupx)^2+4\mupx
\,\, ,\,\,
d_\delta=-2{\cupx(\cupx-\dupx) +2\mupx \over \calPdsq}
\,\, ,\,\,
c_\delta={ \cupx^2\over \calPdsq} 
\end{align}

\subsubsection{Complex $\tQ$}
For complex $\tQ$,  $c$ and $d$ are chosen so that $[c|\tQ|d\ra=0$ while $[d|\tQ|c\ra\neq 0$.
With  $c$ and $d$ fixed there are now special cases if either $[BC]$ or $\spa{A}.d$ vanish: 
The massive momentum is taken to be
\begin{align}
\tQ=P+\bar\lambda_c \lambda_X\,.
\end{align}

\begin{align}
G_{111}=\begin{cases}
         G_{111}^{cc} & [B|P|A\ra \neq 0, \la A|P\tQ|A\ra \neq 0, [B|P\tQ|B] \neq 0 , [Bc] \neq 0, \spa{A}.d \neq 0 \\
         G_{111}^{cd} & [B|P|A\ra   =  0, \la A|P\tQ|A\ra \neq 0, [B|P\tQ|B] \neq 0 , [Bc] \neq 0, \spa{A}.d \neq 0\\
         G_{111}^{ce} & [B|P|A\ra \neq 0, \la A|P\tQ|A\ra  =   0, [B|P\tQ|B] \neq 0 , [Bc] \neq 0, \spa{A}.d \neq 0\\
         G_{111}^{cx} & [B|P|A\ra \neq 0, \la A|P\tQ|A\ra \neq 0, [B|P\tQ|B]   =  0 , [Bc] \neq 0, \spa{A}.d \neq 0\\
         G_{111}^{cs} & [B|P|A\ra \neq 0, \la A|P\tQ|A\ra \neq 0, [B|P\tQ|B]   =  0 , [Bc]  =   0, \spa{A}.d \neq 0 \\
         G_{111}^{ct} & [B|P|A\ra \neq 0, \la A|P\tQ|A\ra  =   0, [B|P\tQ|B]   =  0 , [Bc]  =   0, \spa{A}.d \neq 0 \\
         G_{111}^{ca} & [B|P|A\ra \neq 0, \la A|P\tQ|A\ra  =   0, [B|P\tQ|B] \neq 0 , [Bc] \neq 0, \spa{A}.d  =   0 \\
        \end{cases}
\end{align}

\def\Cnp0J{C_{np:0}^J}
\def\CnpgJ{C_{np:\g}^J}
\def\Cap0J{C_{ap:0}^J}
\def\CapaJ{C_{ap:a}^J}
\def\CapgJ{C_{ap:\g}^J}
\def\Csp0J{C_{sp:0}^J}
\def\CspsJ{C_{sp:s}^J}
\def\CspgJ{C_{sp:\g}^J}
\def\Cdp0J{C_{dp:0}^J}
\def\CdpaJ{C_{dp:a}^J}
\def\CdpsJ{C_{dp:s}^J}
\def\CdpgJ{C_{dp:\g }^J}

\def\spur{ {^*} }
\begin{align}
G_{111}^{cc}=(  \Cnp0J
                                              +    \CnpgJ{\spur}
                                              +    \Cap0J
                                              +    \CapaJ
                                              +    \CapgJ{\spur}
                                              +    \Csp0J
                                              +    \CspsJ
                                              +    \CspgJ{\spur}
                                              +    \Cdp0J
                                              +    \CdpsJ
                                              +    \CdpaJ
                                              +    \CdpgJ{\spur} 
                                              )   
\end{align}

\def\CdpayJ{C_{dp:ay}}
\def\CdpsyJ{C_{dp:sy}}

\begin{align}
G_{111}^{cd}=(  \Cnp0J
                                              +    \CnpgJ{\spur}
                                              +    \Cap0J
                                              +    \CapaJ
                                              +    \CapgJ{\spur}
                                              +    \Csp0J
                                              +    \CspsJ
                                              +    \CspgJ{\spur}
                                              +    \Cdp0J
                                              +   \CdpayJ
                                              +   \CdpsyJ
                                              +    \CdpgJ{\spur}                                         
                                             )
\end{align}                                            
\def\CapaxJ{C_{ap:ax}^J}
\def\CapgxJ{C_{ap:\g x}^J}
\def\CapgIx{C_{ap:\g Ix}}
\def\CdpaxJ{C_{dp:ax}^J}
\def\CdpgzJ{C_{dp:\g z}^J}

\begin{align}
G_{111}^{ce}=(  \Cnp0J
                                              +    \CnpgJ{\spur}
                                              +    \Cap0J
                                              +    \CapaxJ
                                              +    \CapgxJ{\spur}
                                              +    \Csp0J
                                              +    \CspsJ
                                              +    \CspgJ{\spur}
                                              +    \Cdp0J
                                              +    \CdpsJ
                                             +     \CdpaxJ
                                              +    \CdpgzJ{\spur}                                              
                                             )
\end{align}                                            
  
\def\CspsxJ{C_{sp:sx}^J}
\def\CspgxJ{C_{sp:\g x}^J}
\def\CdpsxJ{C_{dp:s x}^J}
\def\CdpgxJ{C_{dp:\g x}^J}

\begin{align}
G_{111}^{cx}=(\Cnp0J
                                              +    \CnpgJ{\spur}
                                              +    \Cap0J
                                              +    \CapaJ
                                              +    \CapgJ{\spur}
                                              +    \Csp0J
                                              +    \CspsxJ
                                              +    \CspgxJ{\spur}
                                              +    \Cdp0J
                                              +    \CdpsxJ
                                              +    \CdpaJ
                                              +    \CdpgxJ{\spur}                                               
                                             )
\end{align}                                            

\def\Cnp0H{C_{np:0}^H}
\def\CnpgH{C_{np:\g}^H}
\def\Cap0H{C_{ap:0}^H}
\def\CapaH{C_{ap:a}^H}
\def\CapgH{C_{ap:\g}^H}

\begin{align}
G_{111}^{cs}=(  \Cnp0H 
                                                 +    \CnpgH{\spur}
                                                 +    \Cap0H
                                                 +    \CapaH
                                                 +    \CapgH{\spur} )
\end{align}      

\def\CapaHa{C_{ap:a a}^H}
\def\CapgHa{C_{ap:\g a}^H}

\begin{align}
G_{111}^{ct}=(  \Cnp0H 
                                                 +    \CnpgH{\spur}
                                                 +    \Cap0H
                                                 +   \CapaHa
                                                +    \CapgHa{\spur}
                                    )                                               
\end{align}   
\def\Cnp0Y{C_{np:0}^Y}
\def\CnpgY{C_{np:\g}^Y}
\def\Csp0Y{C_{sp:0}^Y}
\def\CspsY{C_{sp:a}^Y}
\def\CspgY{C_{sp:\g}^Y}

\begin{align}
G_{111}^{ca}=(  \Cnp0Y 
                                                 +    \CnpgY{\spur}
                                                 +    \Csp0Y
                                                 +    \CspsY
                                                 +    \CspgY{\spur}
                                                   )
\end{align}                                            
                                           
\def\CpreJ{-1} 
                                           
The contributions are: from the no bonus pole piece
\begin{align}
C_{np:0}^J&= \frac{1 }{2 {[d|\tQ|c\ra}}                                  
    \left(-\frac{([c|\tQ|c\ra+{[d|\tQ|d\ra})                        
    {\cal A}_0
    {\cal S}_{0}}{{[d|\tQ|c\ra}}+{\cal A}_0 {\cal S}_1+{\cal A}_1
    {\cal S}_{0}\right)
\end{align}
\begin{align}
C_{np:0}^H={ C^x_{pre}\over 2}\biggl(
                             {  {\cal A}_1 T_0 +{\cal A}_0 T_1  \over [d|\tQ|c\ra}
                               -{{\cal A}_0 T_0 ([c|\tQ|c\ra+[d|\tQ|d\ra)\over [d|\tQ|c\ra^2} 
                               \biggr)
\end{align}
with
\begin{align}
C^x_{pre}&={1\over [dB]}
\,\, ,\,\,
T_x=[cb][cf][cx]
\,\, ,\,\,
T_0=[dx] [db] [df]
\notag \\ 
T_1&= [dx][db][cf]+[dx][df][cb]+[db][df][cx]
\,\, ,\,\,
T_2=[dx][cb][cf]+[db][cf][cx]+[df][cb][cx]
\end{align}

\begin{align}
C_{np:0}^Y={C^y_{pre}\over 2}\biggl(
                             {  {\cal S}_1 U_0 +{\cal S}_0 U_1\over [d|\tQ|c\ra}
                               -{{\cal S}_0 U_0 ([c|\tQ|c\ra+[d|\tQ|d\ra)\over [d|\tQ|c\ra^2} \biggr)
\end{align}
with
\begin{align}
C^y_{pre}&={1\over \spa{c}.A}
\,\, ,\,\,
U_x=\spa{d}.{a} \spa{d}.{e} \spa{d}.{y}
\,\, ,\,\,
U_0=\spa{c}.{a} \spa{c}.{e} \spa{c}.{y}
\notag \\ 
U_1&= \spa{d}.{a} \spa{c}.{e} \spa{c}.{y}+\spa{c}.{a}
    \spa{c}.{y} \spa{d}.{e}+\spa{c}.{a} \spa{c}.{e}
    \spa{d}.{y}
\notag \\
U_2&= \spa{d}.{a} \spa{c}.{y} \spa{d}.{e}+\spa{c}.{a}
    \spa{d}.{e} \spa{d}.{y}+\spa{d}.{a} \spa{c}.{e}
    \spa{d}.{y}
\end{align}

\begin{align}
C_{np:\g}^J= \frac{1 }{2 \spb{d}.{c}} 
\biggl(&\frac{(2                    
    \spa{c}.{d}+\spa{X}.{d}) {\cal A}_0               
    {\cal S}_{0}}{\spa{X}.{c}^2}-\frac{{\cal A}_0 
{\cal S}_1}{\spa{X}.{c}}-\frac{(2             
    \spa{c}.{d}+3 \spa{X}.{d}) {\cal A}_0             
    {\cal S}_x}{\spa{X}.{d}^2}
    \notag \\ &
    -\frac{{\cal A}_1 
{\cal S}_{0}}{\spa{X}.{c}}+\frac{(2             
    \spa{c}.{d}+\spa{X}.{d}) {\cal A}_1               
    {\cal S}_1}{\spa{X}.{d}^2}+\frac{\spa{X}.{c} (6   
    \spa{c}.{d}+5 \spa{X}.{d}) {\cal A}_1             
    {\cal S}_x}{\spa{X}.{d}^3}
    \notag \\ &
    +\frac{(\spa{X}.{d}-2
    \spa{c}.{d}) {\cal A}_x
    {\cal S}_{0}}{\spa{X}.{d}^2}-\frac{\spa{X}.{c} (6
    \spa{c}.{d}+\spa{X}.{d}) {\cal A}_x
    {\cal S}_1}{\spa{X}.{d}^3}-\frac{6 \spa{X}.{c}^2
    (2 \spa{c}.{d}+\spa{X}.{d}) {\cal A}_x
    {\cal S}_x}{\spa{X}.{d}^4}
    \biggr)
\end{align}
where
\begin{align}
{\cal A}_x={\spa{d}.a \spa{d}.e \spa{d}.y \over \spa{d}.A}
\,\, ,\,\,
{\cal S}_x={[cb][cf][cx]\over [cB]}
\end{align}

\begin{align}
 C_{np:\g}^H=  &\frac{C^x_{pre}}{6 [c|\tQ|c\ra
    ([c|\tQ|c\ra-{[d|\tQ|d\ra})^5 {[d|\tQ|c\ra}^2}
\notag \\ & \hskip-50pt\times
    \begin{pmatrix}\hskip-330pt
    3
    {\cal A}_0([c|\tQ|c\ra-{[d|\tQ|d\ra})^2
    \notag \\ \hskip +20pt \times
    \begin{pmatrix}
    T_0
    [c|\tQ|c\ra^5-(2 {[d|\tQ|d\ra}
    T_0+{[d|\tQ|c\ra} T_1)
    [c|\tQ|c\ra^4+{[d|\tQ|c\ra} (3 {[d|\tQ|d\ra}
    T_1+{[d|\tQ|c\ra} T_2)
    [c|\tQ|c\ra^3
    \\
    +
    \left(
    2 T_0 {[d|\tQ|d\ra}^3-3
    {[d|\tQ|c\ra} T_1 {[d|\tQ|d\ra}^2-4
    {[d|\tQ|c\ra}^2 T_2
    {[d|\tQ|d\ra}-{[d|\tQ|c\ra}^3 T_x
    \right)
    [c|\tQ|c\ra^2
    \\
    +{[d|\tQ|d\ra} 
    \left(-T_0
    {[d|\tQ|d\ra}^3+{[d|\tQ|c\ra} T_1
    {[d|\tQ|d\ra}^2+3 {[d|\tQ|c\ra}^2 T_2
    {[d|\tQ|d\ra}+5 {[d|\tQ|c\ra}^3 T_x
    \right)
    [c|\tQ|c\ra
    \\
    +2 {[d|\tQ|d\ra}^2 {[d|\tQ|c\ra}^3
    T_x
    \end{pmatrix}    
    \notag \\ \hskip -420pt 
    +{[d|\tQ|c\ra}
    \notag \\ \hskip -0pt \times
    \begin{pmatrix}
    {\cal A}_x {[d|\tQ|c\ra} 
      \begin{pmatrix}
    -9
    T_0 [c|\tQ|c\ra^5+15 (2 {[d|\tQ|d\ra}
    T_0+{[d|\tQ|c\ra} T_1)
    [c|\tQ|c\ra^4
    \\
    -9 
        \left(
    4 T_0 {[d|\tQ|d\ra}^2+3
    {[d|\tQ|c\ra} T_1 {[d|\tQ|d\ra}+2
    {[d|\tQ|c\ra}^2 T_2
        \right)
    [c|\tQ|c\ra^3
    \\
    +
        \left(
       18 T_0 {[d|\tQ|d\ra}^3+9
    {[d|\tQ|c\ra} T_1 {[d|\tQ|d\ra}^2+20
    {[d|\tQ|c\ra}^3 T_x
        \right)
    [c|\tQ|c\ra^2
    \\
    +{[d|\tQ|d\ra} 
        \begin{pmatrix}
        -3 T_0
    {[d|\tQ|d\ra}^3+3 {[d|\tQ|c\ra} T_1
    {[d|\tQ|d\ra}^2\\+18 {[d|\tQ|c\ra}^2 T_2
    {[d|\tQ|d\ra}\\+ 38 {[d|\tQ|c\ra}^3 T_x
        \end{pmatrix}
    [c|\tQ|c\ra
    \\
    \hskip -150pt +2 {[d|\tQ|d\ra}^2 {[d|\tQ|c\ra}^3
    T_x
      \end{pmatrix}
     \notag \\ \hskip-250pt 
    -3 {\cal A}_1
    ([c|\tQ|c\ra-{[d|\tQ|d\ra}) 
    \notag \\ \hskip -00pt \times
      \begin{pmatrix}
    T_0
    [c|\tQ|c\ra^5-(4 {[d|\tQ|d\ra}
    T_0+{[d|\tQ|c\ra} T_1)
    [c|\tQ|c\ra^4
    \\+
        \left(
    6 T_0
    {[d|\tQ|d\ra}^2+{[d|\tQ|c\ra} T_1
    {[d|\tQ|d\ra}+{[d|\tQ|c\ra}^2 T_2
        \right)
    [c|\tQ|c\ra^3
    \\+
        \begin{pmatrix}
    -4 T_0
    {[d|\tQ|d\ra}^3+{[d|\tQ|c\ra} T_1
    {[d|\tQ|d\ra}^2
    \\
    +4 {[d|\tQ|c\ra}^2 T_2
    {[d|\tQ|d\ra}-{[d|\tQ|c\ra}^3 T_x
        \end{pmatrix}
    [c|\tQ|c\ra^2
    \\
    +{[d|\tQ|d\ra} 
        \begin{pmatrix}
    T_0
    {[d|\tQ|d\ra}^3-{[d|\tQ|c\ra} T_1
    {[d|\tQ|d\ra}^2
    \\
    -5 {[d|\tQ|c\ra}^2 T_2
    {[d|\tQ|d\ra}-10 {[d|\tQ|c\ra}^3 T_x
        \end{pmatrix}
    [c|\tQ|c\ra
    \\ \hskip-150pt
    -{[d|\tQ|d\ra}^2 {[d|\tQ|c\ra}^3
    T_x
      \end{pmatrix}
    \end{pmatrix}
    \end{pmatrix}
\end{align}

\begin{align}
 C_{np:\g}^Y= &\frac{C^y_{pre} }{6
    ([c|\tQ|c\ra-{[d|\tQ|d\ra})^5 {[d|\tQ|d\ra}
    {[d|\tQ|c\ra}^2}
\notag \\ & \times    
    \biggl(3
    {\cal S}_0 ([c|\tQ|c\ra-{[d|\tQ|d\ra})^2
\notag \\ &\hskip 10 pt \times    
    \begin{pmatrix}
    {[d|\tQ|d\ra} U_0
    [c|\tQ|c\ra^4
    \\
    -{[d|\tQ|d\ra} (2 {[d|\tQ|d\ra}
    U_0+{[d|\tQ|c\ra} U_1)
    [c|\tQ|c\ra^3
    \\
    +{[d|\tQ|c\ra} \left(3 U_1
    {[d|\tQ|d\ra}^2-3 {[d|\tQ|c\ra} U_2
    {[d|\tQ|d\ra}-2 {[d|\tQ|c\ra}^2 U_x\right)
    [c|\tQ|c\ra^2
    \\
    +{[d|\tQ|d\ra} \left(2 U_0
    {[d|\tQ|d\ra}^3-3 {[d|\tQ|c\ra} U_1
    {[d|\tQ|d\ra}^2+4 {[d|\tQ|c\ra}^2 U_2
    {[d|\tQ|d\ra}-5 {[d|\tQ|c\ra}^3 U_x\right)
    [c|\tQ|c\ra
    \\
    +{[d|\tQ|d\ra}^2 \left(-U_0
    {[d|\tQ|d\ra}^3+{[d|\tQ|c\ra} U_1
    {[d|\tQ|d\ra}^2-{[d|\tQ|c\ra}^2 U_2
    {[d|\tQ|d\ra}+{[d|\tQ|c\ra}^3
    U_x\right)
    \end{pmatrix}
\notag \\ &    
    -{[d|\tQ|c\ra} 
\notag \\ & \hskip -10pt \times    
    \begin{pmatrix}
    3
    {\cal S}_1 ([c|\tQ|c\ra-{[d|\tQ|d\ra})
\notag \\ & \hskip -115pt \times   
    \begin{pmatrix}
    {[d|\tQ|d\ra} U_0
    [c|\tQ|c\ra^4-{[d|\tQ|d\ra} (4 {[d|\tQ|d\ra}
    U_0+{[d|\tQ|c\ra} U_1)
    [c|\tQ|c\ra^3
    \\
    +
    \left(6 U_0
    {[d|\tQ|d\ra}^3+{[d|\tQ|c\ra} U_1
    {[d|\tQ|d\ra}^2-5 {[d|\tQ|c\ra}^2 U_2
    {[d|\tQ|d\ra}-{[d|\tQ|c\ra}^3 U_x
    \right)
    [c|\tQ|c\ra^2
    \\
    +{[d|\tQ|d\ra} 
    \left(-4 U_0
    {[d|\tQ|d\ra}^3+{[d|\tQ|c\ra} U_1
    {[d|\tQ|d\ra}^2+4 {[d|\tQ|c\ra}^2 U_2
    {[d|\tQ|d\ra}-10 {[d|\tQ|c\ra}^3 U_x
    \right)
    [c|\tQ|c\ra
    \\
    +{[d|\tQ|d\ra}^2 
    \left(U_0
    {[d|\tQ|d\ra}^3-{[d|\tQ|c\ra} U_1
    {[d|\tQ|d\ra}^2+{[d|\tQ|c\ra}^2 U_2
    {[d|\tQ|d\ra}-{[d|\tQ|c\ra}^3
    U_x
    \right)
    \end{pmatrix}
\notag \\ & \hskip -520pt        
    +{\cal S}_x
    {[d|\tQ|c\ra} 
\notag \\ & \hskip -120pt \times       
    \begin{pmatrix}
    -3 {[d|\tQ|d\ra} U_0
    [c|\tQ|c\ra^4+3 {[d|\tQ|d\ra} (6 {[d|\tQ|d\ra}
    U_0+{[d|\tQ|c\ra} U_1)
    [c|\tQ|c\ra^3
    \\
    +
    \left(-36 U_0 {[d|\tQ|d\ra}^3+9
    {[d|\tQ|c\ra} U_1 {[d|\tQ|d\ra}^2+18
    {[d|\tQ|c\ra}^2 U_2 {[d|\tQ|d\ra}+2
    {[d|\tQ|c\ra}^3 U_x
    \right)
    [c|\tQ|c\ra^2
    \\
    +\left(
    30 U_0 {[d|\tQ|d\ra}^4-27
    {[d|\tQ|c\ra} U_1 {[d|\tQ|d\ra}^3+38
    {[d|\tQ|c\ra}^3 U_x {[d|\tQ|d\ra}
    \right)
    [c|\tQ|c\ra
    \\
    +{[d|\tQ|d\ra}^2 
    \left(-9 U_0
    {[d|\tQ|d\ra}^3+15 {[d|\tQ|c\ra} U_1
    {[d|\tQ|d\ra}^2-18 {[d|\tQ|c\ra}^2 U_2
    {[d|\tQ|d\ra}+20 {[d|\tQ|c\ra}^3
    U_x
    \right)
    \end{pmatrix}
    \end{pmatrix}
    \biggr)
\end{align}

From the bonus angle  pole piece:

\begin{align}
C_{ap:0}^J=  \frac{\spa{d}.{A} {\cal A}_{p}
    }{2 \spa{c}.{A} {[d|\tQ|c\ra}}
    \left({\cal S}_1-\left(\frac{\spa{d}.{A}}{\spa{c}.{A}}
+\frac{[c|\tQ|c\ra+{[d|\tQ|d\ra}}{{[d|\tQ|c\ra}}\right)
    {\cal S}_{0}\right)
\end{align}
\begin{align}
C_{ap:a}^J=  \frac{\spa{d}.{A}^2 {\cal A}_{p}                        
     }
{\spa{c}.{A}^2 {\cal P}_{\cal A}^J}
    \begin{pmatrix}
    \left(\frac{\spa{d}.{A}^2     
    \spb{A}.{d}^2}{2 [A|P|A\ra^2}-\frac{\spa{d}.{A}            
    \spb{A}.{d} {b}_{\cal A}^J}{[A|P|A\ra}\right)          
    \left(\frac{\spa{c}.{A}                                                 
    {\cal S}_1}{\spa{d}.{A}}+\frac{\spa{c}.{A}^2
    {\cal S}_x}{\spa{d}.{A}^2}+{\cal S}_{0}\right)
    \\
-\frac{\spa{d}.{A} \spb{A}.{d} \left(\frac{\spa{c}.{A}
    {\cal S}_1}{\spa{d}.{A}}+2
    {\cal S}_{0}\right)}{[A|P|A\ra}
    \end{pmatrix}
\end{align}
with
\begin{align}
{\cal P}_{\cal A}^J=-{\spa{c}.{d} \la A|P\tQ|A\ra \over \spa{d}.{A}^2}
\,\, ,\,\,
b_{\cal A}^J=-{\spa{c}.{d} [d|\tQ|A\ra \over \spa{d}.{A} }{1\over {\cal P}_{\cal A}}  
\end{align}

\begin{align}
C_{ap:\g}^J= \frac{ {\cal A}_{p}   }{2 {[d|\tQ|c\ra}^2                      
    {\cal P}_{\cal A}^J}   
    \begin{pmatrix}
    ([c|\tQ|c\ra-{[d|\tQ|d\ra}) ([c|\tQ|c\ra (2             
    {b}_{\cal A}^J-1)-{[d|\tQ|d\ra} (2 {b}_{\cal A}^J+3))  
    {\cal S}_{0}
    \\
    +{[d|\tQ|c\ra} (([c|\tQ|c\ra (1-2   
    {b}_{\cal A}^J)
    +{[d|\tQ|d\ra} (2 {b}_{\cal A}^J+1))    
    {\cal S}_1+{[d|\tQ|c\ra} (2 {b}_{\cal A}^J-1)
    {\cal S}_x)
    \end{pmatrix}
\end{align}

\begin{align}
C_{ap:ax}^J= -\frac{ \spa{d}.{A}^3 {\cal A}_{p}                         
    }{\spa{c}.{A}^2
    \spa{c}.{d} [d|\tQ|A\ra}
    \begin{pmatrix}
    \frac{\spa{d}.{A}^3           
    \spb{A}.{d}^3 \left(\frac{\spa{c}.{A}                               
    {\cal S}_1}{\spa{d}.{A}}+\frac{\spa{c}.{A}^2      
    {\cal S}_x}{\spa{d}.{A}^2}
+{\cal S}_{0}\right)}{3 [A|P|A\ra^3}
\\
-\frac{\spa{d}.{A}^2 \spb{A}.{d}^2
    \left(\frac{\spa{c}.{A} {\cal S}_1}{\spa{d}.{A}}+2
    {\cal S}_{0}\right)}{2
    [A|P|A\ra^2}+\frac{\spa{d}.{A} \spb{A}.{d}
    {\cal S}_{0}}{[A|P|A\ra}
    \end{pmatrix}
\end{align}
\begin{align}
C_{ap:\g x}^J= \frac{ \spa{d}.{A} {\cal A}_{p}                          
    }{6 \spa{c}.{d} [d|\tQ|A\ra  
    {[d|\tQ|c\ra}^2} 
    \begin{pmatrix}
    2 \left([c|\tQ|c\ra        
    {[d|\tQ|d\ra}+[c|\tQ|c\ra^2+{[d|\tQ|d\ra}^2\right)         
    {\cal S}_{0}
    \\
    +{[d|\tQ|c\ra} (2 {[d|\tQ|c\ra}       
    {\cal S}_x-(2 [c|\tQ|c\ra+{[d|\tQ|d\ra})      
    {\cal S}_1)
    \end{pmatrix}
\end{align}

\begin{align}
C_{ap:0}^H&= \frac{\spa{d}.{A} {\cal A}_p                           
    C^x_{pre}}{2 \spa{c}.{A}
    {[d|\tQ|c\ra}}
    \left(T_1-T_0
    \left(\frac{\spa{d}.{A}}{\spa{c}.{A}}+\frac{[c|\tQ|c\ra
+{[d|\tQ|d\ra}}{{[d|\tQ|c\ra}}\right)\right)
\end{align}
\begin{align}
C_{ap:a}^H&= \frac{\spb{A}.{d} {\cal A}_p
    C^x_{pre}}{2 \spa{c}.{A}^2
    {\cal P}_{\cal A}^J (\spa{c}.{A} \spb{A}.{c}+\spa{d}.{A}
    \spb{A}.{d})^2}
\notag \\ & \hskip 50pt \times    
    \begin{pmatrix}
     \spa{d}.{A}
    \begin{pmatrix}
    \spa{d}.{A} (2 c_A^J (\spa{c}.{A}
    \spb{A}.{c}+\spa{d}.{A} \spb{A}.{d}) (\spa{d}.{A}
    T_0+\spa{c}.{A} T_1)
    \\
    +\spa{d}.{A}
    (\spa{c}.{A} \spb{A}.{d} T_1-(2 \spa{c}.{A}
    \spb{A}.{c}+\spa{d}.{A} \spb{A}.{d})
    T_0))
    \\
    +\spa{c}.{A}^2 T_2 (2
    \spa{c}.{A} \spb{A}.{c} (c_A^J+1)+\spa{d}.{A}
    \spb{A}.{d} (2 c_A^J+3))
    \end{pmatrix}
    \\
    +\spa{c}.{A}^3
    T_x (2 \spa{c}.{A} \spb{A}.{c}
    (c_A^J+2)+\spa{d}.{A} \spb{A}.{d} (2
    c_A^J+5))
    \end{pmatrix}
\end{align}
with
\begin{align}
 c_A^J=\frac{\spa{c}.{A} [c|\tQ|c\ra}{\spa{d}.{A}
    {\cal P}_{\cal A}^J}
\end{align}

\begin{align}
C_{ap:aa}^H&=  -\frac{\spa{d}.{A} \spb{A}.{d} {\cal A}_p          
    C^x_{pre} }{6 \spa{c}.{A}^3 [c|\tQ|c\ra
    (\spa{c}.{A} \spb{A}.{c}+\spa{d}.{A} \spb{A}.{d})^3}
\notag \\ & \hskip 50pt \times    
    \begin{pmatrix}
    \spa{d}.{A}^3 
    \begin{pmatrix}
    2 
      \left(3 
    \spa{c}.{A} \spb{A}.{c} \spa{d}.{A} \spb{A}.{d}+3           
    \spa{c}.{A}^2 \spb{A}.{c}^2+\spa{d}.{A}^2
    \spb{A}.{d}^2
      \right) T_0
    \\
    +\spa{c}.{A}
    \spb{A}.{d} (2 \spa{c}.{A} \spb{A}.{d} T_2-(3
    \spa{c}.{A} \spb{A}.{c}+\spa{d}.{A} \spb{A}.{d})
    T_1)
    \end{pmatrix}
    \\
    +\spa{c}.{A}^3 \left(15 \spa{c}.{A}
    \spb{A}.{c} \spa{d}.{A} \spb{A}.{d}+6 \spa{c}.{A}^2
    \spb{A}.{c}^2+11 \spa{d}.{A}^2 \spb{A}.{d}^2\right)
    T_x
    \end{pmatrix}
\end{align}
\begin{align}
C_{ap:\g}^H&=   \frac{{\cal A}_p C^x_{pre}  }{2
    {\cal P}_{\cal A}^J}     
    \begin{pmatrix}
    \frac{2 [c|\tQ|c\ra {[d|\tQ|d\ra} (2                      
    c_A^J+1)                                                   
    T_0}{{[d|\tQ|c\ra}^2}+\frac{{[d|\tQ|c\ra}        
    T_x ([c|\tQ|c\ra (2                                 
    c_A^J+3)-{[d|\tQ|d\ra} (2                             
    c_A^J+5))}{([c|\tQ|c\ra-{[d|\tQ|d\ra})^2}
    \\
    -\frac
    {[c|\tQ|c\ra^2 (2 c_A^J+3)                          
    T_0}{{[d|\tQ|c\ra}^2}+\frac{[c|\tQ|c\ra (2     
    c_A^J+3)                                                   
    T_1}{{[d|\tQ|c\ra}}-\frac{{[d|\tQ|d\ra}^2 (2
    c_A^J-1)
    T_0}{{[d|\tQ|c\ra}^2}-\frac{{[d|\tQ|d\ra} (2
    c_A^J+1) T_1}{{[d|\tQ|c\ra}}-(2
    c_A^J+3) T_2
    \end{pmatrix}
\end{align}
\begin{align}
C_{ap:\g a}^H&= \frac{\spa{d}.{A} {\cal A}_p                           
    C^x_{pre} }{6 \spa{c}.{A} [c|\tQ|c\ra
    {[d|\tQ|c\ra}^2 ([c|\tQ|c\ra-{[d|\tQ|d\ra})^3}
\notag \\ & \hskip 0pt \times    
    \begin{pmatrix}
    -[c|\tQ|c\ra^2 
    \left(3     
    {[d|\tQ|d\ra}^2 {[d|\tQ|c\ra} T_1+6              
    {[d|\tQ|d\ra} {[d|\tQ|c\ra}^2 T_2+2              
    {[d|\tQ|d\ra}^3 T_0+2 {[d|\tQ|c\ra}^3            
    T_x
    \right)
    \\
    +[c|\tQ|c\ra {[d|\tQ|d\ra}           
    \left(
    -{[d|\tQ|d\ra}^2 {[d|\tQ|c\ra} T_1+6       
    {[d|\tQ|d\ra} {[d|\tQ|c\ra}^2 T_2+4              
    {[d|\tQ|d\ra}^3 T_0+7 {[d|\tQ|c\ra}^3            
    T_x
    \right)
    \\
    +[c|\tQ|c\ra^3 
    \left(5 {[d|\tQ|d\ra} 
    {[d|\tQ|c\ra} T_1+2 {[d|\tQ|d\ra}^2
    T_0+2 {[d|\tQ|c\ra}^2 T_2
    \right)
    \\
    -2
    [c|\tQ|c\ra^4 (2 {[d|\tQ|d\ra}
    T_0+{[d|\tQ|c\ra} T_1)
    \\
    +2
    [c|\tQ|c\ra^5 T_0
    \\
    +{[d|\tQ|d\ra}^2
    \left({[d|\tQ|d\ra}^2 {[d|\tQ|c\ra} T_1-2
    {[d|\tQ|d\ra} {[d|\tQ|c\ra}^2 T_2-2
    {[d|\tQ|d\ra}^3 T_0-11 {[d|\tQ|c\ra}^3
    T_x
    \right)
    \end{pmatrix}
\end{align}

\noindent{From} the bonus square  pole piece:

\begin{align}
C_{sp:0}^J= \frac{\spb{c}.{B} {\cal S}_p                          
    }{2 \spb{d}.{B}
    {[d|\tQ|c\ra}}
    \left({\cal A}_1-{\cal A}_0
    \left(\frac{\spb{c}.{B}}{\spb{d}.{B}}+\frac{[c|\tQ|c\ra
+{[d|\tQ|d\ra}}{{[d|\tQ|c\ra}}\right)\right)
\end{align}
\begin{align}
C_{sp:s}^J=  \frac{ \spb{c}.{B}^2 {\cal S}_p                        
    }{\spb{d}.{B}^2 {\cal P}_S^J}
    \begin{pmatrix}
    \left(\frac{\spa{c}.{B}^2     
    \spb{B}.{c}^2}{2 [B|P|B\ra^2}-\frac{\spa{c}.{B}            
    \spb{B}.{c} b_s^J}{[B|P|B\ra}\right)          
    \left(\frac{\spb{d}.{B}
    {\cal A}_1}{\spb{c}.{B}}+\frac{\spb{d}.{B}^2
    {\cal A}_x}{\spb{c}.{B}^2}+{\cal A}_0\right)
    \\
-\frac{\spa{c}.{B} \spb{B}.{c} \left(\frac{\spb{d}.{B}
    {\cal A}_1}{\spb{c}.{B}}+2
    {\cal A}_0\right)}{[B|P|B\ra}
    \end{pmatrix}
\end{align}
with
\begin{align}
{\cal P}_S^J= \frac{\spb{c}.{d} (\spb{B}.{c} [B|\tQ|c\ra+\spb{B}.{d}
    [B|\tQ|d\ra)}{\spb{c}.{B}^2}
\,\, ,\,\,
b_S^J= -\frac{\frac{\spb{d}.{B}
    [c|\tQ|c\ra}{\spb{c}.{B}}-{[d|\tQ|c\ra}}{{\cal P}_S^J}
\end{align}

\begin{align}
C_{sp:\g}^J=  \frac{  {\cal S}_p }{2 {[d|\tQ|c\ra}^2
    {\cal P}_S^J}    
    \begin{pmatrix}
([c|\tQ|c\ra-{[d|\tQ|d\ra}) ([c|\tQ|c\ra (2             
    b_s^J+3)+{[d|\tQ|d\ra} (1-2 b_s^J))  
    {\cal A}_0
    \\
    +{[d|\tQ|c\ra} (([c|\tQ|c\ra (2     
    b_s^J+1)+{[d|\tQ|d\ra} (1-2 b_s^J))  
    {\cal A}_1+{[d|\tQ|c\ra} (2 b_s^J-1)
    {\cal A}_x)
    \end{pmatrix}
\end{align}

\begin{align}
C_{sp:sx}^J= \frac{\spa{c}.{B} \spb{B}.{c} \spb{c}.{B}                       
    {\cal S}_p  }{6 \spb{d}.{B}^2
    [B|P|B\ra^3 \spb{c}.{d} [B|\tQ|c\ra}
\notag \\ & \hskip -160pt\times 
    \begin{pmatrix}
    2    
    \spb{c}.{B}^2 \left(-3 \spa{c}.{B} \spb{B}.{c}                  
    [B|P|B\ra+\spa{c}.{B}^2 \spb{B}.{c}^2+3                    
    [B|P|B\ra^2\right) {\cal A}_0
    \\
    +\spa{c}.{B}
    \spb{B}.{c} \spb{d}.{B} (\spb{c}.{B} (2 \spa{c}.{B}
    \spb{B}.{c}-3 [B|P|B\ra) {\cal A}_1+2
    \spa{c}.{B} \spb{B}.{c} \spb{d}.{B}
    {\cal A}_x)
    \end{pmatrix}
\end{align}

\begin{align}
C_{sp:\g x}^J= - \frac{\spb{c}.{B} {\cal S}_p                           
     }{6 \spb{c}.{d} [B|\tQ|c\ra  
    {[d|\tQ|c\ra}^2}
    \begin{pmatrix}
    2 \left([c|\tQ|c\ra        
    {[d|\tQ|d\ra}+[c|\tQ|c\ra^2+{[d|\tQ|d\ra}^2\right)         
    {\cal A}_0
    \\
    +{[d|\tQ|c\ra} (2 {[d|\tQ|c\ra}       
    {\cal A}_x-([c|\tQ|c\ra+2 {[d|\tQ|d\ra})      
    {\cal A}_1)
    \end{pmatrix}
\end{align}

\begin{align}
C_{sp:0}^Y=  \frac{C^y_{pre}\spb{c}.{B} {\cal S}_p                           
    }{2 \spb{d}.{B}
    {[d|\tQ|c\ra}}
    \left(U_1-U_0
    \left(\frac{\spb{c}.{B}}{\spb{d}.{B}}+\frac{[c|\tQ|c\ra
+{[d|\tQ|d\ra}}{{[d|\tQ|c\ra}}\right)\right)
\end{align}

\begin{align}
C_{sp:s}^Y= \frac{C^y_{pre}\spa{c}.{B} \spb{B}.{c} {\cal S}_p           
     }{2 \spb{c}.{B}
    \spb{d}.{B}^2 {\cal P}_S^J (\spa{c}.{B}
    \spb{B}.{c}+\spa{d}.{B} \spb{B}.{d})^2}
\notag \\ &  \hskip-160pt \times
    \begin{pmatrix}
    \spb{c}.{B}^3                 
    U_0 (\spa{c}.{B} \spb{B}.{c} (2                    
    c_S^J-1)+2 \spa{d}.{B} \spb{B}.{d}                 
    (c_S^J-1))
    \\
    +\spb{d}.{B} 
    \begin{pmatrix}
    \spb{c}.{B}^2
    U_1 (\spa{c}.{B} \spb{B}.{c} (2
    c_S^J+1)+2 \spa{d}.{B} \spb{B}.{d}
    c_S^J)
    \\
    +\spb{d}.{B} (\spb{c}.{B}
    U_2 (\spa{c}.{B} \spb{B}.{c} (2
    c_S^J+3)+2 \spa{d}.{B} \spb{B}.{d}
    (c_S^J+1))
    \\
    +\spb{d}.{B} U_x
    (\spa{c}.{B} \spb{B}.{c} (2 c_S^J+5)+2
    \spa{d}.{B} \spb{B}.{d}
    (c_S^J+2)))
    \end{pmatrix}
    \end{pmatrix}
\end{align}
with
\begin{align}
c_S^J= \frac{\spb{d}.{B} {[d|\tQ|d\ra}}{\spb{c}.{B}
    {\cal P}_S^J}
\end{align}

\begin{align}
C_{sp:\g}^Y= \frac{C^y_{pre}{\cal S}_p  }{2
    {\cal P}_S^J}      
    \begin{pmatrix}
    \frac{2 [c|\tQ|c\ra {[d|\tQ|d\ra} (2                      
    c_S^J+1)                                                   
    U_0}{{[d|\tQ|c\ra}^2}-\frac{{[d|\tQ|c\ra}        
    U_x ([c|\tQ|c\ra (2                                 
    c_S^J+5)-{[d|\tQ|d\ra} (2                             
    c_S^J+3))}{([c|\tQ|c\ra-{[d|\tQ|d\ra})^2}
    \\
    -\frac
    {[c|\tQ|c\ra^2 (2 c_S^J-1)                          
    U_0}{{[d|\tQ|c\ra}^2}-\frac{[c|\tQ|c\ra (2     
    c_S^J+1)                                                   
    U_1}{{[d|\tQ|c\ra}}-\frac{{[d|\tQ|d\ra}^2 (2
    c_S^J+3)
    U_0}{{[d|\tQ|c\ra}^2}+\frac{{[d|\tQ|d\ra} (2
    c_S^J+3) U_1}{{[d|\tQ|c\ra}}-(2
    c_S^J+3) U_2
    \end{pmatrix}
\end{align}

\noindent{Finally,} from the double bonus pole piece:
\def\CpredpaJ{C_{pre:dpa}^J}

\begin{align}
\Cdp0J= -\frac{\spa{d}.{A} \spb{c}.{B} {\cal A}_{p}           
    {\cal S}_p }{2
    \spa{c}.{A} \spb{d}.{B} {[d|\tQ|c\ra}}
    \left(\frac{\spa{d}.{A}}{\spa{c}.{A}}
+\frac{\spb{c}.{B}}{\spb{d}.{B}}+\frac{[c|\tQ|c\ra}{{[d|\tQ|c\ra}}
+\frac{{[d|\tQ|d\ra}}{{[d|\tQ|c\ra}}\right)
\end{align}

\begin{align}
\CdpaJ= -\frac{\CpredpaJ}{2 ({\cal E}_S^J-1)
    ({\cal E}_Q^J-1)} 
    \left(\frac{\spa{c}.{A}
    \spb{A}.{c}}{[A|P|A\ra}-1\right)
\left(
\left( \frac{\spa{c}.{A}                                          
    \spb{A}.{c}}{[A|P|A\ra}+1 \right)
   -   
    \frac{
    2({\cal E}_S^J+{\cal E}_Q^J-2)}{({\cal E}_S^J-1
   ) ({\cal E}_Q^J-1)}   
\right)
\end{align}
with
\begin{align}
\CpredpaJ=& \frac{\spa{a}.{A} \spa{e}.{A} \spa{y}.{A} \spb{b}.{B}
    \spb{f}.{B} \spb{x}.{B} \spa{c}.{d}^3
    \spb{d}.{c}^3}{\spa{c}.{A}^4 \spb{c}.{B}^4 [c|\tQ|c\ra}
\notag \\ 
{\cal E}_S^J=&-\frac{\spa{d}.{A} \spb{d}.{B}}{\spa{c}.{A} \spb{c}.{B}}
\,\, ,\,\,
{\cal E}_Q^J=\frac{ {[d|\tQ|d\ra\spa{c}.{A}}-\spa{d}.{A}
    {[d|\tQ|c\ra}}{[c|\tQ|c\ra\spa{c}.{A}}
\end{align}

\begin{align}
\CdpaxJ= &\frac{\spa{d}.{A} \spb{A}.{d} \CpredpaJ}{6 ({\cal E}_S^J-1)^2
    (\spa{c}.{A} \spb{A}.{c}+\spa{d}.{A} \spb{A}.{d})^3
    ({\cal E}_S^J-{\cal E}_Q^J)}
    \notag \\ &\hskip -50pt \times 
    \left(
    6 \spa{c}.{A}^2
    \spb{A}.{c}^2 \left({{\cal E}_S^J}^2-3
    {\cal E}_S^J+3\right)+3 \spa{c}.{A} \spb{A}.{c}
    \spa{d}.{A} \spb{A}.{d} \left(2 {{\cal E}_S^J}^2-7
    {\cal E}_S^J+9\right)+\spa{d}.{A}^2 \spb{A}.{d}^2 \left(2
    {{\cal E}_S^J}^2-7 {\cal E}_S^J+11\right)
    \right)
\end{align}

\begin{align}
\CdpayJ=  &\frac{\spa{d}.{A} \spb{A}.{d} \CpredpaJ}{6 ({\cal E}_Q^J-1)^2
    (\spa{c}.{A} \spb{A}.{c}+\spa{d}.{A} \spb{A}.{d})^3
    ({\cal E}_Q^J-{\cal E}_S^J)}
    \notag \\ &\hskip -50pt \times     
    \left(6 \spa{c}.{A}^2
    \spb{A}.{c}^2 \left({{\cal E}_Q^J}^2-3
    {\cal E}_Q^J+3\right)+3 \spa{c}.{A} \spb{A}.{c}
    \spa{d}.{A} \spb{A}.{d} \left(2 {{\cal E}_Q^J}^2-7
    {\cal E}_Q^J+9\right)+\spa{d}.{A}^2 \spb{A}.{d}^2 \left(2
    {{\cal E}_Q^J}^2-7 {\cal E}_Q^J+11\right)\right)
\end{align}

\begin{align}
\CdpsJ= \frac{\spb{c}.{B}^2 {\cal A}_{p}                        
    {\cal S}_p }{2 \spb{d}.{B}^2
    {\cal D}_A^J {\cal G}_Q^J}\left(-\frac{2 \spa{c}.{B} \spb{B}.{c} 
    ({\cal E}_A^J+{\cal F}_Q^J)}{[B|P|B\ra}
-\frac{\spa{d}.{B}^2 \spb{B}.{d}^2}{[B|P|B\ra^2}+1\right)          
\end{align}
with
\begin{align}
{\cal D}_A^J&= \frac{\spa{c}.{A}}{\spa{d}.{A}}+\frac{\spb{d}.{B}}{\spb{c}.{B}}
\,\, ,\,\,
{\cal E}_A^J= -\frac{\spa{d}.{A} \spb{d}.{B}}{\spa{c}.{A}
    \spb{c}.{B}+\spa{d}.{A} \spb{d}.{B}}
\notag \\
{\cal F}_Q^J&=  \frac{\spb{c}.{B} \spb{d}.{B} [d|\tQ|d\ra}{\spb{c}.{d}
    (\spb{B}.{c} [B|\tQ|c\ra+\spb{B}.{d} [B|\tQ|d\ra)}
\,\, ,\,\,    
{\cal G}_Q^J=-\frac{\spb{c}.{d} (\spb{B}.{c} [B|\tQ|c\ra+\spb{B}.{d}
    [B|\tQ|d\ra)}{\spb{c}.{B}^2}
\end{align}

\begin{align}
\CdpsxJ=&\frac{\spa{c}.{B}  \spb{c}.{B}^4 {\cal A}_{p} {\cal S}_p }{6 \spb{d}.{B}^3 {[d|\tQ|d\ra}
    {\cal D}_A^J (\spa{c}.{B} \spb{B}.{c}+\spa{d}.{B}
    \spb{B}.{d})^3}
\notag \\ & \hskip -0pt \times
    \begin{pmatrix}
    \spa{c}.{B}^2 \spb{B}.{c}^2 \left(6 {{\cal E}_A^J}^2-3
    {\cal E}_A^J+2\right)+3 \spa{c}.{B} \spb{B}.{c}           
    \spa{d}.{B} \spb{B}.{d} \left(4 {{\cal E}_A^J}^2-3
    {\cal E}_A^J+2\right)
    \\
    +6 \spa{d}.{B}^2 \spb{B}.{d}^2
    \left({{\cal E}_A^J}^2-{\cal E}_A^J+1\right)
    \end{pmatrix}    
\end{align}

\begin{align}
\CdpsyJ= &\frac{\spa{c}.{B} \spb{c}.{B}^4  {\cal A}_{p} {\cal S}_p
    }{6 \spb{d}.{B}^3
    {\cal G}_Q^J (\spa{c}.{B} \spb{B}.{c}+\spa{d}.{B}
    \spb{B}.{d})^3}                    
\notag \\ & \hskip -0pt \times
    \begin{pmatrix}
    \spa{c}.{B}^2 \spb{B}.{c}^2 \left(6 {{\cal F}_Q^J}^2-3
    {\cal F}_Q^J+2\right)+3 \spa{c}.{B} \spb{B}.{c}           
    \spa{d}.{B} \spb{B}.{d} \left(4 {{\cal F}_Q^J}^2-3
    {\cal F}_Q^J+2\right)
    \\
    +6 \spa{d}.{B}^2 \spb{B}.{d}^2
    \left({{\cal F}_Q^J}^2-{\cal F}_Q^J+1\right)
    \end{pmatrix}
\end{align}

\begin{align}
\CdpgJ= \frac{([c|\tQ|c\ra-{[d|\tQ|d\ra})^2 {\cal A}_{p}
    {\cal S}_p  }{{[d|\tQ|c\ra}^2
    {\cal P}_{\cal A}^J {\cal P}_S^J}\left(([c|\tQ|c\ra-{[d|\tQ|d\ra})  
    \left(-c_A^J-b_s^J+\frac{1}{2}\right)-3   
    [c|\tQ|c\ra\right)
\end{align}
\begin{align}
\CdpgxJ=\frac{\spb{c}.{B} ([c|\tQ|c\ra-{[d|\tQ|d\ra}) {\cal A}_{p} {\cal S}_p                   
    }{\spb{c}.{d} [B|\tQ|c\ra
    {[d|\tQ|c\ra}^2 {\cal P}_{\cal A}^J}            
    \begin{pmatrix}
    \frac{3}{2} [c|\tQ|c\ra (2 c_A^J-1)           
    ([c|\tQ|c\ra-{[d|\tQ|d\ra})
    \\
    +\left({c_A^J}^2
-\frac{1}{2} c_A^J+\frac{1}{3}\right)                            
    ([c|\tQ|c\ra-{[d|\tQ|d\ra})^2+3 [c|\tQ|c\ra^2
    \end{pmatrix}    
\end{align}

\begin{align}
\CdpgzJ= \frac{\spa{d}.{A} ([c|\tQ|c\ra-{[d|\tQ|d\ra}){\cal A}_{p} {\cal S}_p                   
    }{\spa{c}.{A} [c|\tQ|c\ra
    {[d|\tQ|c\ra}^2 {\cal P}_S^J}               
    \begin{pmatrix}
    \frac{3}{2} [c|\tQ|c\ra (2 b_s^J-1)           
    ([c|\tQ|c\ra-{[d|\tQ|d\ra})
    \\
    +\left({b_s^J}^2
-\frac{1}{2} b_s^J+\frac{1}{3}\right)                            
    ([c|\tQ|c\ra-{[d|\tQ|d\ra})^2+3 [c|\tQ|c\ra^2
    \end{pmatrix}    
\end{align}

\end{document}